\documentclass[11pt,onecolumn,amssymb,nofootinbib]{revtex4}
\usepackage{amsmath, amsthm, amscd, amssymb}
\usepackage{graphicx, braket}
\usepackage{bm}
\usepackage{bbm}

\begin{document}

\title{\bf   Parameterized Adaptive Multidimensional Integration Routines (PAMIR)\break
  Localization by Repeated $2^p$ Subdivision}
\bigskip
\author{Stephen L. Adler}
\bigskip
\bigskip

\email{adler@ias.edu} \affiliation{Institute for Advanced Study,
Einstein Drive, Princeton, NJ 08540, USA.}

\maketitle
\copyright~ 2010 Stephen L. Adler\hfill\break
\tableofcontents

\section{Introduction}

This book is concerned with numerical integration in general $p$ dimensional spaces.  To
understand why special methods are needed, let us consider for the moment trapezoidal or
center-of-bin integration
on the unit interval in $p=1$ dimension.  Since these are both second order methods, to
achieve an accuracy of one part in $10^4$ one needs a division of the unit interval into roughly
100 subdivisions, with an evaluation of the integrand function at each.  This poses no
problem for numerical evaluation, but suppose instead we wish to integrate a function over a
9 dimensional region, achieving a similar accuracy of one part in $10^4$.  One then
needs 100 divisions per axis, and $(100)^9=10^{18}$ function evaluations, which is a daunting
task even for the fastest current computers.  So a brute force extension of the trapezoidal rule
(or similar higher order methods, such as Simpson's rule) is not a viable approach when the
dimension of the space $p$ is more than around four.

In consequence,  methods for high dimensional spaces have focused on adaptive algorithms, in
which function evaluations are concentrated in regions where the integrand is large and
rapidly varying.  Both Monte Carlo and deterministic algorithms have been proposed and widely
used.  Typically, they start from a base region, and then subdivide or refine on one to three or four sides
along which the integrand is most rapidly varying.  The process is then iterated, leading to
finer subdivisions and an improved estimate of the integrand.  Most authors, however, have considered it
to be computationally prohibitive to proceed at each step by dividing the base region
into $2^p$ subregions, so that the maximal length of $\it each$ side is reduced by a factor of
2 at each step. Such a subdivision would allow localization of isolated integrand peaks in
$p$ dimensions, giving a method with the potential of achieving high accuracy for
integrations in high dimensional spaces, and high resolution in applications such as template-based pattern recognition.

The motivation for this book is the observation that computer speed has dramatically increased
in recent years, while the cost of memory has simultaneously dramatically decreased; our current
laptop speeds, and memories, are characterized by ``giga'' rather than the ``mega'' of two decades
ago.  So it is now timely to address the problem of formulating practical high dimension integration
routines that proceed by $2^p$ subdivision.   We will develop methods for adaptive integration over
both general simplexes, and axis-parallel hypercubes.  Our simplex method is
based on combining  Moore's (1992) algorithm for $2^p$ subdivision of a general simplex, with new formulas for parameterized higher order integration over a general simplex that we derive using the centroid approach of Good and Gaskins (1969, 1971), to give a a fully
localizable adaptive integration procedure for general dimension $p \geq 1$.  In addition to giving a hypercube method based
on partition into simplexes, we also give a simpler, direct method for integration over hypercubes, constructed by analogy
with our methods for simplexes.  We focus specifically
on a few special base region geometries: the standard simplex (relevant for calculating Feynman
parameter integrals in physics),  the Kuhn simplex, which can be used to tile the $p$ dimensional
side 1  hypercube by symmetrization of the integrand, and the half-side 1 hypercube, which for which we
give direct algorithms which are simpler than the simplex-based algorithms.  By changes of variable, any multiple integral
with fixed limits of integration in each dimension can be converted to an integration over the
side 1 or half-side 1 hypercube. In the following sections we develop the theory
behind our methods, and then give a suite of Fortran programs, for both serial and MPI parallel
computation, implementing them.

\section{One dimensional adaptive integration}

As a simple example, let us sketch how to write an adaptive integration program in one dimension
for the integral
\begin{equation}\label{eq:onedim}
I=\int_0^1 dx f(x)~~~.
\end{equation}
A first estimate can be obtained by using  the trapezoidal rule
\begin{equation}\label{eq:trap}
I_a \simeq 0.5 [f(0)+f(1)]~~~,
\end{equation}
and a second estimate obtained by using the center-of-bin rule
\begin{equation}\label{eq:center}
I_b \simeq f(0.5) ~~~.
\end{equation}
These are both first order accurate methods, but since they are applied
to the entire interval ($0,1)$ there will be a significant error, unless
$f(x)$ happens to be a linear function over the interval.  If we want an
evaluation of the integral with an estimated error $\epsilon$, we test
whether $|I_a-I_b| < \epsilon$.  If this condition is satisfied, we output
$I_a$ and $I_b$ as estimates of the integral.  If the condition is not
satisfied, we subdivide the interval $(0,1)$ into two half-sized intervals
$(0,0.5)$ and $(0.5,1)$.  In each subinterval we follow the same procedure.  For a subinterval
with upper limit $x_U$ and lower limit $x_L$, and midpoint $x_M$, we now define
\begin{equation}\label{eq:trap1}
I_a \simeq 0.5 [f(x_U)+f(x_L)]~~~,
\end{equation}
and
\begin{equation}\label{eq:center1}
I_b \simeq f(x_M) ~~~.
\end{equation}
For the two subintervals, we evaluate the trapezoidal and
center-of-bin approximations to the integral, keeping $I_a({\rm
subinterval})$ and $I_b({\rm subinterval})$ for the subinterval,
multiplied by the subinterval width of 1/2, as contributions to the
answer if the ``thinning'' condition
\begin{equation}\label{eq:thinningcondition}
|I_a({\rm
subinterval})-I_b({\rm subinterval})|<\epsilon
\end{equation}
is met,  and subdividing the
interval by half again if this condition is not met. When, after a
sequence of subdivisions, the condition is met for all subintervals,
we have obtained good approximations to both a trapezoidal and
center-of-bin evaluation of the integral,
\begin{align}\label{eq:iab}
I_a \simeq & \sum_{\rm subintervals} L({\rm subinterval}) I_a({\rm subinterval})~~~,\cr
I_b\simeq & \sum_{\rm subintervals} L({\rm subinterval}) I_b({\rm subinterval})~~~.\cr
\end{align}
Here $L({\rm subinterval})$ is the subinterval length, and since the subintervals
are a tiling of the interval $(0,1)$, we clearly have
\begin{equation}\label{eq:lsum}
\sum_{\rm subintervals} L({\rm subinterval})=1~~~.
\end{equation}
From the difference of $I_a$ and $I_b$ we get an estimate of the
error, given by
\begin{equation}\label{eq:err1}
{\rm |outdiff|}\equiv |I_a-I_b|~~~.
\end{equation}
We can also compute the sum of the absolute values of the local
subinterval errors,
\begin{equation}\label{eq:err2}
{\rm errsum}\equiv \sum_{\rm subintervals} L({\rm subinterval}) |I_a({\rm subinterval})
-I_b({\rm subinterval})|\geq {\rm |outdiff|} ~~~.
\end{equation}
When the condition $|I_a({\rm subinterval})-I_b({\rm
 subinterval})|<\epsilon$ is met for all subintervals, then errsum reduces, using Eq. \eqref{eq:lsum}, to
\begin{equation}\label{eq:err21}
{\rm errsum} < \epsilon~~~,
\end{equation}
and if all local subinterval errors have the same sign, then we have
${\rm errsum}={\rm |outdiff|}$.

If the process of subdivision has to be stopped before the condition
$|I_a({\rm subinterval})-I_b({\rm subinterval})|<\epsilon$ is
satisfied for all subintervals, with the remaining subregion
contributions  added to $I_a$ and $I_b$ before the program
terminates, then errsum will typically be larger than $\epsilon$.
Such premature termination can happen for very irregular or singular
functions, or if the parameter $\epsilon$ is made too small, or if
one subdivides without imposing the thinning condition of Eq.
\eqref{eq:thinningcondition}. But for smooth functions $f(x)$ and
thinning with attainable $\epsilon$ the subdivision process will
terminate quite rapidly. The reason is that both the trapezoidal and
center-of-bin methods are accurate to first order with a second
order error, and so the difference $I_a({\rm subinterval})-I_b({\rm
subinterval})$ scales as $[L({\rm subinterval})]^2$ as the
subinterval length $L({\rm subinterval})$ approaches zero.

The adaptive integration method just sketched is easily programmed,
and works well.  One does not have to keep track of the relative
location of the various subintervals, only of their  starting and
ending $x$ values. Thus,  one maintains a list of active
subintervals, stored in any convenient order; when a subinterval is
divided the two resulting halves are added to the list of active
subintervals, while if a subinterval obeys the thinning condition  , its
contributions to $I_a$, $I_b$, and ${\rm errsum}$ are added to an
accumulation register, and the subinterval is removed from the list
of active subintervals.

Even faster termination is obtained if Simpson's rule or an even higher order integration
rule is used; see for example the Wikipedia article on the McKeeman (1962) adaptive Simpson rule, and
references given there.  The idea again is to compute two different evaluations of the integral over each
subinterval, giving an error estimate that is used to determine whether to ``harvest'' the result at that
level of subdivision, or to subdivide further.  In generalizing to higher dimensional integrals, the
same features persist: for each integration subregion, we evaluate a local thinning condition   obtained from the difference of two alternative higher order integration rules.  If the condition   is obeyed, that
subregion is ``harvested'' and deleted from the list of active subregions; if the condition   is not obeyed,
the subregion is further subdivided and the resulting smaller subregions are added to the active list.

\section{Generalizing to higher dimensions: simplexes and hypercubes.  Review of prior work.}

 The first question to decide in generalizing to higher dimensions is the choice of base region geometry.
 There are two natural higher dimensional analogs of the one dimensional interval $(0,1)$.  The first
 is the side 1 hypercube $(0,1)\otimes(0,1)\otimes ...\otimes (0,1)$, and the second is what we will term a {\it standard simplex} with
 vertices $(0,0,...,0),~(0,1,0,0.....0), ~(0,0,1,0,0....,0),....,(0,0,....,0,1)$.  We will also make use of the half-side 1 hypercube, spanning $(-1,1)\otimes (-1,1)\otimes ...\otimes(-1,1)$.
 These three basic regions are illustrated, in two dimensions, in Fig. 1.

\begin{figure}
\begin{center}
\includegraphics{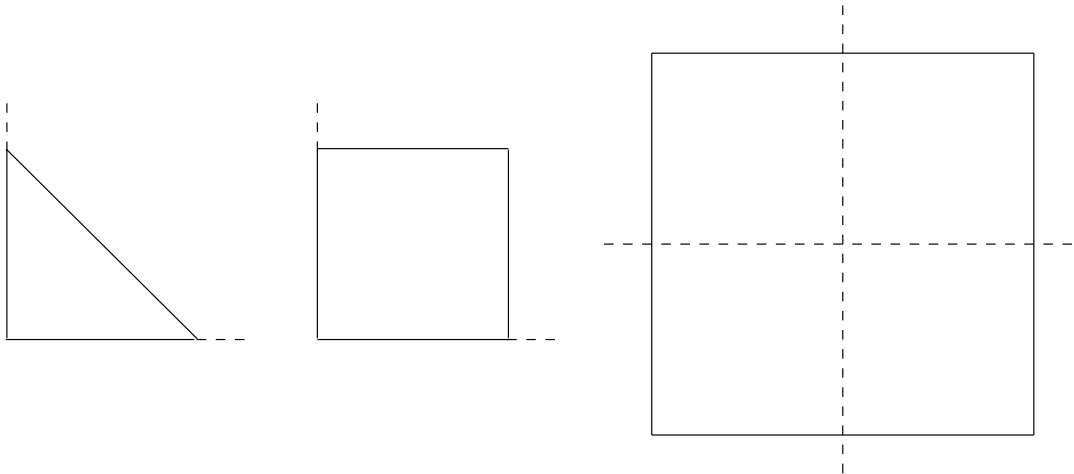}

\end{center}
\caption{ From left to right, the unit standard simplex, the side 1 hypercube, and the half-side 1 hypercube, in 2 dimensions.}
\end{figure}

 Some simple geometric facts are important in setting a strategy.
 For a $p$ dimensional simplex, the number of vertices  is $p+1$ and the number of sides connecting
 vertices is $(p+1)p/2$, both of which have polynomial growth.  Thus, the indexing problem of keeping track of
 vertices which define active regions is relatively simple.  For a $p$ dimensional hypercube, the number of
 $m$-dimensional hypercubes on the boundary (see, e.g.,  the Wikipedia article on hypercubes) is
  $2^{p-m} p\,!/(m\,! (p-m)\,!)$, and so the number of vertices ($m=0$) is $2^p$, and
 the number of sides connecting vertices ($m=1$) is $p \,2^{p-1}$,  both of which grow exponentially with $p$.  Thus, if
 one labels hypercubes in terms of their vertices or sides, an exponentially growing index is required for
 large $p$.  However, for the maximal boundary hypercube, with $m=p-1$, the number given by the above formula
 is just $2p$, which again has polynomial, in fact linear,
 growth. (For example, a square has $2\times 2=4$ lines as sides, and a cube has $2\times 3=6$ squares as faces.)
 So our direct method for hypercubes will use geometric features of the maximal boundary hypercubes for
 indexing, subdivision, and integration, closely following the methods that we develop for simplexes.

 Before getting into further details, let us first give a very brief survey of adaptive methods for higher dimensional
 integration that are currently in the literature.  A method that is widely used by physicists to evaluate Feynman
 parameter integrals is the VEGAS program of Lepage (1978), which uses a hypercube as the base geometry.  This is
 a Monte Carlo method, in which random samplings of the integration volume are done with a separable probability
 density that is a product of one dimensional densities along each axis.  This  probability density is then iterated
 to give a more detailed sampling along axes on which the projection of the integrand is rapidly varying.  A deterministic
 method of Genz and Cools (2003) is based on  simplexes as the base regions.  The algorithm picks the subregion with
 the largest estimated error, and subdivides it into up to four equal volume subregions by cutting edges along which
 the integrand is most rapidly varying.  This, and related adaptive algorithms, are discussed in the survey of CUBPACK
 by Cools and Haegemans (2003).    The CUBA set of algorithms described by Hahn (2005) includes both Monte Carlo methods
 and deterministic methods; the former include refinements of VEGAS and the latter proceed by bisection of the subregion
 with largest error.  A survey of many types of high dimensional integration algorithms, including adaptive algorithms,
 is contained in the HIntLib Manual of Sch\"urer (2008).

Most of the algorithms just described do not proceed directly to a $2^p$ subdivision of the base region (although the possibility
of $2^p$ subdivision is sketched in ``Algorithm 2'' of  Cools and Haegemans (2003)).  An algorithm in the literature which
makes  use of a $2^p$ subdivision was given by Kahaner and Wells (1979).  Unlike the algorithms which we develop below, which work directly from the vertex coordinates of a general simplex, the Kahaner Wells algorithm uses changes of variable  for both
simplex subdivision and integration.  It also rank orders the errors for each subregion (as do most of the algorithms described
in the preceding paragraph), and at each stage subdivides the subregion with the largest contribution to the total error.   While this global method may result in efficiencies in reducing the number of subdivisions needed, it makes parallelization of the algorithm more complicated, since the
computations for the different subregions are not independent of one another.  Also, when many subregions have errors of similar size, which is often the case,  the computational effort involved in rank ordering the errors may not be justified.  In the algorithms developed below, as in the one-dimensional example given in Sec. 1, we use a  local thinning condition   for the subregions, making it easy to turn serial versions of the algorithm into parallel ones. We note, however, that the subdivision and integration methods that we use could also be incorporated into global adaptive algorithms.

\section{Simplex properties and applications}

Any set of $p+1$ points in $p$ dimensional space defines a $p$-simplex, and we will be concerned with integrations over the interior
region defined this way.  Thus, in 1 dimension, 2 points define a 1-simplex that is the line segment joining them, in 2 dimensions,
3 points define a 2-simplex that is a triangle, in three dimensions, 4 points define a 3-simplex that is a tetrahedron, and so forth.
We will refer to the $p+1$ points, that each define a $p$-vector, as the vertices of the simplex, and our strategy will be to express
all operations, both for the subdivision of simplexes and for calculating approximations to integrals over simplexes, directly in
terms of these vertices.  Our convention, both here in the text and in the programs, is that the $p+1$ vertices of a simplex are enumerated from $0$ to $p$, and the $p$ vector components of each vertex are enumerated from $1$ to $p$.  Both will be denoted by subscripts; it should
be clear from context and from the range of the index whether an index is the label of a simplex vertex, as in $x_0,...,x_p$, or the
component index of a general point $x$, as in $x_1,...,x_p$.  In this notation, the $i$th component of the $j$th simplex vertex is
denoted by a double subscript $x_{ji}$.

\subsection{Simplex properties}

A simplex forms a convex set.  This means that for any integer $n\geq 1$ and any set of points $x_1,....,x_n$ lying within (or on the boundary) of a simplex,
and any set of non-negative numbers $\alpha_1,....,\alpha_n$ which sum to unity,
\begin{align} \label{eq:convex}
\alpha_j \geq 0, ~~~j=1,...,n~~~,\cr
\sum_{j=1}^n \alpha_j =1~~~,\cr
\end{align}
the point
\begin{equation}\label{eq:linsum}
x=\sum_{j=1}^n  \alpha_j x_j
\end{equation}
also lies within (or on the boundary) of the simplex (see, e.g., Osborne (2001)).

In constructing integration rules for simplexes, we will be particularly interested in linear combinations of the form of Eq.  \eqref{eq:linsum} in which the points $x_1,...,x_n$ are vertices of the simplex. For such sums, one can state a rule which determines precisely where the point $x$ lies with respect to the boundaries of the simplex. Let $x_0,x_1,...,x_p$ be the vertices
of a simplex, and let $x_c$ denote the centroid of the simplex,
\begin{equation}\label{eq:centroid}
x_c=\frac {1}{p+1} \sum_{j=0}^p x_j~~~.
\end{equation}
Let us denote by $\tilde x_j$ the coordinates of the vertices with respect to the centroid as origin,
\begin{equation}\label{eq:bary}
\tilde x_j= x_j-x_c~~~,
\end{equation}
which obey the constraint following from Eq. \eqref{eq:centroid},
\begin{equation}\label{eq:constraint}
\sum_{j=0}^p \tilde x_j=0~~~.
\end{equation}
Correspondingly, let $x$ denote a general point, and let $\tilde x=x-x_c$ denote the general point referred to the centroid as origin.
Since we are assuming that the simplex is non-degenerate, the vectors $\tilde x_j$ span a linearly independent basis for the $p$-dimensional space, and so we can always expand $\tilde x$ as a linear combination of the $\tilde x_j$,
\begin{equation}\label{eq:expan}
\tilde x = \sum_{j=0}^p \alpha_j \tilde x_j~~~.
\end{equation}

This expansion is not unique, since by Eq. \eqref{eq:constraint} we can add a constant $a$ to all of the coefficients $\alpha_j$,
without changing the sum in Eq. \eqref{eq:expan}.  In particular, we can use this freedom to put the expansion of Eq. \eqref{eq:expan}
in a standard form, which we will assume henceforth, in which the sum of the coefficients $\alpha_j$ is unity,
\begin{equation}\label{eq:unit}
\sum_{j=0}^p \alpha_j =1~~~.
\end{equation}
For coefficients (called barycentric coordinates) obeying this unit sum condition, we can use Eqs. \eqref{eq:centroid} and \eqref{eq:bary} to also write
\begin{equation}\label{eq:expan1}
x=\sum_{j=0}^p \alpha_j x_j~~~.
\end{equation}

In terms of the expansion of Eqs. \eqref{eq:expan} through \eqref{eq:expan1} we can now state a rule \big(see Pontryagin (1952) and the Wikipedia article on barycentric coordinates\big) for determining
where the point $x$ lies with respect to the simplex: (1) If all of the $\alpha_j$ are strictly positive, the point lies inside the
boundaries of the simplex; (2) If a coefficient $\alpha_j$ is zero, the point lies on the boundary plane opposite to the vertex $x_j$, and
if several of the $\alpha_j$ vanish, the point lies on the intersection of the corresponding boundary planes; (3) If any coefficient
$\alpha_j$ is negative, the point lies outside the simplex.

To derive this rule, we observe that a point $x$ lies within the simplex only if it lies on the same side of each boundary plane of
the simplex as the simplex vertex opposite that boundary.  Let us focus on one particular vertex of the simplex, which we label $x_p$, so that the
other $p$ vertices are $x_0,..., x_{p-1}$. These $p$ vertices span an affine hyperplane, which divides the $p$-dimensional space into
two disjoint parts, and constitutes the simplex boundary hyperplane  opposite the simplex vertex $x_p$.  A general parameterization of
this hyperplane takes the form
\begin{equation}\label{eq:hyperplane}
x=x_0+\sum_{j=1}^{p-1} \beta_j (x_j-x_0) ~~~,
\end{equation}
that is, we take $x_0$ as a fiducial point on the hyperplane and add arbitrary multiples of a complete basis of vectors $x_j-x_0$ in
the hyperplane.  Rewriting Eq. \eqref{eq:hyperplane} as
\begin{equation}\label{eq:hyperplane1}
x=\sum_{j=0}^{p-1} \gamma_j x_j~~~,
\end{equation}
with $\gamma_0=1-\sum_{j=1}^{p-1} \beta_j$ and $\gamma_j=\beta_j, ~~j\geq 1$, we see that the $p$ coefficients $\gamma_j$ obey the
condition
\begin{equation}\label{eq:gammasum}
\sum_{j=0}^{p-1} \gamma_j=1~~~.
\end{equation}
By virtue of this condition, we can also write the hyperplane parameterization of Eq. \eqref{eq:hyperplane1} in terms of coordinates with origin at
the simplex centroid,
\begin{equation}\label{eq:hyperplane2}
\tilde x=\sum_{j=0}^{p-1} \gamma_j \tilde x_j~~~.
\end{equation}

We now wish to determine whether the general point $\tilde x$ lies on the same side of this hyperplane as the vertex $\tilde x_p$, or lies  on the hyperplane, or lies on
the opposite side from $\tilde x_p$, by using the expansion of Eqs. \eqref{eq:expan} and \eqref{eq:unit}, which we rewrite in the form
 \begin{align}\label{eq:newexp}
\tilde x=& \sum_{j=0}^{p-1} \alpha_j \tilde x_j + \frac {\alpha_p} {p} \sum_{j=0}^{p-1} \tilde x_j \cr
-&\frac {\alpha_p} {p} \sum_{j=0}^{p-1} \tilde x_j  + \alpha_p \tilde x_p~~~.\cr
\end{align}
The first line on the right hand side of Eq. \eqref{eq:newexp} has the form of the hyperplane parameterization of Eq. \eqref{eq:hyperplane2},
since by construction the coefficients add up to unity, and so this part of the right hand side is a point on the boundary hyperplane opposite
the vertex $\tilde x_p$.  The second line on the right hand side of Eq. \eqref{eq:newexp} can be rewritten, by using Eq. \eqref{eq:constraint},
as
\begin{equation}\label{eq:second}
\alpha_p \frac {p+1} {p} \tilde x_p  ~~~.
\end{equation}
To appreciate the significance of this, we note that the centroid of the $p$ points defining the boundary hyperplane is
\begin{equation}\label{eq:hcentroid}
\tilde x_{h;\,c}= \frac{1}{p} \sum_{j=0}^{p-1} \tilde x_j = -\frac{1}{p} \tilde x_p~~~,
\end{equation}
where we have again used Eq. \eqref{eq:constraint}.  Therefore the vector from the centroid of the
points defining the hyperplane to the vertex $\tilde x_p$ is
\begin{equation}\label{eq:vecxp}
\tilde x_p - \tilde x_{h;\,c}= \frac {p+1}{p} \tilde x_p ~~~.
\end{equation}
So Eq. \eqref{eq:second} tells us that the point $\tilde x$ is displaced from the hyperplane by a vector parallel to that of Eq. \eqref{eq:vecxp}, with its length rescaled by the factor $\alpha_p$.
 Therefore, if $\alpha_p >0$, the point $\tilde x$ lies on the same side of the boundary hyperplane as the opposite vertex $\tilde x_p$.
 If $\alpha_p=0$, the point $\tilde x$ lies on  the boundary hyperplane, and if $\alpha_p<0$, the point $\tilde x$ lies on the opposite side of the
 boundary hyperplane from the vertex $\tilde x_p$.  Applying this argument to all $p+1$ vertices in turn gives the rules stated above.

 In constructing integration rules for simplexes, we will use the following elementary corollary of the result that we have just
 derived.  Consider the sum
 \begin{equation}\label{eq:gensum}
 \tilde X= \sum_{i=1}^N \lambda_i \tilde x_i~~~,
 \end{equation}
 with the coefficients $\lambda_i$ obeying
 \begin{align}\label{eq:lamsum}
 &\lambda_i >0, ~~~i=1,...,N~~~\cr
 &\sum_{i=1}^N \lambda_i < 1~~~,\cr
 \end{align}
 with the points $\tilde x_i$ any vertices of a simplex. Some vertices may be omitted, and
 some used more than once, in the sum of Eq. \eqref{eq:gensum}.  Then the point $\tilde X$ lies inside the simplex.  To see this,
 we note that by adding a positive multiple of zero in the form of Eq. \eqref{eq:constraint}, the sum of Eq. \eqref{eq:gensum} can be
 reduced to the form of Eqs. \eqref{eq:expan} and \eqref{eq:unit}, with all expansion coefficients $\alpha_j$ strictly positive.  By
 the rule stated above, this implies that the point $\tilde X$ lies within the simplex.

\subsection{Simplex applications}

Our programs for $p$-dimensional integration make special use of two kinds of simplexes, the unit ``standard simplex'' introduced above,
and the unit Kuhn simplex. In this subsection, we discuss important applications of these two special types of simplexes.

To recapitulate, the unit standard simplex has vertices given by
\begin{align}\label{eq:stdsimplex}
x_0=&(0,0,0,...,0)~~~,\cr
x_1=&(1,0,0,...,0)~~~,\cr
x_2=&(0,1,0,...,0)~~~,\cr
x_3=&(0,0,1,0,...,0)~~~,\cr
&............\cr
x_{p-1}=&(0,0,0,...,0,1,0)~~~,\cr
x_p=&(0,0,0,...,0,0,1)~~~.\cr
\end{align}
It is bounded by axis-parallel hyperplanes $x_j=0,~~j=1,...,p$ and the diagonal hyperplane $1=x_1+x_2+...+x_p$.
Thus, the integral of a function $f(x_1,...,x_p)$ over the standard simplex can be written as a multiple integral in the
form
\begin{align}\label{eq:stdint}
\int_{\rm standard ~simplex}f(x_1,...,x_p)&dx_1...dx_p = \int_0^1 dx_1 \int_0^{1-x_1} dx_2 \int_0^{1-x_1-x_2} dx_3 ....\cr \times &\int_0^{1-x_1-x_2-...-x_{p-2}}dx_{p-1}\int_0^{1-x_1-x_2-...-x_{p-1}}dx_p
f(x_1,...,x_p)~~~.
\end{align}

An important physics application of this formula is the Feynman-Schwinger  formula for combining perturbation theory denominators,
\begin{equation}\label{eq:feynschw}
\frac {1}{D_0D_1...D_p}=p\,!\int_{\rm standard ~simplex} \frac {1}{[(1-x_1-x_2-...-x_p)D_0+x_1D_1+...+x_pD_p]^{p+1}}~~~,
\end{equation}
which can be proved inductively as follows.  For $p=1$, the Feynman-Schwinger formula reads
\begin{equation}\label{eq:feynschw1}
\frac{1} {D_0D_1}=\int_0^1 dx_1 \frac {1}{[(1-x_1)D_0+x_1D_1]^2} ~~~,
\end{equation}
which is easily verified by carrying out the integral.  Assume now that this formula holds for dimension $p$.
For $p+1$, the formula asserts that
\begin{align}\label{eq:feynschw2}
\frac {1}{D_0D_1...D_pD_{p+1}}=&(p+1)\,!\int_0^1 dx_1....\int_0^{1-x_1-x_2-...-x_{p-1}}dx_p \int_0^{1-x_1-x_2-...-x_{p-1}-x_p}dx_{p+1}  \cr
\times&\frac {1}{[(1-x_1-x_2-...-x_p-x_{p+1})D_0+x_1D_1+...+x_pD_p+x_{p+1}D_{p+1}]^{p+2}}~~~.\cr
\end{align}
Carrying out the integral over $x_{p+1}$, we get
\begin{align}\label{eq:feynschw2a}
&\frac {1}{D_0D_1...D_pD_{p+1}}=p\,!\int_0^1 dx_1....\int_0^{1-x_1-x_2-...-x_{p-1}}dx_p \frac{1}{D_{p+1}-D_0} \cr
\times&\left[ \frac{1}{[(1-\sum_{i=1}^p x_i)D_0+x_1D_1+...+x_pD_p]^{p+1}}-\frac{1}{[(1-
\sum_{i=1}^p x_i)D_{p+1}+x_1D_1+...+x_pD_p]^{p+1}}\right]~~~.\cr
\end{align}
But applying the induction hypothesis for $p$ dimensions, the right hand side of this equation reduces to
\begin{equation}\label{eq:feynschw3}
\frac{1}{D_{p+1}-D_0} \left[ \frac{1}{D_0D_1....D_p}-\frac{1}{D_{p+1}D_1....D_p}\right] =\frac{1}{D_0D_1...D_pD_{p+1}}~~~,
\end{equation}
which is the result to be proved.  In the literature, numerical evaluation of Eq. \eqref{eq:feynschw} is usually
accomplished by first making
changes of variable to convert the simplex integral to an integral over a hypercube, and then using a hypercube-based program such as
VEGAS. Using the methods developed below for direct evaluation of integrals over a standard simplex in arbitrary dimensions, the
formula of Eq. \eqref{eq:feynschw} can also be integrated numerically in its original simplex form.

We next turn to the unit Kuhn (1960) simplex, which has the vertices given by
\begin{align}\label{eq:kuhnsimplex}
x_0=&(0,0,0,...,0)~~~,\cr
x_1=&(1,0,0,...,0)~~~,\cr
x_2=&(1,1,0,...,0)~~~,\cr
x_3=&(1,1,1,0,...,0)~~~,\cr
&............\cr
x_{p-1}=&(1,1,1,...,1,1,0)~~~,\cr
x_p=&(1,1,1,...,1,1,1)~~~,\cr
\end{align}
and which defines a simplex in which $1\geq x_1 \geq x_2 \geq x_3 ....\geq x_{p-1} \geq x_p$.  The unit Kuhn simplex in two dimensions
is illustrated in Fig. 2.

\begin{figure}
\begin{center}
\includegraphics{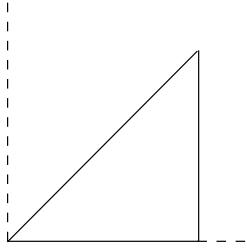}

\end{center}
\caption{ The unit Kuhn simplex in 2 dimensions.}
\end{figure}
The integral of a function $f(x_1,...,x_p)$ over a unit Kuhn simplex can be written as a multiple integral in the
form
\begin{align}\label{eq:Kuhnint}
\int_{\rm unit~ Kuhn ~simplex}f(x_1,...,x_p)&dx_1...dx_p = \int_0^1 dx_1 \int_0^{x_1} dx_2 \int_0^{x_2} dx_3 ....\cr
\times & \int_0^{x_{p-2}}dx_{p-1}\int_0^{x_{p-1}}dx_p
f(x_1,...,x_p)~~~.\cr
\end{align}

Consider now the integral of the function $f(x_1,...,x_p)$ over the unit hypercube,
\begin{equation}\label{eq:hyperintegral}
\int_0^1 dx_1\int_0^1 dx_2 ....\int_0^1 dx_{p-1}\int_0^1 dx_p  f(x_1,...,x_p)~~~.
\end{equation}
This hypercube can be partitioned  into $p\,!$ regions, each congruent to the unit Kuhn simplex, by the
requirement that in the region corresponding to the permutation $P$ of the coordinate labels  $1,...,p$,
the coordinates are ordered according to $x_{P(1)} \geq x_{P(2)}\geq x_{P(3)}....\geq x_{P(p-1)} \geq x_{P(p)}$.
This partitioning or tiling is illustrated for a square in Fig. 3, and for a cube in three dimensions in Fig. 1
of Plaza (2007).

\begin{figure}
\begin{center}
\includegraphics{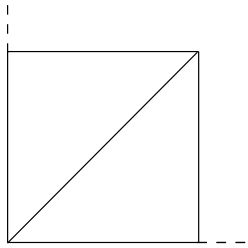}

\end{center}
\caption{Kuhn simplex tiling of a unit hypercube in 2 dimensions.}
\end{figure}
Hence the integral of $f$ over the unit hypercube is equal to the integral of the {\it symmetrized} function
computed from $f$, integrated over the unit Kuhn simplex,
\begin{align}\label{eq:symm}
\int_0^1 dx_1\int_0^1 dx_2& ....\int_0^1 dx_{p-1}\int_0^1 dx_p  f(x_1,...,x_p)\cr =&
\int_{\rm  unit~ Kuhn ~simplex}\sum_{p\,!~ {\rm permutations}~ P} f(x_{P(1)},...,x_{P(p)})dx_1...dx_p ~~~.
\end{align}
We will use this equivalence to construct adaptive programs for integration over a unit hypercube, based on
first reducing it, by symmetrization, to an integral over a unit Kuhn simplex, and then adaptively subdividing
the Kuhn simplex to reduce the integration error as needed.

\section{Simplex subdivision and properties}

\subsection{Simplex subdivision algorithms}

Two very simple algorithms for subdividing simplexes have been given in the computer graphics literature by Moore (1992).
Let us denote the vertices of the starting simplex by $x_0,...,x_p$, each of which is a $p$-vector, and from these
let us form the $p$-vectors $V(k_1,k_2)$ defined by
\begin{equation}\label{eq:vk1k2}
V(k_1,k_2)=\frac{1}{2}(x_{k_1}+x_{k_2})~,~~~k_1\,,k_2=0,...,p.~~~
\end{equation}
Thus, $V(0,0)=x_0$, $V(0,1)=(1/2)(x_0+x_1)$ and so forth, so that the vectors $V(k_1,k_2)$
consist of the original simplex vertices, together with the midpoints of the original simplex edges.
Let $k=0,...,2^p-1$ be an index  which labels the $2^p$ subsimplexes into which the original simplex is divided.
Moore then gives two algorithms, which he terms {\it recursive subdivision} and {\it symmetric subdivision}, for
determining the vertices to be assigned to the subsimplex labelled with $k$.  Both make use of the binary representation
of $k$, and of a function determined by this representation, the  bitcount function  $b(k)$, which is the number of
1 bits appearing in the binary representation of $k$.

{\it The recursive subdivision algorithm} proceeds as follows.  As the 0 vertex of the subsimplex labelled by $k$, take
the vector $V(b(k),b(k))$, that is, $k_1=k_2=b(k)$.  To get the other vertices, scan along the binary representation of
$k$ from right (the units digit) to left. For each 0 encountered, add 1 to $k_2$, and for each 1 encountered, subtract 1 from $k_1$.  The sequence of vectors $V(k_1,k_2)$ obtained this way gives the desired $p+1$ vertices of the $k$th subsimplex.

{\it The symmetric subdivision algorithm} proceeds as follows.  As the 0 vertex of the subsimplex labelled by $k$, take
the vector $V(0,b(k))$, that is, $k_1=0,\,k_2=b(k)$.  To get the other vertices, scan along the binary representation of
$k$ from right (the units digit) to left. For each 0 encountered, add 1 to $k_2$, and for each 1 encountered, add 1 to $k_1$.  The sequence
of vectors $V(k_1,k_2)$ obtained this way gives the desired $p+1$ vertices of the $k$th subsimplex.

\begin{figure}
\begin{center}
\includegraphics{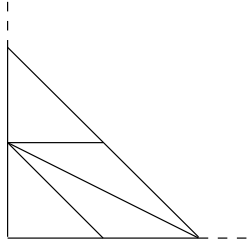}

\end{center}
\caption{Recursive subdivision of a standard simplex.}
\end{figure}

\begin{figure}
\begin{center}
\includegraphics{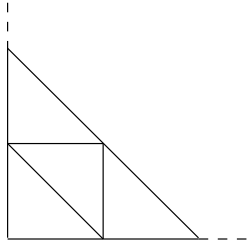}

\end{center}
\caption{Symmetric subdivision of a standard simplex.}
\end{figure}

The application of these algorithms in the $p=2$ case is illustrated
in Tables I and II and Figs. 4--7, and in the $p=3$ case is
illustrated in Tables III and IV, where the notations
$V^{(j)}(k_1,k_2)$ and $x^{(j)}$ both refer to the $j$th vertex,
$j=0,...,p$, of the subdivided simplex labelled by the $k$ in each
row.  After reviewing these tables, it should be easy to follow the
Fortran program for the algorithms given later on. The standard
Fortran library does not include a bitcount function, but it does
include a function $IBITS(I,POS,LEN)$, which gives the value of the
substring of bits of length $LEN$, starting at position $POS$, of
the argument $I$. Thus, $IBITS(k,j,1)$ gives the binary digit (0 or
1) at position $j$ in the binary representation of $k$, which is all
the information needed for the algorithm.
\begin{table} [t]
\caption{Recursive subdivision of a triangle~($p=2$)}
\centering
\begin{tabular}{c c c c c c c c}
\hline\hline
$k$ & $b(k)$ & $V^{(0)}(k_1,k_2)$ & $V^{(1)}(k_1,k_2)$ & $V^{(2)}(k_1,k_2)$&  $x^{(0)}$ & $x^{(1)}$ & $x^{(2)}$ \\
\hline
0=00 & 0 & (0,0) & (0,1) & (0,2) & $x_0$ &$\frac{1}{2}(x_0+x_1)$ & $\frac{1}{2}(x_0+x_2)$ \\
1=01 & 1 & (1,1) & (0,1) & (0,2) & $x_1$ &$\frac{1}{2}(x_0+x_1)$ & $\frac{1}{2}(x_0+x_2)$ \\
2=10 & 1 & (1,1) & (1,2) & (0,2) & $x_1$ &$\frac{1}{2}(x_1+x_2)$ & $\frac{1}{2}(x_0+x_2)$ \\
3=11 & 2 & (2,2) & (1,2) & (0,2) & $x_2$ &$\frac{1}{2}(x_1+x_2)$ & $\frac{1}{2}(x_0+x_2)$ \\
\hline
\end{tabular}
\label{table:recursive}
\end{table}
\begin{table} [t]
\caption{Symmetric subdivision of a triangle ~($p=2$)}
\centering
\begin{tabular}{c c c c c c c c}
\hline\hline
$k$ & $b(k)$ & $V^{(0)}(k_1,k_2)$ & $V^{(1)}(k_1,k_2)$ & $V^{(2)}(k_1,k_2)$&  $x^{(0)}$ & $x^{(1)}$ & $x^{(2)}$ \\
\hline
0=00 & 0 & (0,0) & (0,1) & (0,2) & $x_0$ &$\frac{1}{2}(x_0+x_1)$ & $\frac{1}{2}(x_0+x_2)$ \\
1=01 & 1 & (0,1) & (1,1) & (1,2) & $\frac{1}{2}(x_0+x_1)$ &$x_1$ & $\frac{1}{2}(x_1+x_2)$ \\
2=10 & 1 & (0,1) & (0,2) & (1,2) & $\frac{1}{2}(x_0+x_1)$ &$\frac{1}{2}(x_0+x_2)$ & $\frac{1}{2}(x_1+x_2)$ \\
3=11 & 2 & (0,2) & (1,2) & (2,2) & $\frac{1}{2}(x_0+x_2)$ &$\frac{1}{2}(x_1+x_2)$ & $x_2$ \\
\hline
\end{tabular}
\label{table:symmetric}
\end{table}
\begin{table} [t]
\caption{Recursive subdivision of a tetrahedron~($p=3$)}
\centering
\begin{tabular}{c c c c c c c c c c}
\hline\hline
$k$ & $b(k)$ & $V^{(0)}(k_1,k_2)$ & $V^{(1)}(k_1,k_2)$ & $V^{(2)}(k_1,k_2)$&$V^{(3)}(k_1,k_2)$&  $x^{(0)}$ & $x^{(1)}$ & $x^{(2)}$ &$x^{(3)}$\\
\hline
0=000 & 0 & (0,0) & (0,1) & (0,2) &(0,3)& $x_0$ &$\frac{1}{2}(x_0+x_1)$ & $\frac{1}{2}(x_0+x_2)$ &$\frac{1}{2}(x_0+x_3)$\\
1=001 & 1 & (1,1) & (0,1) & (0,2) &(0,3)& $x_1$ &$\frac{1}{2}(x_0+x_1)$ & $\frac{1}{2}(x_0+x_2)$ &$\frac{1}{2}(x_0+x_3)$\\
2=010 & 1 & (1,1) & (1,2) & (0,2) &(0,3)& $x_1$ &$\frac{1}{2}(x_1+x_2)$ & $\frac{1}{2}(x_0+x_2)$ &$\frac{1}{2}(x_0+x_3)$\\
3=011 & 2 & (2,2) & (1,2) & (0,2) &(0,3)& $x_2$ &$\frac{1}{2}(x_1+x_2)$ & $\frac{1}{2}(x_0+x_2)$ &$\frac{1}{2}(x_0+x_3)$\\
4=100 & 1 & (1,1) & (1,2) & (1,3) &(0,3)& $x_1$ &$\frac{1}{2}(x_1+x_2)$ & $\frac{1}{2}(x_1+x_3)$ &$\frac{1}{2}(x_0+x_3)$\\
5=101 & 2 & (2,2) & (1,2) & (1,3) &(0,3)& $x_2$ &$\frac{1}{2}(x_1+x_2)$ & $\frac{1}{2}(x_1+x_3)$ &$\frac{1}{2}(x_0+x_3)$\\
6=110 & 2 & (2,2) & (2,3) & (1,3) &(0,3)& $x_2$ &$\frac{1}{2}(x_2+x_3)$ & $\frac{1}{2}(x_1+x_3)$ &$\frac{1}{2}(x_0+x_3)$\\
7=111 & 3 & (3,3) & (2,3) & (1,3) &(0,3)& $x_3$ &$\frac{1}{2}(x_2+x_3)$ & $\frac{1}{2}(x_1+x_3)$ &$\frac{1}{2}(x_0+x_3)$\\

\hline
\end{tabular}
\label{table:recursivea}
\end{table}
\begin{table} [t]
\caption{Symmetric subdivision of a tetrahedron~($p=3$)}
\centering
\begin{tabular}{c c c c c c c c c c}
\hline\hline
$k$ & $b(k)$ & $V^{(0)}(k_1,k_2)$ & $V^{(1)}(k_1,k_2)$ & $V^{(2)}(k_1,k_2)$&$V^{(3)}(k_1,k_2)$&  $x^{(0)}$ & $x^{(1)}$ & $x^{(2)}$ &$x^{(3)}$\\
\hline
0=000 & 0 & (0,0) & (0,1) & (0,2) &(0,3)& $x_0$ &$\frac{1}{2}(x_0+x_1)$ & $\frac{1}{2}(x_0+x_2)$ &$\frac{1}{2}(x_0+x_3)$\\
1=001 & 1 & (0,1) & (1,1) & (1,2) &(1,3)& $\frac{1}{2}(x_0+x_1)$ &$x_1$ & $\frac{1}{2}(x_1+x_2)$ &$\frac{1}{2}(x_1+x_3)$\\
2=010 & 1 & (0,1) & (0,2) & (1,2) &(1,3)& $\frac{1}{2}(x_0+x_1)$ &$\frac{1}{2}(x_0+x_2)$ & $\frac{1}{2}(x_1+x_2)$ &$\frac{1}{2}(x_1+x_3)$\\
3=011 & 2 & (0,2) & (1,2) & (2,2) &(2,3)& $\frac{1}{2}(x_0+x_2)$ &$\frac{1}{2}(x_1+x_2)$ & $x_2$ &$\frac{1}{2}(x_2+x_3)$\\
4=100 & 1 & (0,1) & (0,2) & (0,3) &(1,3)& $\frac{1}{2}(x_0+x_1)$ &$\frac{1}{2}(x_0+x_2)$ & $\frac{1}{2}(x_0+x_3)$ &$\frac{1}{2}(x_1+x_3)$\\
5=101 & 2 & (0,2) & (1,2) & (1,3) &(2,3)& $\frac{1}{2}(x_0+x_2)$ &$\frac{1}{2}(x_1+x_2)$ & $\frac{1}{2}(x_1+x_3)$ &$\frac{1}{2}(x_2+x_3)$\\
6=110 & 2 & (0,2) & (0,3) & (1,3) &(2,3)& $\frac{1}{2}(x_0+x_2)$&$\frac{1}{2}(x_0+x_3)$ & $\frac{1}{2}(x_1+x_3)$ &$\frac{1}{2}(x_2+x_3)$\\
7=111 & 3 & (0,3) & (1,3) & (2,3) &(3,3)& $\frac{1}{2}(x_0+x_3)$ &$\frac{1}{2}(x_1+x_3)$ & $\frac{1}{2}(x_2+x_3)$ &$x_3$\\

\hline
\end{tabular}
\label{table:symmetrica}
\end{table}
\subsection{Subdivision properties}

This subdivision algorithm has a number of properties that will be useful in applying it to $p$ dimensional integration.
\begin{enumerate}

\item  As noted by Moore, the subdivided simplexes all have equal volume, equal to the initial simplex volume divided
by $2^p$.  This follows from the fact that the general formula for the volume of a simplex with vertices
$x_0,x_1,...,x_p$ is
\begin{equation}\label{eq:volume}
V=\frac{1}{p\,!}|\det(x_1-x_0,x_2-x_0,...,x_p-x_0)|~~~.
\end{equation}
Applying this to the vertices for the subdivided simplexes in Tables I-IV verifies this statement for $p=2,3$,
while a proof in the general case is given in  Edelsbrunner and Grayson (2000).

\begin{figure}
\begin{center}
\includegraphics{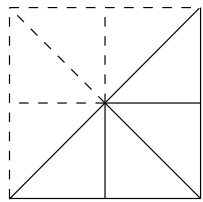}

\end{center}
\caption{Recursive subdivision of a Kuhn simplex.}
\end{figure}

\begin{figure}
\begin{center}
\includegraphics{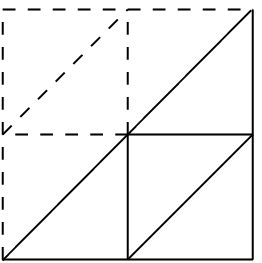}

\end{center}
\caption{Symmetric subdivision of a Kuhn simplex.}
\end{figure}

\item Again as noted by Moore, both the recursive and symmetric algorithms subdivide Kuhn simplexes into Kuhn
simplexes, which however do not all have the same orientation, as illustrated in Fig. 6 and Fig.  7.  This follows from
the fact that Kuhn simplexes are a tiling of hypercubes, which are divided into hypercubes by axis parallel planes that
intersect the midpoints of the hypercube edges.  Adding additional diagonal slices intersecting the midpoints of the hypercube edges gives
Kuhn tilings of both the original and the subdivided hypercubes.  However, as also noted by Moore, when the algorithms are applied to  general simplexes, the resultant subdivided simplexes can have different shapes, and are not isomorphic.  For $p=2$, Fig. 4 shows that recursive subdivision applied to the standard simplex leads to subsimplexes of different shapes,
while Fig. 5 shows that symmetric subdivision applied to the standard simplex leads to subsimplexes that are all standard simplexes with dimension
reduced by half.  However, an examination of the vertices in Table IV shows that already at $p=3$, symmetric subdivision of a standard
simplex does not
lead to subsimplexes that are all half size standard simplexes.  For example, for $k=2$  in Table IV, there are vertices $\frac{1}{2}(x_0+x_2)$ and
$\frac{1}{2}(x_1+x_3)$, the edge joining which has length $\sqrt{3} /2$, whereas the maximum side length of a half size $p=3$ standard simplex is $\sqrt{2}/2$.

\item  An important question is whether the maximum side length of the subdivided simplexes decreases at each stage of subdivision.
For Kuhn simplexes, the answer is immediate, since subdivision results in Kuhn simplexes of half the dimension.  Since the longest
side of a unit Kuhn simplex in dimension $p$ has length $\sqrt{p}$, after $\ell$ subdivisions the maximum side length will be
\begin{equation}\label{eq:lkuhn}
L_{\rm max}^{\rm Kuhn} =\sqrt{p}/2^{\ell}~~~,
\end{equation}
irrespective of whether recursive or symmetric subdivision is used.  For standard simplexes, we can get an upper
bound on the maximum side length by noting that a unit standard simplex on axes $y_1,...,y_p$ is obtained from a unit Kuhn simplex
on axes $x_1,...,x_p$ by the linear transformation
$y_p=x_p,\,y_{p-1}=x_{p-1}-x_p,\,y_{p-2}=x_{p-2}-x_{p-1},...,y_2=x_2-x_3,\,y_1=x_1-x_2$, since this maps the components of the Kuhn simplex
vertices given in Eq. \eqref{eq:kuhnsimplex} to the corresponding components of the standard simplex vertices given in Eq. \eqref{eq:stdsimplex}.   By linearity,  this relation also holds between vertices of corresponding
subdivided simplexes obtained from the initial unit standard and Kuhn simplexes by applying the same midpoint subdivision method (either symmetric or recursive) successively to each.  Consequently, the length $L^{\rm standard}$ of an edge with components $E_{1,...,p}^{\rm S}$ of a subdivided standard simplex can be expressed in terms of the components $E_{1,...,p}^{\rm K}$ of the corresponding edge of the
related Kuhn simplex by
\begin{align}\label{eq:correspondence}
L^{\rm standard}\equiv& [\sum_{j=1}^p (E_j^S)^2]=[\sum_{j=1}^{p-1}(E_j^K-E_{j+1}^K)^2 + (E_p^K)^2]^{\frac{1}{2}}\cr
\leq & 2 [\sum_{j=1}^p (E_j^K)^2]^{\frac{1}{2}} = 2L^{\rm Kuhn} ~~~.\cr
\end{align}
Thus the length $L^{\rm standard}$ is bounded from above by
 twice the maximum length corresponding to a subdivided Kuhn simplex, and so
\begin{equation}\label{eq:lstand}
L_{\rm max}^{\rm standard} \leq \sqrt{p}/2^{\ell-1}~~~.
\end{equation}
We have verified this inequality numerically for both the recursive and symmetric subdivision algorithms.  The numerical results
suggest that the symmetric subdivision algorithm is in fact a factor of 2 better than the bound of Eq. \eqref{eq:lstand}, so that
\begin{equation}\label{eq:lstand1}
L_{\rm max}^{\rm standard;~ symmetric} \leq \sqrt{p}/2^{\ell}~~~,
\end{equation}
but we do not have a proof of this.  We already see evidence of this difference between the symmetric and recursive algorithms in
Tables III and IV.  As noted above, from Table IV we saw that symmetric subdivision of a $p=3$ standard simplex gives an edge of length $\sqrt{3} /2$, and it is easy to see that this is the longest edge. However, from Table III we see that for $k=5$ there are vertices
$x_2$ and $\frac{1}{2}(x_1+x_3)$, the edge joining which, for an initial standard simplex, has length $\sqrt{6}/2$.
\item  The result of Eqs. \eqref{eq:lstand} and\eqref{eq:lstand1} suggests the stronger conjecture, that after any number $\ell$ of symmetric (recursive) subdivisions of a standard simplex, the resulting subsimplexes each fit within a hypercube of side $1/2^{\ell}$ ($1/2^{\ell-1}$).
A simple argument shows this to be true for $\ell=1$ in any dimension $p$. Although we do not have a proof for general $\ell$, we will use this conjecture in certain of the algorithms constructed below.  For Kuhn simplexes, an analogous statement with a
hypercube of side  $1/2^{\ell}$ is
true for both symmetric and recursive subdivision, as noted above in the discussion preceding Eq. \eqref{eq:lkuhn}.

\item Finally, we note that although the symmetric algorithm gives the same
simplex subdivision  after permutation of the starting vertices
in dimension $p=2$, as can be verified from Table II, it is not
permutation symmetric in dimension $p=3$, as can be verified
from Table IV.  For example, interchanging the labels 0 and 1 in the $k=2$ line of Table IV gives a set of vertices that is
not in the table.
This means that with symmetric (as well as
recursive) subdivision in dimension $p\geq 3$,  inequivalent
simplex subdivisions can be generated by permuting the labels of
the starting vertices. However, we have not incorporated this
feature into our programs.
\end{enumerate}

The properties just listed show that the symmetric and recursive subdivision algorithms are well suited to adaptive higher dimensional
integration.  They are easily computable in terms of the vertex coordinates for a general simplex, and give subsimplexes of equal volume, so that it is not necessary to calculate a determinant to obtain the volume. Additionally,  the bound on the maximum side length decreases as a constant times
 $1/2^{\ell}$ with increasing order of subdivision $\ell$, so that the application of high order integration formulas gives errors that decrease rapidly
with $\ell$.

\section{Hypercube subdivision and properties}

We have discussed simplexes first, because as noted in Sec. III, our direct approach to hypercube integration will be based on following
as closely as possible the methods that we develop for simplex integration.  In our direct hypercube programs (i.e., the ones not based on
tiling a side 1 hypercube with Kuhn simplexes), we will start from a half-side 1 hypercube with base region
\begin{equation}\label{eq:hyperbase}
(-1,1)\otimes (-1,1) \otimes ...\otimes(-1,1)~~~.
\end{equation}
This region has inversion symmetry around the origin, and consequently the only monomials that have non-vanishing
integrals over this region are ones in which {\it each} coordinate appears with an even exponent, considerably simplifying the calculations
needed to construct  higher order integration rules.

Since we restrict ourselves to axis-parallel hypercubes, only $p+1$ real numbers are needed to uniquely specify a hypercube:  the $p$ coordinates
of the centroid $x_c$  and the half-side length $S$.  For example, for the region of Eq. \eqref{eq:hyperbase}, the centroid is $x_c=(0,0,...,0)$
and the half-side is 1.   Once we have adopted this labelling, we can give a very simple subdivision algorithm for hypercubes, constructed in
direct analogy with Moore's simplex subdivision algorithms.

{\it  The hypercube subdivision algorithm} proceeds as follows.  Start from a hypercube with centroid $x_c$ and half-side $S$, with sides parallel to the $p$ unit axis vectors
\begin{align}\label{eq:hyperaxis}
\hat u_1=&(1,0,0,...,0)\cr
\hat u_2=&(0,1,0,...,0)\cr
&............\cr
\hat u_{p-1}=&(0,0,...,1,0)\cr
\hat u_p=&(0,0,...,0,1)~~~.\cr
\end{align}
To subdivide it into $2^p$ subhypercubes, take the new half-side as $S/2$.  To get the new centroids $x_{c;k}$, labelled by $k=0,...,2^p-1$, scan along the binary representation of
$k$ from right (the units digit) to left. Denoting the $p$ digits in this representation by $1\leq j\leq p$,  let us label the units digit as $j=1$, the
power of 2 digit as $j=2$, the power of 4 digit as $j=3$, and so forth. For all $1\leq j\leq p$,   if the  $j\,$th digit  is 0, add $\frac{1}{2}S \hat u_j$ to $x_c$, and if the $j\,$th digit  is 1, add $-\frac{1}{2}S \hat u_j$ to $x_c$.  For each given $k$, this gives the centroid of the $k$th subhypercube.  This
algorithm is illustrated for the case of a cube ($p=3$) in Table V.

This algorithm is simpler than the ones for subdividing simplexes, since it only needs the Fortran IBITS function, but does not require
subsequent computation of the bitcount function.  It evidently has properties analogous to those of the simplex subdivision algorithms: each
subhypercube has the same volume, equal to the original hypercube volume divided by $2^p$, and every linear dimension of each subhypercube is
a factor of 2 smaller than the corresponding linear dimension of the hypercube that preceded it in the subdivision chain.
This latter implies that after $\ell$ subdivisions, the resulting subhypercubes all have dimension reduced by a factor $1/2^{\ell}$.

\begin{table} [t]
\caption{Subdivision of a cube of half-side $S$ and centroid $x_c$~($p=3$)}
\centering
\begin{tabular}{c c}
\hline\hline
$k$ & $x_{c;k}-x_c$ \\
\hline
0=000 & $(S/2,S/2,S/2)$\\
1=001 & $(-S/2,S/2,S/2)$\\
2=010 & $(S/2,-S/2,S/2)$\\
3=011 & $(-S/2,-S/2,S/2)$\\
4=100 & $(S/2,S/2,-S/2)$\\
5=101 & $(-S/2,S/2,-S/2)$\\
6=110 & $(S/2,-S/2,-S/2)$\\
7=111 & $(-S/2,-S/2,-S/2)$\\

\hline
\end{tabular}
\label{table:cubedivide}
\end{table}

For a hypercube with centroid $x_c$ and half-side $S$, and for a general point $x$, let us define the coordinate relative to the centroid as
$\tilde x=x-x_c$, as we did in the simplex case in Eq. \eqref{eq:bary}.  Consider now the set of $2p$ points $\tilde x_j~,~~j=1,...,2p$ defined by
\begin{align}\label{eq:cubepoints}
\tilde x_1=&(S,0,0,...,0)\cr
\tilde x_2=&(0,S,0,...,0)\cr
&............\cr
\tilde x_p=&(0,0,...,S)\cr
\tilde x_{p+1}=&(-S,0,0,...,0)\cr
\tilde x_{p+2}=&(0,-S,0,...,0)\cr
&............\cr
\tilde x_{2p}=&(0,0,...,-S)~~~~.\cr
\end{align}
These points are the centroids of the maximal boundary hypercubes, and will play a role in the direct hypercube algorithm analogous to that
played by the simplex vertices in the simplex adaptive algorithm.  For future use, we need the following result, analogous to that of
Eqs. \eqref{eq:gensum} and \eqref{eq:lamsum} for the simplex case.    Consider the sum
\begin{equation}\label{eq:gensum1}
 \tilde X= \sum_{i=1}^N \lambda_i \tilde x_i~~~,
\end{equation}
 with the coefficients $\lambda_i$ obeying
\begin{align}\label{eq:lamsum2}
 &\lambda_i >0, ~~~i=1,...,N~~~\cr
 &\sum_{i=1}^N \lambda_i < 1~~~,\cr
 \end{align}
 with the points $\tilde x_i$ any of the hypercube boundary points of Eq. \eqref{eq:cubepoints}. Some of these points may be omitted (in which case
 the corresponding coefficient $\lambda_i$ is 0), and
 some used more than once, in the sum of Eq. \eqref{eq:gensum1}.  Then the point $\tilde X$ lies inside the hypercube.  To see this,
 we note that the projection of $\tilde X$ along any axis $j$ is of the form $\tilde X_j = S(\mu_+-\mu_-)$, with $\mu_{\pm}$ each a sum of some subset of
 the coefficients $\lambda_i$, and hence $0\leq \mu_{\pm}<1$.  Therefore $-S <  -S \mu_- \leq \tilde X_j \leq   S \mu_+  <S$ for each axis component  $\tilde X_j$, and
 thus $X$ lies within the hypercube.  This proof, again, is simpler than the corresponding result in the simplex case.

\section{Parameterized higher order integration formulas for a general simplex}

We turn next to deriving higher order integration formulas for a general simplex, which are expressed directly in terms
of the set of simplex vertices, and which involve parameters that can be changed to sample the function over the simplex in different
ways.  Two different choices of the parameters then give two different integration rules of the same order, which can be
compared to give a local error estimate for use in adaptive integration.

Since we want to derive integration rules up to ninth order in accuracy,  we start from an expansion of a general function
$f(\tilde x)$ up to ninth order, with $\tilde x$ as before the $p$ dimensional coordinate referred to the simplex centroid as
origin.  The expansion reads,
\begin{align}\label{eq:expansion}
f(\tilde x)=& A + B_{i_1}\tilde x_{i_1} + C_{i_1i_2}\tilde x_{i_1}\tilde x_{i_2} +D_{i_1i_2i_3}\tilde x_{i_1}\tilde x_{i_2}\tilde x_{i_3}
+E_{i_1i_2i_3i_4}\tilde x_{i_1}\tilde x_{i_2}\tilde x_{i_3}\tilde x_{i_4}+F_{i_1...i_5}\tilde x_{i_1}...\tilde x_{i_5}\cr
+&G_{i_1...i_6}\tilde x_{i_1}...\tilde x_{i_6}+H_{i_1...i_7}\tilde x_{i_1}...\tilde x_{i_7}
+I_{i_1...i_8}\tilde x_{i_1}...\tilde x_{i_8}+J_{i_1...i_9}\tilde x_{i_1}...\tilde x_{i_9}+...~~~.\cr
\end{align}
We next need expressions for the integral of the monomials appearing in the expansion of Eq. \eqref{eq:expansion} over a general simplex with vertices $x_0,...,x_p$.  A general formula for these integrals has been given by Good and Gaskins (1969, 1971).  They define $m(\nu)$ as the generalized moment
\begin{equation}\label{eq:moment}
m(\nu)=\int_{\rm simplex} dx_1...dx_p \tilde x_1^{\nu_1}...\tilde x_p^{\nu_p}~~~,
\end{equation}
and show that $m(\nu)$ is equal to the coefficient of $t_1^{\nu_1}...t_p^{\nu_p}$ in the expansion of
\begin{equation}\label{eq:expansion1}
\frac{V p\,! \nu_1\,!...\nu_p\,!}{(p+\nu_1+...+\nu_p)\,!} \exp[\sum_{s=2}^{\infty} \frac{1}{s}W_s]~~~.
\end{equation}
Here $W_s$ is a double sum over $i$th components of the $p+1$ simplex vertices labelled by $a=0,...,p$, given by
\begin{equation}\label{eq:wdef}
W_s=\sum_{a=0}^p [\sum_{i=1}^p \tilde x_{ai}t_i]^s ~~~,
\end{equation}
and $V$ is the simplex volume.
Good and Gaskins derive this formula by first transforming the original simplex to a standard simplex, followed by lengthy algebraic
manipulations to express the resulting formula symmetrically in terms of standard simplex vertices.  We give in Sec. VIII  below a derivation that
proceeds directly, and with manifest symmetry, from the vertices of the original simplex.

To proceed to 9th order we need an expansion of the exponential
in Eq. \eqref{eq:expansion1} through 9th order. Since each $W_s$ is of degree  $s$ in the coordinates,  the terms in this expansion are as follows:
\begin{align}\label{eq:terms}
{\rm second~ order:~~~~~}& \frac{W_2}{2}\cr
{\rm third ~order:~~~~~}& \frac{W_3}{3} \cr
{\rm fourth~order:~~~~~}& \frac{W_2^2}{8} + \frac{W_4}{4} \cr
{\rm fifth~order:~~~~~}&\frac{W_2W_3}{6}+\frac{W_5}{5} \cr
{\rm sixth~order:~~~~~}&\frac{W_2^3}{48}+\frac{W_3^2}{18}+\frac{W_2W_4}{8}+\frac{W_6}{6}\cr
{\rm seventh~order:~~~~~}&\frac{W_2^2W_3}{24}+\frac{W_3W_4}{12}+\frac{W_2W_5}{10}+\frac{W_7}{7}\cr
{\rm eighth~order:~~~~~}&\frac{W_2^4}{384}+\frac{W_2W_3^2}{36}+\frac{W_2^2W_4}{32}+\frac{W_4^2}{32}
+\frac{W_3W_5}{15}+\frac{W_2W_6}{12}+\frac{W_8}{8}\cr
{\rm ninth~order:~~~~~}&\frac{W_2^3W_3}{144}+\frac{W_3^3}{162}+\frac{W_2W_3W_4}{24}+\frac{W_2^2W_5}{40}
+\frac{W_4W_5}{20}+\frac{W_3W_6}{18}+\frac{W_2W_7}{14}+\frac{W_9}{9}~~~.\cr
\end{align}
We are interested in integrals of monomials of the form $\tilde x_{i_1}...\tilde x_{i_n}$, with $n$ ranging from 1 to 9.
Good and Gaskins note that it suffices to consider the case in which
all the indices $i_1,...,i_n$ are distinct (which is always possible for $p \geq n$), since the combinatoric factors are such
that this gives a result that also applies to the case when some of the component indices are equal, as must necessarily
be the case when $p<n$.  So we can take $\nu_i=1\,,~i=1,...,n$, and $\sum_i\nu_i=n$, with $n$ the order of the
monomial.  We now can infer from Eq. \eqref{eq:terms} the moment integrals
\begin{equation}\label{eq:moment1}
\frac{1}{V}\int_{\rm simplex} dx_1...dx_p \tilde x_{i_1}...\tilde x_{i_n}=\frac{p\,!}{(p+n)\,!}{\cal S}_n~~~,
\end{equation}
with the quantities ${\cal S}_n$ (with tensor indices suppressed) given in terms of tensors $S_{i_1...i_n}$ defined by
sums over the vertices,
\begin{equation}\label{eq:sdef}
S_{i_1...i_n}=\sum_{j=0}^p \tilde x_{ji_1}...\tilde x_{ji_n}~~~,
\end{equation}
as follows:
\begin{align}\label{eq:tensors}
{\cal S}_2=&S_{i_1i_2}\cr
{\cal S}_3=&2S_{i_1i_2i_3} \cr
{\cal S}_4=&S_{i_1i_2}S_{i_3i_4}+S_{i_1i_3}S_{i_2i_4}+S_{i_1i_4}S_{i_2i_3}+6S_{i_1i_2i_3i_4}=S_{i_1i_2}S_{i_3i_4}+2 \,{\rm terms} +6S_{i_1i_2i_3i_4}\cr
{\cal S}_5=&2(S_{i_1i_2}S_{i_3i_4i_5} + 9 \,{\rm terms})+24 S_{i_1i_2i_3i_4i_5}\cr
{\cal S}_6=&S_{i_1i_2}S_{i_3i_4}S_{i_5i_6}+ 14\,{\rm terms}+4(S_{i_1i_2i_3}S_{i_4i_5i_6}+9\,{\rm terms})
+6(S_{i_1i_2}S_{i_3i_4i_5i_6}+14\,{\rm terms})\cr
+&120 S_{i_1i_2i_3i_4i_5i_6}\cr
{\cal S}_7=&2(S_{i_1i_2}S_{i_3i_4}S_{i_5i_6i_7}+ 104 \,{\rm terms}) +12(S_{i_1i_2i_3}S_{i_4i_5i_6i_7}+ 34\, {\rm terms}) \cr
+&24(S_{i_1i_2}S_{i_3i_4i_5i_6i_7}+20\,{\rm terms})+720S_{i_1i_2i_3i_4i_5i_6i_7}\cr
{\cal S}_8=&S_{i_1i_2}S_{i_3i_4}S_{i_5i_6}S_{i_7i_8}+104\,{\rm terms}+4(S_{i_1i_2}S_{i_3i_4i_5}S_{i_6i_7i_8}+279 \, {\rm terms})\cr
+&6(S_{i_1i_2}S_{i_3i_4}S_{i_5i_6i_7i_8} + 209\,{\rm terms})\cr
+&36(S_{i_1i_2i_3i_4}S_{i_5i_6i_7i_8}+ 34 \, {\rm terms})+48(S_{i_1i_2i_3}S_{i_4i_5i_6i_7i_8} + 55\, {\rm terms})\cr
+&120(S_{i_1i_2}S_{i_3i_4i_5i_6i_7i_8}+ 27\,{\rm terms})+5040 S_{i_1i_2i_3i_4i_5i_6i_7i_8}\cr
{\cal S}_9=&2(S_{i_1i_2}S_{i_3i_4}S_{i_5i_6}S_{i_7i_8i_9} + 1259\, {\rm terms}) + 8(S_{i_1i_2i_3}S_{i_4i_5i_6}S_{i_7i_8i_9}+ 279\,{\rm terms})\cr
+&12(S_{i_1i_2}S_{i_3i_4i_5}S_{i_6i_7i_8i_9}+1259\,{\rm terms})+24(S_{i_1i_2}S_{i_3i_4}S_{i_5i_6i_7i_8i_9}+377\,{\rm terms})\cr
+&144(S_{i_1i_2i_3i_4}S_{i_5i_6i_7i_8i_9}+ 125\, {\rm terms})+240(S_{i_1i_2i_3}S_{i_4i_5i_6i_7i_8i_9}+83\,{\rm terms})\cr
+&720(S_{i_1i_2}S_{i_3i_4i_5i_6i_7i_8i_9}+35\,{\rm terms})+40320S_{i_1i_2i_3i_4i_5i_6i_7i_8i_9}~~~.\cr
\end{align}
The rule for forming terms in Eq. \eqref{eq:tensors} from those in Eq. \eqref{eq:terms} is this:  for each $W_s$ in Eq. \eqref{eq:terms}
there is a tensor factor $S$ with $s$ indices, and the product of such factors appears repeated in all nontrivial index permutations,
giving the ``terms'' not shown explicitly in Eq. \eqref{eq:tensors}.  The numerical coefficient is constructed from the denominator appearing in
Eq. \eqref{eq:terms}, multiplied by a numerator consisting of a factor $s\,!$ for each $W_s$, and for each $W_s^m$ an additional
factor $m\,!$ (that is, for $W_s^m$ there is altogether a factor $(s\,!)^m m\,!$).  For example, a $W_2^3$ in Eq. \eqref{eq:terms} gives rise to a numerator factor of $(2\,!)^3 3\,!=48$ in Eq. \eqref{eq:tensors}, and a $W_2W_3W_4$ in Eq. \eqref{eq:terms} gives rise to a numerator factor of $2\,!3\,!4\,!=288$ in Eq. \eqref{eq:tensors}.  In each case, the product of this numerator factor, times the number of terms in the symmetrized expansion, is
equal to $n\,!$.  For example, $48\times 15=720=6\,!$, and $288 \times 1260=362880=\,9!$.

Our next step is to combine Eqs. \eqref{eq:expansion}, \eqref{eq:moment1}, and \eqref{eq:tensors} to get a formula for the integral
of the function $f$ over a general simplex, expressed in terms of its expansion coefficients.  Since we will always be dealing
with symmetrized tensors, it is useful at this point to condense the notation, by labelling the contractions of the expansion
coefficients with the tensors ${\cal S}$ by the partition of $n$ which appears.  Thus, we will write
\begin{align}\label{eq:abbrev}
&C_{i_1i_2}S_{i_1i_2}=C_2 \cr
&F_{i_1i_2i_3i_4i_5}(S_{i_1i_2}S_{i_3i_4i_5}+9~{\rm terms})=F_{3+2}\cr
&H_{i_1i_2i_3i_4i_5i_6i_7}(S_{i_1i_2}S_{i_3i_4}S_{i_5i_6i_7}+104~{\rm terms})=H_{3+2+2}~~~, \cr
\end{align}
and so forth.  Since the partitions of $n$ that are relevant only involve $n\geq 2$, a complete list
of partitions that appear through ninth order is as follows:
\begin{align}\label{eq:partlist}
&C~~~2\cr
&D~~~3\cr
&E~~~4,~~2+2 \cr
&F~~~5,~~3+2 \cr
&G~~~6,~~4+2,~~2+2+2,~~3+3 \cr
&H~~~7,~~5+2,~~3+2+2,~~4+3\cr
&I~~~8,~~6+2,~~4+2+2,~~2+2+2+2,~~5+3,~~3+3+2,~~4+4\cr
&J~~~9,~~7+2,~~5+2+2,~~3+2+2+2,~~4+3+2,~~6+3,~~5+4,~~3+3+3~~~.\cr
\end{align}
Employing this condensed notation, we now get the following master formula for the integral of $f$ over a general simplex,
\begin{align}\label{eq:genint}
\frac{1}{V}\int_{\rm simplex}& dx_1...dx_p f(\tilde x)=
A+\frac{p\,!}{(p+2)\,!}C_2+\frac{p\,!}{(p+3)\,!}2D_3 \cr
+&\frac{p\,!}{(p+4)\,!}(6E_4+E_{2+2})+\frac{p\,!}{(p+5)\,!}(24F_5+2F_{3+2})\cr
+&\frac{p\,!}{(p+6)\,!}(120G_6+6G_{4+2}+4G_{3+3}+G_{2+2+2})\cr
+&\frac{p\,!}{(p+7)\,!}(720H_7+24H_{5+2}+12H_{4+3}+2H_{3+2+2})\cr
+&\frac{p\,!}{(p+8)\,!}(5040I_8+120I_{6+2}+48I_{5+3}+36I_{4+4}+6I_{4+2+2}+4I_{3+3+2}+I_{2+2+2+2})\cr
+&\frac{p\,!}{(p+9)\,!}(40320J_9+720J_{7+2}+240J_{6+3}+144J_{5+4}+24J_{5+2+2}+12J_{4+3+2}\cr
+&8J_{3+3+3}+2J_{3+2+2+2})\cr
+&...~~~.\cr
\end{align}

Our procedure is now to match this expansion to discrete sums over the function $f$ evaluated at points on the boundary or interior
of the simplex.  We will construct these sums using parameter multiples of the  vertices of the simplex (in which the summation limits for $a$,
$b$, $c$, $d$
are $0$ to $p$ for simplexes, and will be $1$ to $2p$  later on when we apply these formulas to hypercubes),
\begin{align}\label{eq:discretesums}
\Sigma_1(\lambda)=&\sum_a\, f(\lambda \tilde x_a)~,~~0\leq\lambda\leq 1 \cr
\Sigma_2(\lambda,\sigma)=&\sum_{a,b}f(\lambda \tilde x_a +\sigma \tilde x_b)~,~~0\leq\lambda,\sigma~,~~\lambda+\sigma\leq 1 \cr
\Sigma_3(\lambda,\sigma,\mu)=&\sum_{a,b,c}f(\lambda \tilde x_a+\sigma \tilde x_b+\mu \tilde x_c)~,~~0\leq\lambda,\sigma,\mu~,~~\lambda+\sigma+\mu\leq 1 \cr
\Sigma_4(\lambda,\sigma,\mu,\kappa)=&\sum_{a,b,c,d}f(\lambda \tilde x_a+\sigma \tilde x_b +\mu \tilde x_c +\kappa \tilde x_d)~,~~0\leq \lambda,\sigma,\mu,\kappa~,~~\lambda+\sigma+\mu
+\kappa \leq 1~~~, \cr
\end{align}
where the conditions on the parameters $\lambda,\sigma,\mu,\kappa$ guarantee, by our discussion of simplex properties, that the points
summed over do not lie outside the simplex.  Clearly, once we have a formula for $\Sigma_4$, we can get a formula for $\Sigma_3$ by
setting $\kappa=0$ and dividing by $p+1$ (which becomes $2p$ in the hypercube case); we can then get $\Sigma_2$ by further setting $\mu=0$ and dividing out another factor of
$p+1$, and so forth.  Hence we only exhibit here the expansion of $\Sigma_4$ in terms of $f(\tilde 0)=A$ and the contractions $C_2,...,J_{3+2+2+2}$
appearing in Eq. \eqref{eq:genint}.  Abbreviating $\xi=p+1$, we have
\begin{align}\label{eq:sumexp}
&\Sigma_4=\xi^4A+\xi^3(\lambda^2+\sigma^2+\mu^2+\kappa^2)C_2+\xi^3(\lambda^3+\sigma^3+\mu^3+\kappa^3)D_3+\xi^3(\lambda^4+\sigma^4
+\mu^4+\kappa^4)E_4 \cr
+&2\xi^2(\lambda^2\sigma^2+\lambda^2\mu^2+\lambda^2\kappa^2+\sigma^2\mu^2+\sigma^2\kappa^2+\mu^2\kappa^2)E_{2+2}+\xi^3(\lambda^5+\sigma^5+\mu^5+\kappa^5) F_5\cr
+&\xi^2(\lambda^2\sigma^3+\sigma^2\lambda^3+\lambda^2\mu^3+\mu^2\lambda^3+\lambda^2\kappa^3+\kappa^2\lambda^3
+\sigma^2\mu^3+\mu^2\sigma^3+\sigma^2\kappa^3+\kappa^2\sigma^3+\mu^2\kappa^3+\kappa^2\mu^3)F_{3+2}\cr
+&\xi^3(\lambda^6+\sigma^6+\mu^6+\kappa^6)G_6+\xi^2(\lambda^4\sigma^2+\lambda^2\sigma^4+\lambda^4\mu^2+\lambda^2\mu^4+\lambda^4\kappa^2+
\lambda^2\kappa^4+\sigma^4\mu^2+\sigma^2\mu^4\cr
+&\sigma^4\kappa^2+\sigma^2\kappa^4+\mu^4\kappa^2+\mu^2\kappa^4)G_{4+2}+6\xi(\lambda^2\sigma^2\mu^2+\lambda^2\sigma^2\kappa^2+\lambda^2
\mu^2\kappa^2+\sigma^2\mu^2\kappa^2)G_{2+2+2}\cr
+&2\xi^2(\lambda^3\sigma^3+\lambda^3\mu^3+\lambda^3\kappa^3+\sigma^3\mu^3+\sigma^3\kappa^3+\mu^3\kappa^3)G_{3+3}+\xi^3(\lambda^7+\sigma^7
+\mu^7+\kappa^7)H_7 \cr
+&\xi^2(\lambda^5\sigma^2+\lambda^2\sigma^5+\lambda^5\mu^2+\lambda^2\mu^5+\lambda^5\kappa^2+\lambda^2\kappa^5+\sigma^5\mu^2+\sigma^2\mu^5
+\sigma^5\kappa^2+\sigma^2\kappa^5+\mu^5\kappa^2+\mu^2\kappa^5)H_{5+2} \cr
+&2\xi(\lambda^3\sigma^2\mu^2+\lambda^3\sigma^2\kappa^2+\lambda^3\mu^2\kappa^2+\sigma^3\lambda^2\mu^2+\sigma^3\lambda^2\kappa^2+\sigma^3
\mu^2\kappa^2 +\mu^3\lambda^2\sigma^2+\mu^3\lambda^2\kappa^2+\mu^3\sigma^2\kappa^2\cr+&\kappa^3\lambda^2\sigma^2+\kappa^3\lambda^2\mu^2+\kappa^3
\sigma^2\mu^2)H_{3+2+2}+\xi^2(\lambda^4\sigma^3+\lambda^3\sigma^4+\lambda^4\mu^3+\lambda^3\mu^4+\lambda^4\kappa^3+\lambda^3\kappa^4
+\sigma^4\mu^3+\sigma^3\mu^4\cr
+&\sigma^4\kappa^3+\sigma^3\kappa^4+\mu^4\kappa^3+\mu^3\kappa^4)H_{4+3} +\xi^3(\lambda^8+\sigma^8+\mu^8+\kappa^8)I_8+\xi^2(\lambda^6\sigma^2
+\lambda^2\sigma^6+\lambda^6\mu^2+\lambda^2\mu^6\cr
+&\lambda^6\kappa^2+\lambda^2\kappa^6+\sigma^6\mu^2+\sigma^2\mu^6+\sigma^6\kappa^2+\sigma^2\kappa^6+\mu^6\kappa^2+\mu^2\kappa^6)I_{6+2}
+2\xi(\lambda^4\sigma^2\mu^2+\lambda^4\sigma^2\kappa^2+\lambda^4\mu^2\kappa^2\cr
+&\sigma^4\lambda^2\mu^2+\sigma^4\lambda^2\kappa^2+\sigma^4\mu^2
\kappa^2+\mu^4\lambda^2\sigma^2+\mu^4\lambda^2\kappa^2+\mu^4\sigma^2\kappa^2+\kappa^4\lambda^2\sigma^2+\kappa^4\lambda^2\mu^2+\kappa^4\sigma^2
\mu^2)I_{4+2+2}\cr
+&24\lambda^2\sigma^2\mu^2\kappa^2I_{2+2+2+2}+\xi^2(\lambda^5\sigma^3+\lambda^3\sigma^5+\lambda^5\mu^3+\lambda^3\mu^5+\lambda^5\kappa^3
+\lambda^3\kappa^5+\sigma^5\mu^3+\sigma^3\mu^5+\sigma^5\kappa^3\cr
+&\sigma^3\kappa^5+\mu^5\kappa^3+\mu^3\kappa^5)I_{5+3}+2\xi(\lambda^2\sigma^3\mu^3+\lambda^2\sigma^3\kappa^3+\lambda^2\mu^3\kappa^3
+\sigma^2\lambda^3\mu^3+\sigma^2\lambda^3\kappa^3+\sigma^2\mu^3\kappa^3 \cr
+&\mu^2\lambda^3\sigma^3+\mu^2\lambda^3\kappa^3+\mu^2\sigma^3\kappa^3+\kappa^2\lambda^3\sigma^3+\kappa^2\lambda^3\mu^3+\kappa^2\sigma^3\mu^3)
I_{3+2+2}+2\xi^2(\lambda^4\sigma^4+\lambda^4\mu^4+\lambda^4\kappa^4+\sigma^4\mu^4\cr
+&\sigma^4\kappa^4+\mu^4\kappa^4)I_{4+4}+\xi^3(\lambda^9+\sigma^9+\mu^9+\kappa^9)J_9+\xi^2(\lambda^7\sigma^2+\lambda^2\sigma^7+\lambda^7\mu^2
+\lambda^2\mu^7+\lambda^7\kappa^2+\lambda^2\kappa^7\cr
+&\sigma^7\mu^2+\sigma^2\mu^7+\sigma^7\kappa^2+\sigma^2\kappa^7+\mu^7\kappa^2+\mu^2\kappa^7)J_{7+2}+2\xi(\lambda^5\sigma^2\mu^2+\lambda^5\sigma^2\kappa^2
+\lambda^5\mu^2\kappa^2+\sigma^5\lambda^2\mu^2\cr
+&\sigma^5\lambda^2\kappa^2+\sigma^5\mu^2\kappa^2+\mu^5\lambda^2\sigma^2+\mu^5\lambda^2\kappa^2+\mu^5\sigma^2\kappa^2+\kappa^5\lambda^2\sigma^2
+\kappa^5\lambda^2\mu^2+\kappa^5\sigma^2\mu^2)J_{5+2+2}\cr
+&6(\lambda^3\sigma^2\mu^2\kappa^2+\lambda^2\sigma^3\mu^2\kappa^2+\lambda^2\sigma^2\mu^3\kappa^2+\lambda^2\sigma^2\mu^2\kappa^3)J_{3+2+2+2}
+\xi(\lambda^4\sigma^3\mu^2+\lambda^4\sigma^2\mu^3+\lambda^4\sigma^3\kappa^2\cr
+&\lambda^4\sigma^2\kappa^3+\lambda^4\mu^3\kappa^2+\lambda^4\kappa^2\mu^3+\sigma^4\lambda^3\mu^2+\sigma^4\lambda^2\mu^3+\sigma^4\lambda^3\kappa^2
+\sigma^4\lambda^2\kappa^3+\sigma^4\mu^3\kappa^2+\sigma^4\mu^2\kappa^3\cr
+&\mu^4\lambda^3\sigma^2+\mu^4\lambda^2\sigma^3+\mu^4\lambda^3\kappa^2+\mu^4\lambda^2\kappa^3+\mu^4\sigma^3\kappa^2+\mu^4\sigma^2\kappa^3
+\kappa^4\lambda^3\sigma^2+\kappa^4\lambda^2\sigma^3+\kappa^4\lambda^3\mu^2\cr
+&\kappa^4\lambda^2\mu^3+\kappa^4\sigma^3\mu^2+\kappa^4\sigma^2\mu^3)J_{4+3+2}+\xi^2(\lambda^6\sigma^3+\lambda^3\sigma^6+\lambda^6\mu^3+\lambda^3\mu^6
+\lambda^6\kappa^3+\lambda^3\kappa^6+\sigma^6\mu^3+\sigma^3\mu^6\cr
+&\sigma^6\kappa^3+\sigma^3\kappa^6+\mu^6\kappa^3+\mu^3\kappa^6)J_{6+3}+\xi^2(\lambda^5\sigma^4+\lambda^4\sigma^5+\lambda^5\mu^4+\lambda^4\mu^5
+\lambda^5\kappa^4+\lambda^4\kappa^5+\sigma^5\mu^4+\sigma^4\mu^5\cr
+&\sigma^5\kappa^4+\sigma^4\kappa^5+\mu^5\kappa^4+\mu^4\kappa^5)J_{5+4}+6\xi(\lambda^3\sigma^3\mu^3+\lambda^3\sigma^3\kappa^3+\lambda^3\kappa^3\mu^3
+\sigma^3\mu^3\kappa^3)J_{3+3+3}~~~.\cr
\end{align}

 Evidently, for simplexes Eq. \eqref{eq:discretesums} requires $(p+1)^4$ function evaluations to compute $\Sigma_4$, $(p+1)^3$ function evaluations to compute  $\Sigma_3$, etc.  For hypercubes, with the simplex vertices replaced by the $2p$ points of Eq. \eqref{eq:cubepoints}, and $\xi=2p$,  Eq. \eqref{eq:discretesums} requires $(2p)^4$ function evaluations to compute $\Sigma_4$, $(2p)^3$ function evaluations to compute $\Sigma_3$, etc.
Since it is known that the minimal number of function evaluations for a simplex integration method of order $2t+1$ involves $p^{\,t}/t\,!+O(p^{\,t-1})$
function calls (Stroud (1971), Grundmann and M\"oller (1978)), and for a hypercube integration method of order $2t+1$ involves $(2p)^{\,t}/t\,!+
O(p^{\,t-1})$ function calls (Lyness, 1965), we will take the leading $\Sigma$s in our integration formulas to have equal arguments, e.g. $\Sigma_4(\lambda,\lambda,\lambda,\lambda)$,~$\Sigma_3(\lambda,\lambda,\lambda,\lambda)$, etc.  This  allows the parameterized integration formulas
constructed below
to have an optimal leading order power dependence on the space dimension $p$ (but reflecting the parameter freedom, the non-leading power terms
will not in general be minimal).
In the computer program, the following formulas are useful in evaluating the sums using a minimum number of function calls,
\begin{align}\label{eq:symform}
\Sigma_4(\lambda,\lambda,\lambda,\lambda)=&24\sum_{a<b<c<d}f(\lambda(\tilde x_a+\tilde x_b+\tilde x_c+\tilde x_d)) + 12 \sum_a\sum_{b\neq a,\,c\neq a,\,b<c}f(2\lambda \tilde x_a+\lambda
(\tilde x_b+\tilde x_c))\cr
+&6\sum_a \sum_{b<a}f(2\lambda(\tilde x_a+\tilde x_b))+4\sum_a\sum_{b\neq a}f(3\lambda \tilde x_a+\lambda \tilde x_b) +\sum_a f(4 \lambda \tilde x_a)~~~,\cr
\Sigma_3(\lambda,\lambda,\lambda)=&6\sum_{a<b<c}f(\lambda(\tilde x_a+\tilde x_b+\tilde x_c))+3\sum_a \sum_{b\neq a}f(2\lambda \tilde x_a +\lambda \tilde x_b) + \sum_a f(3\lambda \tilde x_a) ~~~,\cr
\Sigma_2(\lambda,\lambda)=&2\sum_a \sum_{b<a} f(\lambda(\tilde x_a+\tilde x_b))+\sum_a f(2 \lambda \tilde x_a) ~~~,\cr
\Sigma_3(2\lambda,\lambda,\lambda)=&2\sum_a\sum_{b\neq a,\,c\neq a,\,b<c}f(2\lambda \tilde x_a+\lambda(\tilde x_b+\tilde x_c))+2\sum_a\sum_{b\neq a}f(3\lambda \tilde x_a +
\lambda \tilde x_b)\cr
+&2 \sum_a\sum_{b<a}f(2\lambda(\tilde x_a+\tilde x_b))+\sum_a f(4\lambda \tilde x_a)~~~,\cr
\Sigma_2(3\lambda,\lambda)=&\sum_a\sum_{b\neq a}f(3\lambda \tilde x_a+\lambda \tilde x_b)+\sum_a f(4\lambda \tilde x_a)~~~,\cr
\Sigma_2(2\lambda,\lambda)=&\sum_a\sum_{b\neq a} f(2\lambda \tilde x_a+\lambda \tilde x_b) +\sum_a f(3\lambda \tilde x_a)~~~.\cr
\end{align}

With these preliminaries in hand, we are now ready to set up integration formulas of first through fourth, fifth, seventh, and ninth order, for integrals over general simplexes.

\subsection{First through third order formulas}

We begin here with integration formulas of first through third order, which
may be more useful than high order formulas for integrating functions that
are highly irregular, or as explained later on, for integrations in low
dimensional spaces.   Two different first order accurate estimates of the integral
of Eq. \eqref{eq:genint} are clearly
\begin{align}\label{eq:first1}
I_a=&\Sigma_1(\lambda)/\xi=A+{\rm second~ order}~~~,\cr
I_b=&f(\tilde 0)=A~~~,\cr
\end{align}
with $\tilde x=\tilde 0$ the simplex centroid.  Evidently
$I_b$ is the dimension $p$ analog of the dimension one center-of-bin rule,
and when $\lambda =1$, $I_a$ is the dimension $p$ analog of the dimension
one trapezoidal rule.

To get a second order accurate formula, we have to match the terms
\begin{equation}\label{eq:second1}
A+\frac{p\,!}{(p+2)!}  C_2
\end{equation}
 in Eq. \eqref{eq:genint}.  Solving $\Sigma_1(\lambda)=\xi A + \lambda^2 C_2+...$
 for $C_2$, we get
 \begin{equation}\label{eq:second2}
 C_2 \simeq [\Sigma_1(\lambda)-\xi A]/\lambda^2~~~,
 \end{equation}
 which when substituted into Eq. \eqref{eq:second1} gives the
 second order accurate formula
 \begin{equation}\label{eq:second3}
 I=\left(1-\frac{p!}{(p+2)!}\frac{\xi}{\lambda^2}\right) f(\tilde 0)
 +\frac{p\,!}{(p+2)!}\frac{1}{\lambda^2}\Sigma_1(\lambda)~~~.
 \end{equation}
 Using two different parameter values $\lambda_{a,b}$ gives two
 different second order accurate estimates $I_{a,b}$  of the integral.

 We give two different methods of getting a third order accurate formula,
 both of which will play a role in the methods for getting higher odd
 order formulas.  We first note that for $\lambda=2/(p+3)$, we have
 \begin{equation}\label{eq:third1}
 \Sigma_1(\lambda)=\xi A+ \lambda^2[C_2 + 2 D_3/(p+3)]~~~,
 \end{equation}
 and so the coefficients of $C_2$ and $D_3$ are in the same ratio
 as appears in Eq. \eqref{eq:genint}.  Hence defining an overall
 multiplicative factor $\kappa_1$ to make both terms match in
 magnitude, and adding a multiple $\kappa_0$ of $A$ to make this
 term match, we get a third order accurate formula
 \begin{align}\label{eq:third2}
 I_a=&\kappa_1 \Sigma_1(\lambda) + \kappa_0 f(\tilde 0)~~~,\cr
 \kappa_1=&\frac{p\,!}{(p+2)!}\lambda^{-2}=\frac{(p+3)^2}{4(p+2)(p+1)}~~~,\cr
 \kappa_0=&1-\xi\kappa_1 ~~~.\cr
 \end{align}

 An alternative method of getting a third order accurate formula is to
 look for a match by writing
 \begin{align}\label{eq:third3}
 I_b=&\bar \kappa_0 f(\tilde 0) + \sum_{i=1}^2 \kappa_1^i \Sigma_1(\lambda_1^i) \cr
 =&\bar \kappa_0 A +\sum_{i=1}^2 \kappa_1^i[\xi A +(\lambda_1^i)^2 C_2 +(\lambda_1^i)^3 D_3]\cr
 =&A +\frac{p\,!}{(p+2)!}C_2+\frac{2p\,!}{(p+3)!}D_3+...~~~.\cr
 \end{align}
 Equating coefficients of $A$ we get
 \begin{equation}\label{eq:third4}
 \bar \kappa_0=1-\xi \sum_{i=1}^2 \kappa_1^i~~~,
 \end{equation}
 while equating coefficients of $C_2$ and $D_3$, we obtain
 a system of two simultaneous equations for $\kappa_1^i\,,~i=1,2$ ,
 \begin{align}\label{eq:third5}
 q_1=&w_1+w_2~~~,\cr
 q_2=&\lambda_1^1 w_1+\lambda_1^2 w_2~~~,\cr
 \end{align}
 where we have abbreviated
 \begin{align}\label{eq:third6}
 q_1=&\frac{p\,!}{(p+2)!}~,~~~q_2=\frac{2p\,!}{(p+3)!}~~~,\cr
 w_i=&\kappa_1^i (\lambda_1^i)^2~,~~~i=1,2~~~~.\cr
 \end{align}
 This set of equations can be immediately solved to give
 \begin{align}\label{eq:third7}
 w_1=&\frac{\lambda_1^2 q_1-q_2}{\lambda_1^2-\lambda_1^1}~~~,\cr
 w_2=&\frac{\lambda_1^1 q_1-q_2}{\lambda_1^1-\lambda_1^2}~~~,\cr
 \end{align}
 giving a second third order accurate formula for any nondegenerate
 $\lambda_1^1$ and $\lambda_1^2$ lying in the interval (0,1).
 We will see later on, in discussing higher odd order integration
 formulas, that this is our first encounter with a Vandermonde system
 of equations.

\subsection{Fourth order formula}

Although we will subsequently focus on odd-order formulas, we next derive a fourth order formula, which follows a different pattern.  Referring to Eq.
\eqref{eq:genint}, we see that to get a fourth order formula we have to use the sums of Eq. \eqref{eq:discretesums} to match the coefficients
of $A$, $C_2$, $D_3$, $E_4$, and $E_{2+2}$.  Since only the final one of these, $E_{2+2}$, involves two partitions of 4, we can use $\Sigma_2(\lambda,
\lambda)$ to extract this, with any positive value of $\lambda \leq \frac{1}{2}$.  Since  the simplex subdivision algorithm  uses the midpoints
$\frac{1}{2}(x_a+x_b)$ as the vertices of the subdivided simplex, an efficient way to proceed in this case is to take $\lambda=\frac{1}{2}$ in
$\Sigma_2(\lambda,\lambda)$, so that what is needed is the function value at the midpoints, and to compute these function values as part of
the subdivision algorithm.  This also yields the function values at the vertices of the subdivided simplex. We can get $A$ from $f(\tilde x_c)$, and we can fit $C_2$, $D_3$, and $E_4$ by evaluating
$\Sigma_1(\lambda)$ with three distinct values of $\lambda$.  One of these values can be taken as $\lambda=1$, corresponding to the function values
at the simplex vertices.   The other two are free parameters, and by making two different choices for one of these, we get two different fourth order evaluations of the integral.

We worked out the fourth order program before proceeding systematically to the odd order cases, and so used a different notation from that
of Eq. \eqref{eq:discretesums}.  Let us write  $f_c$, $f_v$, and $f_s$ for the sums of function values at the centroid, the vertices, and the side
midpoints,
\begin{align}\label{eq:fourthdefs}
f_c=&f(\tilde x_c)~~~,\cr
f_v=&\frac{1}{p+1}\sum_a f(\tilde x_a)~~~,\cr
f_s=&\frac{1}{(p+1)p}\sum_{a\neq b}f(\frac{1}{2}(\tilde x_a+\tilde x_b))~~~.\cr
\end{align}
Let us also introduce, for $n>m$, the definition
\begin{equation}\label{eq:pdef}
p_{\,nm} \equiv (p+m)(p+m+1)...(p+n)~~~.
\end{equation}
 A simple calculation then shows that through fourth order terms, we have
\begin{align}\label{eq:fourth1}
\frac{1}{V}\int_{\rm simplex}& dx_1...dx_p f(\tilde x)-\frac{8p}{p_{42}}(f_s+\frac{1}{p}f_v) \cr
=&k_0 A + k_2 C_2 + k_3 D_3+k_4 E_4~~~,\cr
\end{align}
with  coefficients given by
\begin{align}
k_0=&1-\frac{8(p+1)}{p_{42}}~~~,\cr
k_2=&\frac{1}{p_{21}}-\frac{4}{p_{42}}~~~,\cr
k_3=&\frac{2}{p_{31}}-\frac{2}{p_{42}}~~~,\cr
k_4=&\frac{6}{p_{41}}-\frac{1}{p_{42}}~~~.\cr
\end{align}
Defining now
\begin{equation}\label{eq:flam}
f_{\lambda}=\frac{1}{p+1}\sum_af(\lambda \tilde x_a)~~~,
\end{equation}
so that $f_{1}=f_v$, we find that through fourth order,
\begin{align}\label{eq:finalfourth}
E_4=&\frac{1}{\lambda_1-\lambda_2}(t_1-t_2),\cr
t_j=&\frac{p+1}{1-\lambda_j}[f_v-f_c-\frac{1}{\lambda_j^2}(f_{\lambda_j}-f_c)]~,~~j=1,2~~~~,\cr
D_3=&\frac{1}{2}\sum_{j=1}^{2}[t_j-(\lambda_j+1)E_4]~~~,\cr
C_2=&(p+1)(f_v-f_c)-D_3-E_4~~~.\cr
\end{align}
When substituted into Eq. \eqref{eq:fourth1}, this gives a fourth order formula for the integral, with a second
evaluation of the integral obtained by replacing $\lambda_2$ by a third, distinct value $\lambda_3$.

\subsection{Fifth order formula}

We turn next to deriving a fifth order formula.   Referring to Eq.
\eqref{eq:genint}, we see that to get a fifth order formula we have to use the sums of Eq. \eqref{eq:discretesums} to match the coefficients
of $A$, $C_2$, $D_3$, $E_4$, $E_{2+2}$, $F_5$, and $F_{3+2}$.  Since at most two partitions appear, we can still get the leading two-partition terms from $\Sigma_2(\lambda,\lambda)$, but we must now impose a condition on $\lambda$ to guarantee that $E_{2+2}$ and $F_{3+2}$ appear with coefficients
in the correct ratio.  From Eq. \eqref{eq:genint} we see that the ratio of the coefficient of $F_{3+2}$ to that of $E_{2+2}$ must be
$2/(p+5)$, and from Eqs. \eqref{eq:discretesums} and \eqref{eq:sumexp} with $\lambda=\sigma$ and $\mu=\kappa=0$, we see that this is obtained with
\begin{equation}\label{eq:fixlambda5}
\lambda=\frac{2}{p+5}~~~,
\end{equation}
which for any $p\geq 1$ obeys the condition $2\lambda < 1$.  The overall coefficient of $\Sigma_2$ needed to fit $E_{2+2}$ and
$F_{3+2}$ is easily seen to be
\begin{equation}\label{eq:kappa2}
\kappa_2=\frac{p\,!}{2\,!(p+4)\,!}\lambda^{-4}=\frac{(p+5)^4}{32 p_{\,41}}~~~,
\end{equation}
where we have used the abbreviated notation of Eq. \eqref{eq:pdef}.  Thus we have, again from Eq. \eqref{eq:sumexp},
\begin{equation}\label{eq:fifth1}
\kappa_2 \Sigma_2(\lambda,\lambda)=\frac{p\,!}{(p+4)\,!}E_{2+2}+2\frac{p\,!}{(p+5)\,!}F_{3+2}
+\kappa_2[\xi^2 A + \xi(2\lambda^2 C_2 +2 \lambda^3 D_3 + 2\lambda^4 E_4 + 2 \lambda^5 F_5)]~~~,
\end{equation}
with $\xi = p+1$.  Since there are four single partition terms, we look for an integration  formula of the form
\begin{equation}\label{eq:fifth2}
\kappa_2\Sigma_2(\lambda,\lambda)+\sum_{i=1}^4 \kappa_1^i\Sigma_1(\lambda_1^i)  + \kappa_0 A~~~,
\end{equation}
which is to be equated to the sum of terms through fifth order in Eq. \eqref{eq:genint}.

The equation for matching the coefficient of $A$ can immediately be solved in terms of the coefficients $\kappa_1^i$, giving
\begin{equation}\label{eq:kappa0fifth}
\kappa_0=1-R_0~,~~~R_0=\xi^2 \kappa_2+\xi\sum_{i=1}^4 \kappa_1^i~~~.
\end{equation}
 The four equations for matching the coefficients of $C_2$, $D_3$, $E_4$, and $F_5$ give a $N=4$ Vandermonde system that determines
 the four coefficients $\kappa_1^i$.  Writing an order $N$ Vandermonde system in the standard form
 \begin{equation}\label{eq:vandermonde}
 \sum_{i=1}^N x_i^{k-1} w_i = q_k~,~~~k=1,...,N~~~~,
 \end{equation}
 the equations determining the $\kappa_1^i$ take this form with
 \begin{align}\label{eq:kappa1}
 x_i=&\lambda_1^i~,~~~~w_i=\kappa_1^i (\lambda_1^i)^2~~~~,\cr
 q_1=&\frac{1}{p_{21}}-2\xi\kappa_2\lambda^2 ~~~,\cr
 q_2=&\frac{2}{p_{31}}-2\xi\kappa_2\lambda^3~~~,\cr
 q_3=&\frac{6}{p_{41}}-2\xi\kappa_2\lambda^4~~~,\cr
 q_4=&\frac{24}{p_{51}}-2\xi\kappa_2\lambda^5~~~.\cr
 \end{align}
 Solving this system of linear equations, for any  nondegenerate values of the parameters $0<\lambda_1^i<1$, gives the coefficients
 $\kappa_1^i$ and completes specification of the integration formula.

\subsection{Vandermonde solvers}

Since we will repeatedly encounter Vandermonde equations in setting up parameterized higher order integration formulas,
both for simplexes and for hypercubes, we digress at this point to discuss methods of solving a Vandermonde system.  The
explicit inversion of the Vandermonde system is well known (see, e.g. Neagoe (1996), Heinen and Niederjohn (1997)), and takes the form
\begin{equation}\label{eq:vandersoln}
w_1=\frac{q_N-S_1(x_2,...,x_N)q_{N-1}+S_2(x_2,...,x_N)q_{N-2}-...+(-1)^{N-1}x_2....x_N q_1}{(x_1-x_2)(x_1-x_3)....(x_1-x_N)}~~~,
\end{equation}
with  $S_j(x_2,...,x_N)$ the sum of $j$-fold products of $x_2,...,x_N$,
\begin{align}\label{eq:sdef1}
S_1(x_2,...,x_N)=&x_2+...+x_N~~~,\cr
S_2(x_2,...,x_N)=&x_2x_3+...+x_2x_N+x_3x_4+...+x_3x_N+...+x_{N-1}x_N~~~,\cr
\end{align}
and so forth.  The remaining unknowns $w_2$ through $w_N$ are obtained from this formula by cyclic permutation of the indices $i=1,...,N$ on the $w_i$ and the $x_i$, with the $q_k$ held fixed.  For $N$ not too large it is straightforward to program this solution, and we include subroutines for the
$N=2,3,4,6,8$ cases in the programs.  This suffices to solve the Vandermonde equations appearing in the fifth through ninth order simplex formulas, and in the
fifth through ninth order hypercube formulas derived below.

For large $N$, programming the explicit solution becomes inefficient and a better procedure is to use a compact algorithm for solving the Vandermonde equations
for general $N$, based on polynomial operations, which has running time proportional to $N^{\,2}$.   A good method of this type, that we have tested,
is the algorithm vander.for given in the book {\it Numerical Recipes in Fortran}
by Press et al. (1992). A similar algorithm for inverting the Vandermonde matrix (that we have not tested)  can be found in an on-line paper of Dejnakarintra
and Banjerdpongchai, searchable under the title ``An Algorithm for Computing the Analytical Inverse of the Vandermonde Matrix''.

 \subsection{Seventh order formula}

To get a seventh order formula, we use the sums of Eq. \eqref{eq:discretesums} to match the coefficients appearing in Eq.
\eqref{eq:genint} through the term $H_{3+2+2}$.  Since at most three partitions appear, we can get the leading three-partition terms $G_{2+2+2}$
and $H_{3+2+2}$ from $\Sigma_3(\lambda,\lambda,\lambda)$ by imposing the condition
\begin{equation}\label{eq:fixlambda7}
\lambda=\frac{2}{p+7}~~~,
\end{equation}
which guarantees that their coefficients are in the correct ratio, and which for any $p\geq 1$ obeys the condition $3\lambda < 1$.  The overall
coefficient of $\Sigma_3$ needed to fit $G_{2+2+2}$ and $H_{3+2+2}$ is
\begin{equation}\label{eq:kappa3}
\kappa_3=\frac{p\,!}{3\,!(p+6)\,!}\lambda^{-6}=\frac{(p+7)^6}{384 p_{\,61}}~~~.
\end{equation}
We now look for an integration formula of the form
\begin{equation}\label{eq:seventh1}
\kappa_3\Sigma_3(\lambda,\lambda,\lambda)+ \kappa_2' \Sigma_2(2\lambda,\lambda) +\sum_{i=1}^U\kappa_2^i\Sigma_2(\lambda_2^i,\lambda_2^i)
+\sum_{i=1}^6\kappa_1^i\Sigma_1(\lambda_1^i)+\kappa_0 A~~~,
\end{equation}
with $U\leq 6$ since there are 6 two-partition terms to be matched.  Equating coefficients of the two-partition terms, we find that
the equations for $G_{4+2}-G_{3+3}$ and $H_{5+2}-H_{4+3}$ are both automatically satisfied by taking
\begin{equation}\label{eq:kappa2prime}
\kappa_2'=3\kappa_3~~~.
\end{equation}
This leaves only the two-partition terms $E_{2+2}$, $F_{3+2}$, $G_{4+2}$, and $H_{5+2}$ to be matched, so we can take the upper limit
in the $\Sigma_2$ summation as $U=4$.  The four coefficients $\kappa_2^i$ are then determined by solving an $N=4$ Vandermonde system with
inhomogeneous terms $q2_i~,~~~i=1,...,4$,
\begin{align}\label{eq:kappa2a}
x_i=&\lambda_2^i~,~~~w_i=2\kappa_2^i(\lambda_2^i)^4~~~,\cr
q2_1=&\frac{1}{p_{41}}-(6\xi+24)\kappa_3\lambda^4~~~,\cr
q2_2=&\frac{2}{p_{51}}-(6\xi+36)\kappa_3\lambda^5~~~,\cr
q2_3=&\frac{6}{p_{61}}-(6\xi+60)\kappa_3\lambda^6~~~,\cr
q2_4=&\frac{24}{p_{71}}-(6\xi+108)\kappa_3\lambda^7~~~.\cr
\end{align}

We next have to match the 6 single partition terms, using  $\Sigma_1$ sums.  To save function calls, we take four of the parameters
$\lambda_1^i$ to be equal to $2\lambda_2^i$, with the other two $\lambda_1^i$ taken as new, independent parameters.  Equating the coefficients
of the single partition terms $C_2$ through $H_7$ then gives a $N=6$ Vandermonde system determining the coefficients $\kappa_1^i$, with
inhomogeneous terms $q1_i~,~~~i=1,...,6$,
\begin{align}\label{eq:kappa1a}
x_i=&\lambda_1^i~,~~~w_i=\kappa_1^i(\lambda_1^i)^2~~~,\cr
q1_1=&\frac{1}{p_{21}}-2\xi \sum_{i=1}^4 \kappa_2^i (\lambda_2^i)^2-(3\xi^2+15\xi)\lambda^2\kappa_3~~~,\cr
q1_2=&\frac{2}{p_{31}}-2\xi \sum_{i=1}^4 \kappa_2^i (\lambda_2^i)^3-(3\xi^2+27\xi)\lambda^3\kappa_3~~~,\cr
q1_3=&\frac{6}{p_{41}}-\xi q2_1 - (3\xi^2+51\xi)\lambda^4\kappa_3~~~,\cr
q1_4=&\frac{24}{p_{51}}-\xi q2_2-(3\xi^2+99\xi)\lambda^5\kappa_3~~~,\cr
q1_5=&\frac{120}{p_{61}}-\xi q2_3-(3\xi^2+195\xi)\lambda^6\kappa_3~~~,\cr
q1_6=&\frac{720}{p_{71}}-\xi q2_4-(3\xi^2+387\xi)\lambda^7\kappa_3~~~.\cr
\end{align}
Note that in $q1_3,...,q1_6$, the sums $2 \sum_{i=1}^4\kappa_2^i (\lambda_2^i)^j~,~~j=4,...,7$ have been eliminated in terms
of $q2_1,...,q2_4$ by using the Vandermonde system of Eq. \eqref{eq:kappa2a}.
Finally, matching the coefficient of $A$ we get, using Eq. \eqref{eq:kappa2prime}
\begin{equation}\label{eq:kappa0seventh}
\kappa_0=1-R_0~,~~~R_0=(\xi^3+3\xi^2)\kappa_3+\xi^2\sum_{i=1}^4 \kappa_2^i + \xi \sum_{i=1}^6 \kappa_1^i~~~.
\end{equation}

\subsection{Ninth order formula}

To get a ninth order formula, we use the sums of Eq. \eqref{eq:discretesums} to match the coefficients appearing in Eq.
\eqref{eq:genint} through the final exhibited term $J_{3+2+2+2}$.  Since at most four partitions appear, we can get the leading four-partition terms $J_{3+2+2+2}$
and $I_{2+2+2+2}$ from $\Sigma_4(\lambda,\lambda,\lambda,\lambda)$ by imposing the condition
\begin{equation}\label{eq:fixlambda9}
\lambda=\frac{2}{p+9}~~~,
\end{equation}
which guarantees that their coefficients are in the correct ratio, and which for any $p\geq 1$ obeys the condition $4\lambda <1$.  The overall
coefficient of $\Sigma_4$ needed to fit $J_{3+2+2+2}$ and $I_{2+2+2+2}$ is
\begin{equation}\label{eq:kappa4}
\kappa_4=\frac{p\,!}{4\,!(p+8)\,!}\lambda^{-8}=\frac{(p+9)^8}{6144 p_{\,81}}~~~.
\end{equation}
We now look (with benefit of hindsight) for an integration formula of the form
\begin{align}\label{eq:ninth1}
&\kappa_4\Sigma_4(\lambda,\lambda,\lambda,\lambda)+ \kappa_3' \Sigma_3(2\lambda,\lambda,\lambda)  +\kappa_2'' \Sigma_2(3\lambda,\lambda)
+\sum_{i=1}^4\kappa_3^i\Sigma_3(\lambda_3^i,\lambda_3^i,\lambda_3^i) +\sum_{i=1}^4\kappa_2^i\Sigma_2(2\lambda_3^i,\lambda_3^i)\cr
+&\sum_{i=1}^6\bar\kappa_2^i\Sigma_2(\lambda_2^i,\lambda_2^i)+\sum_{i=1}^4\kappa_1^i\Sigma_1(3\lambda_3^i) +\sum_{i=1}^4 \bar\kappa_1^i
\Sigma_1(2\lambda_3^i)+\kappa_0 A~~~,\cr
\end{align}
with four of the $\lambda_2^i$ taken equal to the four $\lambda_3^i$, and the other two $\lambda_2^i$ additional parameters.  (Again, we reuse parameters
wherever similar structures are involved in Eq. \eqref{eq:symform}, so as to save function calls.)

We proceed to sketch the remaining calculation, without writing down the detailed form of the resulting Vandermonde equations (which can
be read off from the programs, and is fairly complicated).  We begin with the three-partition terms.  The equations for $J_{5+2+2}-J_{4+3+2}$,
$J_{4+3+2}-J_{3+3+3}$, and $I_{4+2+2}-I_{3+3+2}$ are all automatically satisfied by taking
\begin{equation}\label{eq:kappa3prime}
\kappa_3'=6 \kappa_4~~~.
\end{equation}
This leaves four independent matching conditions for $G_{2+2+2}$, $H_{3+2+2}$, $I_{4+2+2}$, and $J_{5+2+2}$, which lead to a $N=4$
Vandermonde system determining the coefficients $\kappa_3^i$.  We turn next to the two-partition terms.  We find that the equations for
$I_{6+2}-4.5 I_{5+3}+ 3.5 I_{4+4}$ and $J_{7+2}-3.5J_{6+3}+2.5 J_{5+4}$ are automatically satisfied by taking
\begin{equation}\label{eq:kappa2doubleprime}
\kappa_2''=8 \kappa_4~~~.
\end{equation}
The four equations for $G_{4+2}-G_{3+3}$, $H_{5+2}-H_{4+3}$, $I_{6+2}-I_{4+4}$, and $J_{7+2}+J_{6+3}-2J_{5+4}$ then give a $N=4$
Vandermonde system determining the coefficients $\kappa_2^i$.  The remaining independent equations matching two-partition terms,
for $E_{2+2}$, $F_{3+2}$, $G_{4+2}$, $H_{5+2}$, $I_{6+2}$, and $J_{7+2}$, then give a $N=6$ Vandermonde system determining the
coefficients $\bar \kappa_2^i$.

Turning to the single partition terms, the eight equations obtained by matching coefficients for $C_2$, $D_3$, $E_4$, $F_5$, $G_6$,
$H_7$, $I_8$, and $J_9$ give a $N=8$ Vandermonde system determining simultaneously the four coefficients $\kappa_1^i$ and the
four coefficients $\bar \kappa_1^i$.  Finally, equating coefficients of $A$ gives
\begin{equation}\label{eq:kappa0ninth}
\kappa_0=1-R~,~~~R=(\xi^4+6\xi^3+8\xi^2)\kappa_4+\xi^3 \sum_{i=1}^4 \kappa_3^i+\xi^2(\sum_{i=1}^4 \kappa_2^i + \sum_{i=1}^6 \bar \kappa_2^i)
+\xi \sum_{i=1}^4 (\kappa_1^i+\bar \kappa_1^i)~~~.
\end{equation}

\subsection{Leading term in higher order}

We have not systematically pursued constructing integration formulas of orders higher than ninth, but this should be possible by the
same method.  One can, however, see what the pattern will be for the leading term in such formulas.  An integration formula of order $2t+1$
will have a leading term $\Sigma_t(\lambda,...,\lambda)$, with $t$ arguments $\lambda$.   The only partition $t$ terms appearing in
the continuation of Eq. \eqref{eq:genint} will be $2+2+....+2$, containing $t$  terms 2, and $3+2+...+2$, with one $3$ and $t-1$ terms 2.
Requiring these to have coefficients in the correct ratio restricts $\lambda$ to be
\begin{equation}\label{eq:lambdagen}
\lambda=\frac{2}{p+2t+1}~~~,
\end{equation}
and the leading term in the integration formula will be $\kappa_t \Sigma_t(\lambda,...,\lambda)$,
with $\kappa_t$ given by
\begin{equation}\label{kappagen}
\kappa_t=\frac{p\,!}{t\,!(p+2t)\,!\lambda^{2t}}~~~.
\end{equation}

Where nonleading terms give multiple equations of the same order, corresponding to inequivalent partitions of $2t+1$, $2t$, ...,
one has to include terms proportional to $\Sigma_{t-1}(2\lambda,\lambda,...,\lambda)$, $\Sigma_{t-2}(3\lambda,\lambda,...,\lambda)$, and other
such structures with asymmetric arguments summing to $t\lambda$,
for the differences of these multiple equations to have consistent solutions.  Once such multiplicities have been taken care of, the remaining
independent equations will form a number of sets of Vandermonde equations.

\section{Derivation of the simplex generating function}

We give here a simple proof of the simplex generating function formulas of Eqs. \eqref{eq:expansion1} and \eqref{eq:wdef}, using the standard simplex integral
\begin{equation}\label{eq:multibeta1a}
\int_{\rm standard ~simplex} dx_1...dx_p\,(1-x_1-x_2-...-x_p)^{\nu_0}x_1^{\nu_1}...x_p^{\nu_p}=\frac{\prod_{a=0}^p\nu_a\,!}
{(p+\sum_{a=0}^p \nu_a)\,!}~~~,
\end{equation}
\big(which we obtain later on as a specialization of the multinomial beta function integral of Eq. \eqref{eq:multibeta}\big),
the simplex volume formula of Eq. \eqref{eq:volume}, and the expansion formulas of Eqs. \eqref{eq:expan} through \eqref{eq:expan1}.
We start from
\begin{equation}\label{eq:momentstart}
\int_{\rm simplex}dx_1...dx_p \sum_{\nu_1...\nu_p=0}^{\infty}\frac{\prod_{i=1}^p (\tilde x_it_i)^{\nu_i}}{\prod_{i=1}^p \nu_i\,!}=\int_{\rm simplex}dx_1...dx_p e^{\sum_{i=1}^p \tilde x_it_i}~~~,
\end{equation}
and substitute the expansion of Eq. \eqref{eq:expan} on the right hand side, giving
\begin{equation}\label{eq:momentnext}
\int_{\rm simplex}dx_1...dx_p e^{\sum_{a=0}^p \alpha_a \sum_{i=1}^p \tilde x_{ai}t_i}~~~.
\end{equation}
We now express the integral over the general simplex in terms of an integral over its barycentric coordinates $\alpha_a$.  Since
$\sum_{a=0}^p \alpha_a=1$, we can rewrite Eq. \eqref{eq:expan1}, by subtraction of $x_0$ from both sides, as
\begin{equation}\label{eq:expannew}
x-x_0=\sum_{a=0}^p (x_a-x_0)\alpha_a = \sum_{a=1}^p (x_a-x_0) \alpha_a~~~.
\end{equation}
From this we immediately find for the Jacobian
\begin{equation}\label{eq:jacob}
\left| \det \left( \frac{\partial x_1...\partial x_p}{\partial \alpha_1...\partial \alpha_p}\right)\right|=|\det(x_a-x_0)_i|=V p\,!~~~,
\end{equation}
with $V$ the volume of the simplex.  Since the $\alpha_a~,~~a=1,...,p$ span a standard simplex, we have transformed the integral of Eq. \eqref{eq:momentnext} to the form
\begin{equation}\label{eq:momentnext1}
V p\,! \int_{\rm standard~simplex} d\alpha_1...d\alpha_p  e^{\sum_{a=0}^p \alpha_a \sum_{i=1}^p \tilde x_{ai}t_i}~~~.
\end{equation}
Expanding the exponential on the right in a power series, we have
\begin{equation}\label{eq:momentnext2}
V p\,! \int_{\rm standard~simplex} d\alpha_1...d\alpha_p \sum_{\nu_1...\nu_p=0}^{\infty}\frac{\prod_{a=0}^p (\alpha_a)^{\nu_a}(\sum_{i=1}^p\tilde x_{ai}t_i)^{\nu_a}}{\prod_{a=0}^p \nu_a\,!} ~~~,
\end{equation}
and then recalling that $\alpha_0=1-\sum_{a=1}^p\alpha_a$, and using Eq. \eqref{eq:multibeta1a} to evaluate the integral over the standard simplex, we get
\begin{equation}\label{eq:momentnext3}
V p\,!\sum_{\nu_1...\nu_p=0}^{\infty}\frac{\prod_{a=0}^p (\sum_{i=1}^p\tilde x_{ai}t_i)^{\nu_a}}{(p+\sum_{a=0}^p \nu_a)\,!} ~~~.
\end{equation}
Let us now define $P_n$ as the projector on  terms with a total of $n$ powers of the parameters $t_i$, since this is the projector that extracts
the $n$th order moments.  Applying $P_n$ to Eq. \eqref{eq:momentnext3}, the denominator is converted to $(p+n)\,!$, which can then be pulled outside
the sum over the $\nu_i$, permitting these sums to be evaluated as geometric series,
\begin{align}\label{eq:momentsemifinal}
P_n&V p\,!\sum_{\nu_1...\nu_p=0}^{\infty}\frac{\prod_{a=0}^p (\sum_{i=1}^p\tilde x_{ai}t_i)^{\nu_a}}{(p+\sum_{a=0}^p \nu_a)\,!} \cr
=& \frac{V p\,!}{(p+n)\,!}P_n\sum_{\nu_1...\nu_p=0}^{\infty}\prod_{a=0}^p (\sum_{i=1}^p\tilde x_{ai}t_i)^{\nu_a} \cr
=&\frac{V p\,!}{(p+n)\,!}P_n \prod_{a=0}^p [1-\sum_{i=1}^p\tilde x_{ai}t_i)]^{-1}~~~.\cr
\end{align}
Finally, applying  to  each factor in the product over $a$  the rearrangement
\begin{equation}\label{eq:exponentiate}
(1-y)^{-1}=\exp[-\log(1-y)]= \exp[\sum_{s=1}^{\infty}\frac{y^s}{s}]~~~,
\end{equation}
we get
\begin{equation}\label{eq:momentfinal}
\frac{V p\,!}{(p+n)\,!}P_n \exp[\sum_{s=2}^{\infty}\frac{\sum_{a=0}^p[\sum_{i=1}^p\tilde x_{ai}t_i]^{s}}{s}]~~~,
\end{equation}
where we have used the fact that the  $s=1$ term in the sum vanishes  because $\sum_a \tilde x_{ai}=0$.  Comparing Eq. \eqref{eq:momentfinal}
with the starting equation Eq. \eqref{eq:momentstart}, we get  Eqs. \eqref{eq:expansion1} and \eqref{eq:wdef}.

\section{Parameterized higher order integration formulas for axis-parallel hypercubes}

We turn in this section to the problem of deriving higher order integration formulas for axis-parallel hypercubes, in analogy with
our treatment of the simplex case.  Our formulas can be viewed as a generalization of those obtained by Lyness (1965) and McNamee and Stenger (1967). We consider an axis-parallel hypercube of half-side $S$, and denote by $\tilde x$ coordinates referred
to the centroid of the hypercube.  Through ninth order, the expansion of a general function $f(\tilde x)$ is given as before by Eq.
\eqref{eq:expansion}.  Consider now the moment integrals
\begin{equation}\label{eq:momentcube}
m(\nu)=\int_{\rm hypercube} dx_1...dx_p \tilde x_1^{\nu_1}...\tilde x_p^{\nu_p}~~~.
\end{equation}
Since the limits of integration for each axis are $-S$, $S$, the moment integral factorizes and can be immediately evaluated as
\begin{align}\label{eq:momentcube1}
m(\nu)= &\prod_{\ell=1}^p \frac{S^{\nu_{\ell}+1}}{\nu_{\ell}+1}[1+(-1)^{\nu_{\ell}}]\cr
=&0 ~~~{\rm any } ~ \nu_{\ell} ~{\rm odd}~~~~,\cr
=&\prod_{\ell=1}^p \frac{2S\, S^{\nu_{\ell}}}{\nu_{\ell}+1}~~~{\rm all} ~\nu_{\ell} ~{\rm even}\cr
=&V\prod_{\ell=1}^p \frac{\, S^{\nu_{\ell}}}{\nu_{\ell}+1}~~~{\rm all} ~\nu_{\ell} ~{\rm even}~~~,\cr
\end{align}
with $V=(2S)^p$ in the final line the hypercube volume.

We now reexpress this moment integral in terms of sums over the set of $2p$ hypercube points $\tilde x_j$ given in Eq. \eqref{eq:cubepoints},
which will play a role in hypercube integration analogous to that played by simplex vertices in our treatment of simplex integration.   In analogy with Eq. \eqref{eq:sdef},
we define the  sum
\begin{equation}\label{eq:hypsum}
S_{i_1...i_n} = \sum_{j=1}^{2p}  \tilde x_{ji_1}...\tilde x_{ji_n}~~~.
\end{equation}
Since $\tilde x_{ji}=S\delta_{ji}$ for $1\leq j\leq p$ and $\tilde x_{ji}=-S\delta_{ji}$ for $p+1\leq j\leq 2p$, this sum
vanishes unless $n$  is even and all of the indices $i_1$,...,$i_n$ are equal, in which case it is equal to $2S^{n}$.
The tensors of Eq. \eqref{eq:hypsum} and their direct products form a complete basis on which we can expand moment integrals over the hypercube.
We have carried out this calculation two different ways.  First, by matching the non-vanishing moment
integrals through eighth order, we find
\begin{align}\label{eq:matching}
\frac{1}{V}\int_{\rm hypercube}dx_1...dx_p \tilde x_{i_1}\tilde x_{i_2}=&\frac{1}{6}S_{i_1i_2}~~~,\cr
\frac{1}{V}\int_{\rm hypercube}dx_1...dx_p \tilde x_{i_1}\tilde x_{i_2}\tilde x_{i_3}\tilde x_{i_4}=&\frac{1}{36}(S_{i_1i_2}S_{i_3i_4}
+S_{i_1i_3}S_{i_2i_4}+S_{i_1i_4}S_{i_2i_3})-\frac{1}{15}S_{i_1i_2i_3i_4}\cr
=&\frac{1}{36}(S_{i_1i_2}S_{i_3i_4}+2\,{\rm terms})-\frac{1}{15}S_{i_1i_2i_3i_4}~~~,\cr
\frac{1}{V}\int_{\rm hypercube}dx_1...dx_p \tilde x_{i_1}\tilde x_{i_2}\tilde x_{i_3}\tilde x_{i_4}\tilde x_{i_5}\tilde x_{i_6}=&
\frac{1}{216}(S_{i_1i_2}S_{i_3i_4}S_{i_5i_6}+ 14\,{\rm terms})\cr
-&\frac{1}{90}(S_{i_1i_2}S_{i_1i_2i_3i_4} + 14\,{\rm terms})
+\frac{8}{63}S_{i_1i_2i_3i_4i_5i_6}~~~,\cr
\frac{1}{V}\int_{\rm hypercube}dx_1...dx_p \tilde x_{i_1}\tilde x_{i_2}\tilde x_{i_3}\tilde x_{i_4}\tilde x_{i_5}\tilde x_{i_6}
\tilde x_{i_7}\tilde x_{i_8}=&\frac{1}{1296}(S_{i_1i_2}S_{i_3i_4}S_{i_5i_6}S_{i_7i_8}+ 104\,{\rm terms})\cr
+&\frac{1}{225}(S_{i_1i_2i_3i_4}S_{i_5i_6i_7i_8}+34\,{\rm terms})\cr
-&\frac{1}{540}(S_{i_1i_2}S_{i_3i_4}S_{i_5i_6i_7i_8}+209\,{\rm terms})\cr
+&\frac{4}{189}(S_{i_1i_2}S_{i_3i_4i_5i_6i_7i_8}+27\,{\rm terms})\cr
-&\frac{8}{15}S_{i_1i_2i_3i_4i_5i_6i_7i_8}~~~~.\cr
\end{align}
Combining these formulas with  Eq. \eqref{eq:expansion}, and using a similar condensed notation to that used in the simplex case (but with the
contractions referring now referring to the sums over the hypercube of Eq. \eqref{eq:hypsum}), we have for
the integral of a general function over the hypercube, through ninth order,
\begin{align}\label{eq:hypintegral}
\frac{1}{V}\int_{\rm hypercube} dx_1...dx_p f(\tilde x)=&A +\frac{1}{6}C_2 +\frac{1}{36}E_{2+2}-\frac{1}{15} E_4 \cr
+&\frac{1}{216}G_{2+2+2}-\frac{1}{90}G_{4+2}+\frac{8}{63} G_6 \cr
+&\frac{1}{1296}I_{2+2+2+2}+\frac{1}{225}I_{4+4}-\frac{1}{540}I_{4+2+2}+\frac{4}{189}I_{6+2}-\frac{8}{15}I_8\cr
+&...~~~.\cr
\end{align}

A second, and more general way, to obtain these results is to construct a generating function, analogous to that of Good and Gaskins used in the simplex case.
We start from the formula
\begin{equation}\label{eq:hypgen}
V^{-1}\int_{-S}^S dx_1...\int_{-S}^S dx_p\, e^{\,t_1 x_1+...+t_px_p}=\prod_{\ell=1}^p \frac{\sinh St_{\ell}}{St_{\ell}}~~~,
\end{equation}
and recall the power series expansion for the logarithm of $\frac{\sinh x}{x}$,
\begin{align}\label{eq:logexpan}
\log\big(\frac{\sinh x}{x}\big)=&\frac{1}{6}x^2-\frac{1}{180}x^4+\frac{1}{2835}x^6-\frac{1}{37800}x^8+...\cr
=&\sum_{n=1}^{\infty}\frac{(-1)^{n+1}2^{2n-1}B_{2n-1}}{n (2n)\,!} x^{2n}~~~,\cr
\end{align}
with $B_{2n-1}$ the Bernoulli numbers $B_1=\frac{1}{6}~,~~B_3=\frac{1}{30}~,~~B_5=\frac{1}{42}~,~~B_7=\frac{1}{30}~,~~B_9=\frac{5}{66}~,...$~~.
Defining, in analogy with the simplex case (with $\tilde x_{ai}$ now the components of the $2p$ vectors $\tilde x_a$ of Eq. \eqref{eq:cubepoints}),
\begin{align}\label{eq:cubewdef}
W_u=&\sum_{a=1}^{2p}\left(\sum_{i=1}^p\tilde x_{ai}t_i\right)^u\cr
=& 0~,~~{\rm u~odd}~~~,\cr
=& 2S^u \sum_{i=1}^p t_i^u~,~~{\rm u~even}~~~,\cr
\end{align}
we rewrite Eq. \eqref{eq:hypgen}, using Eq. \eqref{eq:logexpan}, as
\begin{equation}\label{eq:hypgen1}
V^{-1}\int_{-S}^S dx_1...\int_{-S}^S dx_p\, e^{\,t_1 x_1+...+t_px_p}=e^{\sum_{n=1}^{\infty}K_n W_{2n}}~~~,
\end{equation}
with
\begin{equation}\label{eq:kaydef}
K_n=\frac{(-1)^{n+1}2^{2n-2}B_{2n-1}}{n (2n)\,!} ~~~.
\end{equation}
Through eighth order, the right hand side of Eq. \eqref{eq:hypgen1} is
\begin{align}\label{eq:cubegenexpan}
e^{W_2/12-W_4/360+W_6/5670-W_8/75600+...}=&1\cr
+&\frac{W_2}{12}\cr
-&\frac{W_4}{360}+\frac{W_2^2}{288}\cr
+&\frac{W_6}{5670}-\frac{W_2W_4}{4320}+\frac{W_2^3}{10368}\cr
-&\frac{W_8}{75600}+\frac{W_4^2}{259200}+\frac{W_2W_6}{68040}-\frac{W_2^2W_4}{103680}+\frac{W_2^4}{497664}+...~~~.\cr
\end{align}
Applying the same rule as in the simplex case, of multiplying the coefficient of each term by a combinatoric factor $(s\,!)^t t\,!$ for
each factor $W_s^t$, we recover the numerical coefficients in the expansions of Eqs. \eqref{eq:matching} and \eqref{eq:hypintegral}.  This method
can be readily extended to the higher order terms of these expansions.

We now follow the procedure used before in the simplex case.  We match the expansion of Eq. \eqref{eq:hypintegral} to sums over
the function $f$ evaluated at points within the hypercube,  this time constructing these sums using parameter multiples of the $2p$ points
of Eq. \eqref{eq:cubepoints}, which are the centroids of the maximal boundary hypercubes.  The formulas of Eqs. \eqref{eq:discretesums},
\eqref{eq:sumexp}, and \eqref{eq:symform} still apply, with sums that extended from 0 to $p$ in the simplex case extending now from
1 to $2p$, and with $\xi=p+1$ in Eq. \eqref{eq:sumexp} replaced by $\xi=2p$.

We proceed to set up integration formulas of first, third, fifth, seventh, and ninth order, for integrals over an axis-parallel hypercube.  Since
all odd order terms in the expansion of Eq. \eqref{eq:expansion} integrate to zero by symmetry, to achieve this accuracy it suffices to
perform a matching of the non-vanishing terms through zeroth, second, fourth, sixth, and eighth order, respectively.  We will see that as a result of the
absence of odd order terms, the higher order hypercube formulas are considerably simpler than their general simplex analogs.

\subsection{First and third order formulas}

 We begin our derivation of odd order hypercube integration formulas with examples
 of first and third order accuracy, obtained by matching the first two
 terms in the expansion of Eq. \eqref{eq:hypintegral},
 \begin{equation}\label{eq:firstcube}
 I=A+\frac{1}{6}C_2+...~~~.
 \end{equation}
 Proceeding in direct analogy with the first order formulas of Eq. \eqref{eq:first1}
 in the simplex case, we get
 \begin{align}\label{eq:firstcube2}
I_a=&\Sigma_1(\lambda)/\xi~~~,\cr
I_b=&f(\tilde 0)~~~,\cr
\end{align}
 with $\tilde x=\tilde 0$ the centroid of the hypercube, and $\xi=2p$. These are
 again analogs of the trapezoidal and center-of-bin methods for the one dimensional
 case.

 Similarly, in analogy with the second order accurate formula
 of Eq. \eqref{eq:second3} for the simplex,  we get the third order accurate
 hypercube formula
 \begin{equation}\label{eq:thirdcube1}
 I=\left(1-\frac{\xi}{6\lambda^2}\right)f(\tilde 0)
 +\frac{1}{6 \lambda^2}\Sigma_1(\lambda)~~~.
 \end{equation}
 For any two nondegenerate values $\lambda_{a,b}$ in
 the interval (0,1)  this gives two different third
 order accurate estimates $I_{a,b}$ of the hypercube integral.

\subsection{Fifth order formula}

To get a fifth order formula, we have to use the sums of Eq. \eqref{eq:discretesums} to match the coefficients of $A$, $C_2$, $E_{2+2}$,
and $E_4$ appearing on the first line of Eq. \eqref{eq:hypintegral}.  Since there is now only a single two-partition term, $E_{2+2}$, we
can extract it from $\Sigma_2(\lambda,\lambda)$ for any $\lambda$ in the interval $(0,\frac{1}{2})$.  So we look for a fifth order formula of the form
\begin{equation}\label{eq:fifthhyp1}
\kappa_2\Sigma_2(\lambda,\lambda)+\sum_{i=1}^2\kappa_1^i\Sigma_1(\lambda_1^i) +\kappa_0 A~~~.
\end{equation}
Matching the coefficient of $E_{2+2}$ gives
\begin{equation}\label{eq:kappa2hyp}
\kappa_2=\frac{1}{72 \lambda^4}~~~,
\end{equation}
while matching the coefficient of $A$ gives
\begin{equation}
\kappa_0=1-R_0~,~~~R_0=\xi^2\kappa_2+\xi\sum_{i=1}^2 \kappa_1^i~~~.
\end{equation}
Matching the coefficients of $ C_2$ and $E_4$ gives a $N=2$ Vandermonde system \big(c.f. Eq. \eqref{eq:vandermonde}\big) with
\begin{align}\label{eq:fifthhyp2}
x_i=&(\lambda_1^i)^2~,~~~w_i=\kappa_1^i(\lambda_1^i)^2~~~,\cr
q_1=&\frac{1}{6}-\frac{\xi}{36 \lambda^2}~~~,\cr
q_2=&\frac{-1}{15}-\frac{\xi}{36}~~~.\cr
\end{align}

\subsection{Seventh order formula}

To get a seventh order formula, we have to match the coefficients appearing on the first two lines of Eq. \eqref{eq:hypintegral}.  Since there
is only one three-partition term, $G_{2+2+2}$, we can extract it from $\Sigma_3(\lambda,\lambda,\lambda)$ for any $\lambda$ in the
interval $(0,\frac{1}{3})$.  We look for a seventh order formula of the form
\begin{equation}\label{eq:seventhhyp1}
\kappa_3\Sigma_3(\lambda,\lambda,\lambda)+\sum_{i=1}^2 \kappa_2^i\Sigma_2(\lambda_2^i,\lambda_2^i) +\sum_{i=1}^3 \kappa_1^i \Sigma_1(\lambda_1^i)
+\kappa_0 A~~~,
\end{equation}
with matching the coefficient of $G_{2+2+2}$ requiring
\begin{equation}\label{eq:kappa3hyp}
\kappa_3=\frac{1}{1296\lambda^6}~~~.
\end{equation}
To reduce the number of function calls, we take $\lambda_1^{i+1}=2\lambda_2^i~,~~i=1,2$, with only $\lambda_1^1$ an additional parameter.
Matching the coefficients of the two-partition terms $E_{2+2}$ and $G_{4+2}$ gives a $N=2$ Vandermonde system with
\begin{align}\label{eq:seventhhyp2}
x_i=&(\lambda_2^i)^2~,~~~w_i=2\kappa_2^i(\lambda_2^i)^4~~~,\cr
q_1=&\frac{1}{36}-\frac{\xi}{216 \lambda^2}~~~,\cr
q_2=&-\frac{1}{90}-\frac{\xi}{216}~~~,\cr
\end{align}
while matching the coefficient of $A$ gives
\begin{equation}\label{eq:seventhhyp3}
\kappa_0=1-R_0~,~~~R_0=\xi^3\kappa_3 + \xi^2\sum_{i=1}^2 \kappa_2^i + \xi \sum_{i=1}^3 \kappa_1^i~~~.
\end{equation}
Matching coefficients of the single-partition terms $C_2$, $E_4$, and $G_6$ gives the $N=3$ Vandermonde system with
\begin{align}\label{eq:seventhhyp4}
x_i=&(\lambda_1^i)^2~,~~~w_i=\kappa_1^i(\lambda_1^i)^2~~~,\cr
q_1=&\frac{1}{6}-\frac{\xi^2}{432\lambda^4}-2\xi\sum_{i=1}^2\kappa_2^i(\lambda_2^i)^2&~~~,\cr
q_2=&-\frac{1}{15}-\frac{\xi}{36}+\frac{\xi^2}{432\lambda^2}~~~,\cr
q_3=&\frac{8}{63}+\frac{\xi}{90}+\frac{\xi^2}{432}~~~.\cr
\end{align}

\subsection{Ninth order formula}

To get a ninth order formula, we have to match the coefficients of all three lines of Eq. \eqref{eq:hypintegral}.  Since there is only
one four-partition term $I_{2+2+2+2}$, we can extract it from $\Sigma_4(\lambda,\lambda,\lambda,\lambda)$ for any $\lambda$ in the
interval $(0,\frac{1}{4})$.  We look for a ninth order formula of the form
\begin{equation}\label{eq:ninthhyp1}
\kappa_4\Sigma_4(\lambda,\lambda,\lambda,\lambda)+\sum_{i=1}^2 \kappa_3^i\Sigma_3(\lambda_3^i,\lambda_3^i,\lambda_3^i)
 +\kappa_2'\Sigma_2(3\lambda,\lambda)+\sum_{i=1}^3 \bar \kappa_2^i \Sigma_2(\lambda_2^i,\lambda_2^i)+\sum_{i=1}^4 \kappa_1^i\Sigma_1(\lambda_1^i)
+\kappa_0 A~~~,
\end{equation}
with matching the coefficient of $I_{2+2+2+2}$ requiring
\begin{equation}\label{eq:kappa4hyp}
\kappa_4=\frac{1}{31104\lambda^8}~~~.
\end{equation}
To reduce the number of function calls, we take $\lambda_2^i=\lambda_3^i~,~~i=1,2$, and $\lambda_1^i=3\lambda_2^i~,~~i=1,2,3$, with only $\lambda_2^3$
and $\lambda_1^4$  additional parameters.
Matching the coefficients of the three-partition terms $G_{2+2+2}$ and $I_{4+2+2}$ gives a $N=2$ Vandermonde system with
\begin{align}\label{eq:ninthhyp2}
x_i=&(\lambda_3^i)^2~,~~~w_i=6\kappa_3^i(\lambda_3^i)^6~~~,\cr
q3_1=&\frac{1}{216}-\frac{\xi}{1296\lambda^2}~~~,\cr
q3_2=&-\frac{1}{540}-\frac{\xi}{1296}~~~.\cr
\end{align}
Taking the difference of the matching equations for $I_{6+2}$ and $I_{4+4}$ determines $\kappa_2'$ to be
\begin{equation}\label{eq:kappa2prime1}
\kappa_2'=\frac{237}{8164800 \lambda^8}=\frac{237}{262.5}\kappa_4~~~,
\end{equation}
while matching the coefficient of $A$ gives
\begin{equation}\label{eq:ninthhyp3}
\kappa_0=1-R_0~,~~~R_0=\xi^4\kappa_4 + \xi^3\sum_{i=1}^2 \kappa_3^i +\xi^2(\kappa_2'+\sum_{i=1}^3 \bar \kappa_2^i) + \xi \sum_{i=1}^4 \kappa_1^i~~~.
\end{equation}
Matching the remaining independent two-partition terms $E_{2+2}$, $G_{4+2}$, and $I_{4+4}$ gives a $N=3$ Vandermonde system with
\begin{align}\label{eq:ninthhyp4}
x_i=&(\lambda_2^i)^2~,~~~w_i=2\bar\kappa_2^i(\lambda_2^i)^4~~~,\cr
q2_1=&\frac{1}{36}-\kappa_2'18\lambda^4-\frac{\xi^2}{2592\lambda^4}-6\xi\sum_{i=1}^2\kappa_3^i(\lambda_3^i)^4&~~~,\cr
q2_2=&-\frac{1}{90}-\kappa_2'90\lambda^6-\frac{\xi}{216}+\frac{\xi^2}{2592\lambda^2}~~~,\cr
q2_3=&\frac{1}{225}-\kappa_2'162\lambda^8+\frac{\xi}{540}+\frac{\xi^2}{2592}~~~.\cr
\end{align}
Finally, matching coefficients of the single-partition terms $C_2$, $E_4$, $G_6$, and $I_8$ gives a $N=4$ Vandermonde system with
\begin{align}\label{eq:ninthhyp5}
x_i=&(\lambda_1^i)^2~,~~~w_i=\kappa_1^i(\lambda_1^i)^2~~~,\cr
q1_1=&\frac{1}{6}-\frac{\xi^3}{7776\lambda^6}-\kappa_2'10\xi\lambda^2-3\xi^2\sum_{i=1}^2\kappa_3^i(\lambda_3^i)^2-2\xi\sum_{i=1}^3\bar\kappa_2^i
(\lambda_2^i)^2~~~,\cr
q1_2=&-\frac{1}{15}-\frac{\xi^3}{7776\lambda^4}-\kappa_2'82\xi\lambda^4-3\xi^2\sum_{i=1}^2\kappa_3^i(\lambda_3^i)^4-\xi \,q2_1~~~,\cr
q1_3=&\frac{8}{63}-\frac{\xi^3}{7776\lambda^2}-\kappa_2'730\xi\lambda^6-\frac{1}{2}\xi^2\,q3_1-\xi \, q2_2~~~,\cr
q1_4=&-\frac{8}{15}-\frac{\xi^3}{7776}-\kappa_2'6562\xi\lambda^8-\frac{1}{2}\xi^2\, q3_2-\xi\, q2_3~~~.\cr
\end{align}

\subsection{Leading term in higher order}

As in the simplex analysis, in the hypercube case we  have not systematically pursued constructing integration formulas of orders higher than ninth, but this should be possible by the
same method.  Again, one can see what the pattern will be for the leading term in such formulas.  An integration formula of order $2t+1$
will have a leading term $\Sigma_t(\lambda,...,\lambda)$, with $t$ arguments $\lambda$.   The only $t$-partition term appearing in
the continuation of Eq. \eqref{eq:hypintegral} will be $2+2+....+2$, containing $t$  terms 2.  So $\lambda$ can be taken to have
any value in the interval $(0,\frac{1}{t})$, and the leading term in the integration formula will be $\kappa_t \Sigma_t(\lambda,...,\lambda)$,
with $\kappa_t$ given by
\begin{equation}\label{kappagenhyp}
\kappa_t=\frac{1}{t\,!6^t\lambda^{2t}}~~~.
\end{equation}
Where nonleading terms give multiple equations of the same order, corresponding to inequivalent partitions of  $2t$,...
one has to include terms proportional to  $\Sigma_{t-2}(3\lambda,\lambda,...,\lambda)$, and other such structures with asymmetric arguments
summing to $t\lambda$,
for the differences of these multiple equations to have consistent solutions.  Once such multiplicities have been taken care of, the remaining
independent equations will form a number of sets of Vandermonde equations.

\subsection{One dimension revisited: comparison of Vandermonde moment fitting to standard one dimensional methods}

In this subsection we address several related issues.  We first set up a one-dimensional
analog of the moment fitting method that we have used in general $p$ dimensions to
develop higher order integration formulas.  The one dimensional moment fitting equations
can be satisfied by leaving the sampling points as free parameters, giving in any order a
Vandermonde equation system to determine the weights assigned to the sampling points.  Alternatively,
the moment fitting equations  can be
satisfied by restricting the sampling points, which is what is done in Gaussian quadrature, which reduces
the number of function calls. We then show that the fifth (and higher) order direct hypercube integration
formulas for $p$ dimensions, when restricted to one dimension,
involve a larger number of function calls than needed for the case when the sampling points
are all free parameters. We interpret this as resulting from linear dependencies in low dimension among the
various terms appearing in the generating function of Eq. \eqref{eq:cubegenexpan}.
We study these linear dependencies as a function of dimension $p$, and suggest that the
integration rule of order $2t+1$ has redundant parameters when the spatial dimension $p<t$.

Let $f(x)=f_0+f_1x+f_2x^2+f_3x^3+f_4x^4+...$ be a function that is power series expandable on the interval
(-1,1), and consider the one dimensional integral
\begin{align}\label{eq:onedim1}
I=&\frac{1}{2}\int_{-1}^1 f(x)dx \cr
=&f_0+\frac{f_2}{3}+\frac{f_4}{5}+\frac{f_6}{7}+\frac{f_8}{9}+...~~~.\cr
\end{align}
Defining $\Sigma_1(\lambda)$ by
\begin{equation}\label{eq:onedim2}
\Sigma_1(\lambda)=f(\lambda)+f(-\lambda)\,,~~~0<\lambda \leq 1~~~,
\end{equation}
we look for an integration formula of the form
\begin{align}\label{eq:intform1}
I=&\kappa_0 f(0)+\sum_{i=1}^n \kappa^i \Sigma_1(\lambda^i)~~~\cr
=&f_0+\sum_{i=1}^n \kappa^i [f_0+f_2 (\lambda^i)^2 +f_4 (\lambda^i)^4
+f_6 (\lambda^i)^6 +f_8 (\lambda^i)^8+...]~~~.\cr
\end{align}
Matching the coefficients of  $f_{\ell}$ between Eq. \eqref{eq:onedim1} and
Eq. \eqref{eq:intform1}, we get the system of equations
\begin{align}\label{eq:system1}
1=&\kappa_0+2\sum_{i=1}^n \kappa^i~~~,\cr
\frac{1}{3}=&2\sum_{i=1}^n \kappa^i (\lambda^i)^2~~~,\cr
\frac{1}{5}=&2\sum_{i=1}^n \kappa^i (\lambda^i)^4~~~,\cr
\frac{1}{7}=&2\sum_{i=1}^n \kappa^i (\lambda^i)^6~~~,\cr
\frac{1}{9}=&2\sum_{i=1}^n \kappa^i (\lambda^i)^8~~~,\cr
\end{align}
and similarly if one wishes to go to higher order than ninth.

There are now two ways to proceed to solve the matching equations, to give
a discrete approximation to the integral to a given order of accuracy.  The
first, which is what we have done in getting simplex and hypercube integration
formulas, is to regard all of the $\lambda^i$ as adjustable parameters, and
to determine the coefficients $\kappa^i$ to satisfy the system of equations
of Eq. \eqref{eq:system1} to the needed order.  Thus, to get a first order
accurate formula, we take $I \simeq f(0)$ with all the $\kappa^i$ equal to zero, which is the center-of-bin rule.
To get a third order accurate formula we must take $\kappa^1$ as nonzero and solve the system
\begin{align}\label{eq:system2}
1=&\kappa_0+2 \kappa^1~~~,\cr
\frac{1}{3}=&2\kappa^1(\lambda^1)^2~~~.\cr
\end{align}
To get a fifth order accurate formula we must take both $\kappa^1$ and $\kappa^2$ as nonzero
and solve the system
\begin{align}\label{eq:system3}
1=&\kappa_0+2 (\kappa^1+\kappa^2)~~~,\cr
\frac{1}{3}=&2\kappa^1(\lambda^1)^2+2\kappa^2(\lambda^2)^2~~~,\cr
\frac{1}{5}=&2\kappa^1(\lambda^1)^4+2\kappa^2(\lambda^2)^4~~~,\cr
\end{align}
to get a seventh order accurate formula we must take $\kappa^{1,2,3}$ as nonzero
and solve the system
\begin{align}\label{eq:system4}
1=&\kappa_0+2 (\kappa^1+\kappa^2+\kappa^3)~~~,\cr
\frac{1}{3}=&2\kappa^1(\lambda^1)^2+2\kappa^2(\lambda^2)^2+2\kappa^3(\lambda^3)^2~~~,\cr
\frac{1}{5}=&2\kappa^1(\lambda^1)^4+2\kappa^2(\lambda^2)^4+2\kappa^3(\lambda^3)^4~~~,\cr
\frac{1}{7}=&2\kappa^1(\lambda^1)^6+2\kappa^2(\lambda^2)^6+2\kappa^3(\lambda^3)^6~~~,\cr
\end{align}
and so forth. Evidently, to get an order $2t+1$ formula, we must take $n=t$, so that there are $t$
distinct positive sampling points
$\lambda^{1,...,t}$,and to determine the coefficients $\kappa^{1,...,t}$ we must solve an order $N=t$ Vandermonde
system. The resulting order $2t+1$ integration formula uses $2t+1$ function values.

An alternative way to proceed is to adjust the values of the sampling points so that fewer of them are needed
to satisfy the matching conditions.  This is what is done in the well-known Gaussian integration method, which
gives a more efficient scheme, in terms of the number of function calls, starting with third order.  Referring
to Eq. \eqref{eq:system2}, we can evidently achieve a third order match by taking
\begin{align}\label{eq:gaussian3}
2\kappa_0=&0~,~~~2\kappa^1=1~~~,\cr
\lambda^1=&\frac{1}{\sqrt 3}~~~.\cr
\end{align}
Similarly, referring to Eq. \eqref{eq:system3}, we can evidently achieve a fifth order match by taking
\begin{align}\label{eq:gaussian4}
2\kappa_0=&\frac{8}{9}~,~~~2\kappa_1=\frac{5}{9}~~~,\cr
\lambda^1=&\frac{\sqrt 3}{\sqrt 5}~~~.\cr
\end{align}
Proceeding in this way, we can obtain the general Gaussian integration formula, which for order $2t+1$ integration
involves $t$ points.  Of course, the usual derivation of the Gaussian integration rule does not proceed this way,
but instead uses an argument based on one dimensional polynomial long division to relate the special points $\lambda^i$ to zeros of the Legendre polynomials.  Since in higher dimensions there is no analogous polynomial division rule, there
is no universal higher dimensional analog of the Gaussian integration rule, although there are a multitude of special
formulas using specially chosen sampling points in higher dimensions (see, e.g., Stroud (1971)).  On the other hand,
as we have seen, the method of keeping all the sampling points $\lambda^i$ as free parameters, and solving a set of Vandermonde equations to get the coefficients $\kappa^i$, readily extends to higher dimensions.

Let us now examine the number of function evaluations required by our general moment fitting formulas for hypercubes,
when restricted to one dimension.  The first order center-of-bin formula requires just the one function evaluation
$f(0)$, and so is the same in all methods.  The third order formula of Eq. \eqref{eq:thirdcube1} involves $f(0)$ and
one $\Sigma_1(\lambda)$, and so uses 3 function values, in agreement with Eq. \eqref{eq:system2}, whereas the
Gaussian formula needs 2 function values.  Turning to the fifth order formula of Eq. \eqref{eq:fifthhyp1},
calculation of $\Sigma_2(\lambda,\lambda)$ requires 3 function evaluations, calculation  of each of the two
$\Sigma_1(\lambda^i)$ requires 2 function evaluations, and evaluation of $f(0)$ requires one function evaluation,
for a total of 8 function evaluations.  This is to be compared to the one dimensional moment fitting formula of
Eq. \eqref{eq:system3} which requires 5 function evaluations, and the Gaussian method, which requires 3.

The reason that the fifth order integration formula for general $p$, when specialized to one dimension, requires more function
evaluations than the moment fitting method of Eq. \eqref{eq:system3}, is that whereas in two and higher dimensions
$W_4$ and $W_2^2$ are linearly independent, in one dimension they are proportional to one another by virtue of
the identity $t_1^4=(t_1^2)^2$. Hence the term $\Sigma_2(\lambda,\lambda)$ in the general integration formula is not
needed to get a match, and when this is dropped one has a formula identical in form to that of Eq. \eqref{eq:system3},
requiring only 3 function calls.   Turning to the higher order hypercube formulas, we see that the seventh order hypercube
formula of Eq. \eqref{eq:seventhhyp1} has redundant parameters and function calls for dimension $p<3$, since in 2 dimensions $W_6$, $W_2W_4$ and
$W_2^3$ are linearly dependent by virtue of the algebraic identity
\begin{equation}\label{eq:idenseventh}
0=(t_1^2+t_2^2)^3-3(t_1^2+t_2^2)(t_1^4+t_2^4)+2(t_1^6+t_2^6)~~~.
\end{equation}
Similarly, the ninth order hypercube formula of Eq. \eqref{eq:ninthhyp1} has redundant parameters and function calls for dimension $p<4$, since in 3 dimensions $W_8$, $W_4^2$, $W_2W_6$, $W_2^2W_4$, and $W_2^4$ are linearly dependent by virtue of the identity
\begin{align}\label{eq:idenninth}
0=&(t_1^2+t_2^2+t_3^2)^4-6(t_1^8+t_2^8+t_3^8)+3(t_1^4+t_2^4+t_3^4)^2 \cr
+&8(t_1^2+t_2^2+t_3^2)(t_1^6+t_2^6+t_3^6)-6(t_1^2+t_2^2+t_3^2)^2(t_1^4+t_2^4+t_3^4)~~~.\cr
\end{align}

These results suggest the conjecture that the hypercube formula of order $2t+1$ will involve redundant parameters and
function calls for dimension $p<t$, and we expect an analogous statement to apply for the simplex formulas derived
by the moment fitting method in Sec. VII.  This redundancy for small $p$ is a consequence of the fact that the
integration formulas that we have derived for simplexes and hypercubes are universal, in the sense that they involve
the same number of parameters irrespective of the dimension $p$.  As $p$ increases, the number of sampling points
increases, but the number of parameters, and the size of the Vandermonde systems needed to find coefficients, remains
fixed.

\section{Function calls needed for integration routines of various orders }

We summarize in this section the number of function calls needed for a single call
to the integration routines of various orders. These are obtained by running the programs
to integrate the function fcn=1, in which case the programs exit without subdividing the
base region, giving the desired
function call count for two samplings of the integral at the indicated order of accuracy,
as well as unity as the output integral (since the programs all compute the integral over
the base region, divided by the base region volume).

In Table VI
we give the function call counting for the simplex integration programs of first through fourth, fifth,
seventh, and ninth order.  For comparison, in Table VII we give a similar table from the paper
of Genz and Cools (2003), which gives the function call counting for one evaluation of the indicated order,
plus a second evaluation at a lower order used for error estimation.  Unlike our method, which proceeds directly
from the vertices of a general simplex, the Genz and Cools program uses integration rules for a
standard $p$-simplex, with an affine transformation needed to treat more general simplexes.
 Although not directly comparable,  the two tables
show that the strategy we have used, of incorporating a number of free parameters into the integration
which can be used to give different samplings of the integrand, does not lead to an inefficiency of more
than a factor of 2 to 3 compared to the method used by Genz and Cools.

\begin{table} [t]
\caption{Function calls by for simplex integration of order $n$ in dimension $p$ by method of Sec. VII}
\centering
\begin{tabular}{c c c c c c c c c c c}
\hline\hline
$n$&$ p \rightarrow$ & $1$ & $2$ & $3$ & $4$ & $5$ & $6$ & $7$ & $8$ & $9$ \\
\hline
1& & 3 & 4 & 5 & 6 &7& 8 &9 &10 &11 \\
2& & 5 & 7 & 9 & 11 &13& 15 &17 &19 &21 \\
3& & 7 & 10 & 13 & 16 &19& 22 &25 &28 &31 \\
4& & 10 & 16 & 23 & 31 &40& 50 &61 &73 &86 \\
5& & 20 & 31 & 43 & 56 &70& 85 &101 &118 &136 \\
7& & 37 & 71 & 117 & 176 &249& 337 &441 &562 &701 \\
9& & 74 & 168 & 316 & 531 &827& 1219 &1723 &2356 &3136 \\

\hline
\end{tabular}
\label{table:fcncalls_simplex1}
\end{table}

\begin{table} [t]
\caption{Function calls for simplex integration of order $n$ in dimension $p$ from Genz and Cools (2003)}
\centering
\begin{tabular}{ c c c c c c c c c c}
\hline\hline
$n$&$ p \rightarrow$  & $2$ & $3$ & $4$ & $5$ & $6$ & $7$ & $8$ & $9$ \\
\hline

3& &  7 & 9 & 11 &13& 15 &17 &19 &21 \\
5& &  16 & 23 & 31 &40& 50 &61 &73 &86 \\
7& &  32 & 49 & 86 &126& 176 &237 &310 &396 \\
9& &  65 & 114 & 201&315& 470 &675 &940 &1276 \\

\hline
\end{tabular}
\label{table:fcncalls_simplex2}
\end{table}

In Table VIII we give the function call counting for the direct
hypercube programs of first, third, fifth, seventh, and ninth order.
For comparison, in Table IX we have tabulated $t^p+(t+1)^p$, with
the odd order of integration $n$ related to $t$ by $n=2t+1$; this is
the number of function calls needed if one uses a $p$-fold direct
product of Gaussian integrations of indicated order, together with a
$p$-fold direct product of Gaussian integrations of the next higher
odd order to get an error estimate. One sees from these tables that
for $t=1,2,3$ our parameterized method is more efficient than direct
product Gaussian for dimension $p\geq 4$, and for $t=4$ the
parameterized method is more efficient for $p\geq 5$. Since the
number of function calls in the parameterized method is
asymptotically polynomial of order $(2p)^t/t\,!$, whereas in the
direct product Gaussian method it is exponential in $p$, the
parameterized method becomes markedly more efficient for large
dimension $p$.

\begin{table} [t]
\caption{Function calls for hypercube integration of order $n$ in dimension $p$ by method of Sec. VIII}
\centering
\begin{tabular}{c c c c c c c c c c c}
\hline\hline
$n$&$ p \rightarrow$ & $1$ & $2$ & $3$ & $4$ & $5$ & $6$ & $7$ & $8$ & $9$ \\
\hline
1& & 3 & 5 & 7 & 9 &11& 13 &15 &17 &19 \\
3& & 5 & 9 & 13 & 17 &21& 25 &29 &33 &37 \\
5& & 12 & 27 & 46 & 69 &96& 127 &162 &201 &244\\
7& & 21 & 69 & 153 & 281 &461& 701 &1009 &1393 &1861 \\
9& & 48 & 192 & 501 & 1059 &1966& 3338 &5307 &8021 &11644 \\

\hline
\end{tabular}
\label{table:fcncalls_cube1}
\end{table}

\begin{table} [t]
\caption{Function calls for hypercube integration of order $n$ in dimension $p$ by comparison of
two product Gaussian rules}
\centering
\begin{tabular}{c c c c c c c c c }
\hline\hline
$n$&$ p \rightarrow$ & $1$ & $2$ & $3$ & $4$ & $5$ & $6$ & $7$  \\
\hline
3& & 3 & 5 & 9 & 17 &33& 65 &129  \\
5& & 5 & 13 & 35 & 97 &275& 793 &2315 \\
7& & 7 & 25 & 91 & 337 &1267& 4825 &18571 \\
9& & 9 & 41 & 189 & 881 &4149& 19721 &94509 \\

\hline
\end{tabular}
\label{table:fcncalls_cube2}
\end{table}

These results reinforce the indication from the previous section
that, as a very rough rule of thumb, in using integration routines
with $n=2t+1$  in dimension $p$, one should avoid high order
routines with $t>p$. This is true both because in low dimension the higher order
routines have redundant function calls, and because the extra
computation involved in using a high order routine is justified only
when the $2^p$ scaling in the number of subregions, as the program
subdivides from level to level, becomes large enough.  However, this
is only a very general criterion, since the optimum choice or
choices of integration routine order will  depend on the nature of
the function being integrated.  Moreover, in dimension $p=1$ the
programs are so fast on current computers that use of the fifth or
seventh order integration routines, while not as efficient as
Gaussian integration, still gives good results.

\section{Putting it all together -- sketch of the algorithms}

We are now ready to give a sketch of the adaptive algorithms
incorporating the elements described above.  The basic algorithm
starts from a base region, which acts as the initial level
subregion, which is either a standard simplex, a Kuhn simplex (for
hypercube integration treated by tiling with Kuhn simplexes), or a
half-side 1 hypercube. It then proceeds recursively through higher
levels of subdivision, by evaluating the integral using an
integration method of order specified by the user with two different
parameter choices, giving two estimates of the integral over the
subregion divided by the subregion volume, which we denote by
$I_a({\rm subregion})$ and $I_b({\rm subregion})$. (Dividing out the
volume is convenient because of the $1/V$ factor appearing on the
left hand side of Eqs. \eqref{eq:genint} and
\eqref{eq:hypintegral}.) If the level number exceeds a
user-specified value {\it ithinlev} which determines when thinning
begins, then a thinning condition   is applied. When the
user-specified thinning function parameter {\it ithinfun} is given
the value 1, the thinning condition   used is
\begin{equation}\label{eq:thinning}
|I_a({\rm subregion})-I_b({\rm subregion})|<\epsilon~~~,
\end{equation}
with $\epsilon$ an error measure specified by the user. (Further
thinning options will be discussed shortly.)  If this condition   is
met, the results are retained as contributions to the $I_a$ and
$I_b$ estimates of the integral divided by the base region volume,
and the subregion is not further subdivided.  If this condition is
not met, then the subregion is subdivided into $2^p$ subregions, and
the process is repeated.  The process terminates when either the
thinning condition is met for all subregions, or a limit to the number
of levels of subdivision set by the user is reached.  In the latter
case, the contributions of the remaining subregions that have not
satisfied the thinning condition are added to the $I_a$ and $I_b$
totals, as well as to the sum of the absolute values of the local subinterval errors.

With either termination, we get the final estimates of the integral
divided by the base region volume,
\begin{align}\label{eq:iabvol}
I_a \simeq & \sum_{\rm subregions} V({\rm subregion}) I_a({\rm subregion})~~~,\cr
I_b\simeq & \sum_{\rm subregions} V({\rm subregion}) I_b({\rm subregion})~~~.\cr
\end{align}
Here $V({\rm subregion})$ is the subregion volume divided by the base region volume, and since the subregions
are a tiling of the initial base region,  we  have
\begin{equation}\label{eq:lsumvol}
\sum_{\rm subregions} V({\rm subregion})=1~~~.
\end{equation}
From the difference of $I_a$ and $I_b$ we get an estimate of the
error, given by
\begin{equation}\label{eq:err1vol}
{\rm |outdiff|}\equiv |I_a-I_b|~~~.
\end{equation}
We can also (as in the one dimensional illustration) compute the sum
of the absolute values of the local subinterval errors,
\begin{equation}\label{eq:err2vol}
{\rm errsum}\equiv \sum_{\rm subregions} V({\rm subregion}) |I_a({\rm subregion})
-I_b({\rm subregion})| ~~~.
\end{equation}
Comparing Eqs. \eqref{eq:iabvol}, \eqref{eq:err1vol}, and
\eqref{eq:err2vol}, we see that errsum and ${\rm |outdiff|}$  obey
the inequality
\begin{equation}\label{eq:errineq}
{\rm errsum} \geq {\rm |outdiff|}~~~,
\end{equation}
with equality holding if  $I_a-I_b$ has the same sign in
all subregions. When the condition $|I_a({\rm subregion})-I_b({\rm
subregion})|<\epsilon$ is met for all subregions, errsum reduces,
using Eq. \eqref{eq:lsumvol}, to
\begin{equation}\label{eq:err21vol}
{\rm errsum} < \epsilon~~~.
\end{equation}
Hence to evaluate the integral to a relative error $\delta$, one should  choose
\begin{equation}\label{eq:pickeps}
\epsilon \sim \delta |I_a|~~~.
\end{equation}
Since $I_a$ and $I_b$ give the integral over the base region divided by the base region volume,
to get the value of the integral without normalization by the base region volume,
one must multiply these outputs by the base region volume $V_0$.  For a standard simplex, $V_0=1/p\,!$, for a side 1
hypercube, $V_0=1$, while for a half-side 1 hypercube, $V_0=2^p$.

Note that the thinning condition determining whether to subdivide a subregion does {\it not} include
a factor of the subregion volume; we are testing variances of the integrand as sampled over the subregion, not variances of the
net contribution to the integral.  This may seem counter-intuitive, but is motivated  by
the formulas of Eqs. \eqref{eq:lsumvol}--\eqref{eq:err21vol}, by simplicity, and by the fact that it works well in practice.
 The problem with including a subregion
volume weighting factor in the thinning condition is that at a very fine level of subdivision, there are many subregions,
and so small error contributions from each can add up to a large error in the total.  Since the local test does not involve
comparisons of the errors from different regions, the calculation in each subregion proceeds independently from that in all the others.
The local thinning condition that we use is equivalent to the ``Local Subdivision Strategy'' described in the monographs of Krommer and Ueberhuber (1991)
and Ueberhuber (1995) using a parameter $\epsilon_{abs}$, which plays the role of our $\epsilon$.

Using $|I_a({\rm subregion})-I_b({\rm subregion})|$ as the basis for
a thinning decision is only one possibility of many.  More
generally, given $A \equiv I_a({\rm subregion})$ and $B \equiv
 I_b({\rm subregion})$, one can take as the thinning function any
 function $f(A,B)$ with the properties $f(A,B)\geq 0$ and $f(A,B)=0 ~{\rm iff}
 ~ A=B$,  imposing now the thinning condition $f(A,B)< \epsilon$.
 In the programs, we have included three options,
 (1) $f(A,B)=|A-B|$ as in the discussion above, (2)
 $f(A,B)=|A-B|/|A+B|$, and (3) $f(A,B)=(A-B)^2$.  In many cases,
 and in particular for polynomial integrals, we found their
 performance (with appropriate $\epsilon$) to be similar, but for the singular integral $\int_0^1
 dx
 \frac{1}{\surd 1-x^2}$ we found choice (3) to perform considerably
 better than the other two.

Three versions of the basic algorithm are presented in each of the
directories of programs.  In the first, the algorithm subdivides
until all subregions obey the thinning condition, or until a preset
limit on the level of subdivisions is reached, which is dictated by
the available memory.  Typically, for simple integrands and moderate
dimension $p$, this happens rather quickly, in other words, the
algorithm has saturated capabilities of the machine memory, but not
of the machine speed. In a second version labelled ``r'', the
algorithm is ``recirculated'' by keeping, at a level limit set by
the user which is
 chosen to avoid exceeding machine memory capabilities,
 all the subintervals that do not obey the thinning condition.
These are then treated one at a time by the same algorithm, up to a
second level limit again set by the user.  This can take hours
or days for high accuracy, high $p$ computations, with a practical limit set
by the speed capabilities of the machine.  Finally, a third version
labelled ``m'' takes the ``recirculating'' algorithm and
parallelizes it using the MPI (message passing interface) protocol,
by distributing to each process of a cluster a large number of the
subintervals that do not obey the thinning condition  , each of which is
then processed by the algorithm sequentially.  This speeds up the
computation by a factor of the number of processes available.
All routines are coded in double precision, but since the  ninth
order integration formulas involve large numbers in computing coefficients, double
precision computation is not enough to give double precision
accuracy results, so for the fifth, seventh, and ninth order
routines in both the simplex and direct hypercube cases, we also
give a quadruple precision \big(real(16)\big) version of the
programs.

The programs present the user with various options.  By an
appropriate choice of {\it ithinlev}, thinning can be delayed, or
even suppressed entirely so that all subdivisions take place to the
specified subdivision limits. This can give a check that subregions
with large contributions, but accidentally small error estimates,
have not been harvested prematurely, and when the programs are
modified, gives a useful check that the tiling condition of Eq.
\eqref{eq:lsumvol} is obeyed.  By a choice of {\it ithinfun}, the
user can choose which of three preset thinning functions to use, or
by modifying the subroutine containing these functions, the user can
make another choice of thinning function.  For simplex integration,
the user can choose whether to use the recursive or the symmetric
subdivision algorithm. The user can choose the accuracy of the
integration method used: first through fourth, fifth, seventh, or
ninth for simplex based routines, and first, third, fifth, seventh,
and ninth for the direct hypercube routines. Finally, the user can
modify the free parameters in the integration routines, so as to get
different samplings of the integrand, which can give a useful
assessment of whether the error estimates from the initially used
sampling are realistic.

\section{Test integrals; false positives and their avoidance}

For verifying the higher order integration programs, and for checking the operation of the adaptive programs, it is essential to have
test integrals with known answers.
For the standard simplex (c.f. Eqs. \eqref{eq:stdsimplex} and \eqref{eq:stdint}), a useful formula is the multinomial beta function integral,
\begin{equation}\label{eq:multibeta}
\int_{\rm standard ~simplex} dx_1...dx_p\,(1-x_1-x_2-...-x_p)^{\alpha_0-1}x_1^{\alpha_1-1}...x_p^{\alpha_p-1}=\frac{\prod_{a=0}^p\Gamma(\alpha_a)}
{\Gamma(\sum_{a=0}^p \alpha_a)}~~~,
\end{equation}
with $\Gamma$ the usual gamma function (see the Wikipedia article on Dirichlet distributions).  When $\alpha_a-1=\nu_a~,~~~a=0,...,p$ with $\nu_a$ an integer, this can be rewritten as
 \begin{equation}\label{eq:multibeta1}
\int_{\rm standard ~simplex} dx_1...dx_p\,(1-x_1-x_2-...-x_p)^{\nu_0}x_1^{\nu_1}...x_p^{\nu_p}=\frac{\prod_{a=0}^p\nu_a\,!}
{(p+\sum_{a=0}^p \nu_a)\,!}~~~.
\end{equation}
The $\nu_0=0$ case of this formula is the formula given by Stroud (1971) \big(see also Grundmann and M\"oller (1978)\big  ) for the integral of a general monomial over the standard simplex.

For a unit hypercube, the
corresponding formula is
\begin{equation}\label{eq:hyptest1}
\int_0^1 dx_1...\int_0^1 dx_p x_1^{\nu_1}...x_p^{\nu_p}=\prod_{\ell=1}^p\frac{1}{\nu_{\ell}+1}~~~,
\end{equation}
while for a half-side $1$ hypercube the corresponding monomial integrals are (c.f. Eqs. \eqref{eq:momentcube} and \eqref{eq:momentcube1})
\begin{equation}\label{eq:momentcubetest}
\int_{-1}^1dx_1...\int_{-1}^1dx_p \, x_1^{\nu_1}... x_p^{\nu_p}=2^p\prod_{\ell=1}^p \frac{1}{\nu_{\ell}+1}~~~{\rm for~all} ~\nu_{\ell} ~{\rm even,~ and~ zero~ otherwise}~~~.
\end{equation}

Testing the simplex programs with the integral of Eq. \eqref{eq:multibeta1}, and starting thinning at level 1, shows that when the order of the monomial is less than or equal to the order
of the integration formula used, the iteration terminates at the initial level, and the difference between $I_a$ and $I_b$ is of order the computer truncation error. When a monomial is integrated that is of higher order than the integration formula used, with a small enough error measure $\epsilon$, the adaptive program starts to subdivide the base region.

However, a more complicated pattern is seen for the hypercube integrals when evaluated by the  direct hypercube algorithms, and this brings us to
the issue of false positives.  As in the simplex case, when thinning is started at level 1 and the order of the test monomial is less than or equal to the order of the integration formula
used, the iteration terminates again at the initial level, and $I_a-I_b$ is of order the truncation error.  However, when a monomial is integrated
that is of higher order than the integration formula used, the adaptive program does not always start to iterate.  For example, using the
fifth order hypercube formula in dimension $p=4$, the program iterates for the integrand $x_1^6$, but not for the integrand $x_1^2x_2^2x_3^2$.
The reason is that the latter function, although of higher order than that of the integration formula, vanishes on the hyperplanes spanning the axes where the fifth order
integration formula samples the integrand, and so the $I_a$ and $I_b$ evaluations give the same answer (zero), and the thinning condition is obeyed for
arbitrarily small $\epsilon$.  This is an example of a {\it false positive}, in which the thinning condition is obeyed even though the actual error
is large.  Any sampling program for evaluating integrals is subject to false positives for functions that take special values (in our case zero, or a constant) on the sampling points.  Since the sampling points in the simplex integration formulas are on oblique, rather than axis-parallel, lines or
planes, this problem is not so readily seen with the multinomial test functions of Eq. \eqref{eq:multibeta1}, but we have nonetheless found examples of false positives.  For example, using fifth order integration and symmetric subdivision, the $p=5$ monomial $x(1)x(2)x(3)x(4)^2x(5)$, when computed with thinning starting at any level below 3, develops a false positive at level 2 and gives an answer that is wrong in the fourth decimal place, even though the output error measures suggest much higher
accuracy.

There are several general ways to guard against false positives. The
simplest is to use the freedom of choosing the parameter {\it
ithinlev} to delay thinning until several subdivisions have taken
place.  False positives are most dangerous if they occur in the
initial few levels, since these have the largest subregions, and if
a subregion is prematurely harvested, there is a possibility of
significant error.  On the other hand, thinning becomes most
important after several subdivisions have taken place, when the
number of subregions is large.  So there can be a useful tradeoff
between starting thinning early and starting it late.  If computer
time permits, one can always do an {\it a posteriori} check by
choosing {\it ithinlev} greater than the limit on the number of
levels, which suppresses thinning altogether, and gives the
approximate Riemann sum corresponding to the level of subdivision
attained.

A second general way to guard against false positives is to compute the integral using alternative options, for example, using
integration programs of several different orders, or where allowed as an option for simplex integrals, to use recursive instead
of symmetric subdivision.  In the fifth order $p=5$ example noted above, changing to seventh order integration, or changing
from symmetric to recursive subdivision while maintaining fifth order integration, both eliminate the false positive at
level two.

A third way  is to add a function with known integral to the integrand, which has significantly different local behavior, and
to subtract its known integral from the total at the end.
For example, in the hypercube case, consider the integral
\begin{equation}\label{eq:addfn}
0=\int_{-1}^1 \phi_q(x)~,~~~\phi_q(x)=\frac{1}{(q+x)^2}-\frac{1}{q^2-1}~~~,
\end{equation}
which exists for any $q>1$.  Adding a multiple of
\begin{equation}\label{eq:addfn1}
\prod_{\ell=1}^p \phi_q(x_{\ell})
\end{equation}
to the test monomial integrands does not change the expected answer, but forces the adaptive program to start to subdivide at level 1 (for small enough $\epsilon$) in all monomial
cases. It is of course not necessary for the added function to have an integral that can be evaluated in closed form.
In the $p=5$ simplex case discussed above, we eliminated the false positive at level 2 by numerically integrating the
function $(1+x(1))^{-1}$, and then adding a multiple of this function to the integrand and subtracting its integral from
the answer.  When adding such an auxiliary function, it is probably a good idea to rescale it so that its order of magnitude
is similar to that of the integral being evaluated.   Clearly there is an infinite variety of such auxiliary functions that can be added to the integrand,
each of which shifts the false positive problem to a different part of integrand function space.  Even when one is dealing with generic
integrands, in which the program starts to subdivide as expected, adding such functions will alter the pattern of subdivision, and can be used
(in addition to changing the integration formula parameters) to give further estimates of the errors in the output values $I_{a,b}$ provided
by the integration algorithm.

We do not recommend just changing the integration formula parameters
as a way of eliminating false positives.  The reason is that the
samplings in both the simplex and hypercube cases take place on
hyperplanes that are determined by the general structure of the
integration formulas, but do not vary as the parameters in the
integration formulas are changed.  So if  a false positive is
associated with a zero or constant integrand value on one of these
hyperplanes, it will not be eliminated by changing the parameter
values.  Similar remarks apply to changing the thinning function as
a way of eliminating false positives.

For related reasons we have not written into the programs another way of creating a criterion for thinning, the
comparison of results from integration programs of different orders (say, of fifth and seventh order).  In the simplex example discussed above,
doing this would eliminate the false positive, since the seventh order routine uses sampling points that avoid the problematic hyperplanes
sampled by the fifth order routine.  However, in this case one may as well do two seventh order samplings to set up the thinning condition  , and thus benefit from the higher accuracy  accruing from use of the seventh order routine for smooth integrands.

\section{Description of programs in the seven directories}

\subsection{General description}

The Fortran programs are grouped into 7 directories, named
simplex123, simplex4, simplex579, simplex579\_16, cube13, cube579,
and cube579\_16. All programs are valid for arbitrary dimension
$p\geq 1$

The simplex programs all perform adaptive integration over a standard simplex or a Kuhn simplex with one vertex at the origin,
using real(8) precision \big(except for simplex579\_16, which uses real(16)\big).  The programs in simplex123 perform first through third order integration, the programs in simplex4
perform fourth order integration, and the programs in simplex 579 perform fifth, seventh, or ninth order integration.

The same adaptive program treats both the standard and Kuhn simplex
cases, with a subroutine argument ``i\_\,init'' determining which
initialization is used.    Included in all the simplex packages are
programs for integration over a side 1 hypercube with one vertex at
the origin, by tiling with Kuhn simplexes followed by adaptive
simplex integration.

The programs in simplex4 perform fourth order adaptive integration using a different subdivision strategy from that used in all the other cases.  In simplex4 the simplex vertices are used as sampling points, with the side midpoints giving the vertices
at the next level of subdivision.  In all the other programs, only interior points of the simplex are used for sampling.  Hence,
the simplex4 programs cannot be used to integrate functions which have integrable singularities at the base simplex boundary,
whereas the other programs can be used in this case.

The programs in cube13 perform first or third order adaptive integration, and those in cube579 perform fifth, seventh, or ninth order
adaptive integration, over half-side 1 hypercubes centered on the origin, with real(8) precision. These programs use less memory (by roughly a factor $1/p$) than the hypercube tiling programs. The cube programs are valid for arbitrary dimension $p\geq 1$.

The programs in simplex579\_16 are real(16) re-writings of those in simplex579, and the programs in cube579\_16 are real(16) re-writings of those in cube579.  The real(16) versions are obtained from the corresponding real(8) programs by making
the following global substitutions:  (1) Replace ``d0'' by ``q0'', (2) replace ``implicit real(8)'' by ``implicit real(16)'',
(3) replace ``dabs'' by ``qabs'', (4) replace ``d20.13'' by ``d32.36''.  These changes can be made using a ``replace all'' utility, since the strings that have to be modified do not occur anywhere else in the programs.   Note that explicit data type declarations that override the implicit ones  are {\it not} changed.

Each directory contains a package of subprograms, labeled
respectively simplexsubs123.for, simplexsubs4.for,
simplexsubs579.for, simplexsubs579\_16.for, cubesubs123.for,
cubesubs579.for, and cubesubs579\_16.for.  The subroutines in these
packages do not have to be accessed by the user in normal operation
of the adaptive programs. If they are accessed to alter the
programs, we strongly recommend doing several test integrals before
and after the changes, to make sure they still operate correctly.
Each directory also contains a series of main program files,  and
each main program file contains the main program proper, as well as
a subroutine setting up the function to be integrated, subroutines
setting up the free parameters used in the parameterized
integrations, a subroutine setting up three options for the thinning
function,  and in the case of the Kuhn tiling treatment of
hypercubes, a subroutine symmetrizing the function to be integrated
over all its variables. Each program that requires user setting of
input parameters contains comment statements giving instructions. To
run the programs, the user must compile and link the subroutine
package in a directory with the appropriate main program file in the
same directory.

As noted in the section on Vandermonde solvers, all programs are self-contained, since their subroutine packages include  Vandermonde solvers that
compute the explicit solution of the Vandermonde system for the relevant values of $N$.
 Because the ninth order simplex integration routines and associated
Vandermonde equations involve large numbers in computing
 coefficients, use of real(16) is recommended if one wants to get answers with real(8) accuracy.  Solving the Vandermonde equations to get the coefficient parameters for the integrations need be done only once before adaptive integration begins; this is done in the subroutines with names beginning with ``ext'', the output of which is then fed to the integration programs that are used repeatedly in the adaptive integration process.

There are three generic types of main programs in each directory.  Those with names not ending in ``r'' or ``m'' execute adaptive integration
to a subdivision level set by the user (and limited by machine memory).  Those with names ending in ``r'' execute the ``recirculating'' routines, in which after the first stage of subdivision, the remaining subregions are subdivided sequentially in a second stage to a second level of subdivision set by the user.
Those with names ending in ``m'' execute an MPI parallel version of the ``recirculating'' routines, in which after the first stage of subdivision,
the remaining subregions are farmed out to the available processes for a second stage of subdivision to the second level of subdivision set by the user.

In order to conserve memory, the labeling of simplex and cube points and the simplex subdivision routines use a lattice built on integer(2) arithmetic. This allows 14 levels of subdivision in the initial stage, since $2^{14}$=16384, which is half the maximum integer representable in integer(2).  In  order to go beyond 14 levels of subdivision in one stage, say to 30 levels of subdivision, one would have to replace 16384 in the subroutines by $2^{30}=1,073,741,824$, which is half the maximum integer representable in integer(4),  replace all integer(2) data type declarations by integer(4), and enlarge the level number limits in the programs. The explicit limits in the programs on the number of levels correspond to the requirement that the minimum integer(2) lattice spacing must not be smaller than 1, since in integer arithmetic 1/2 is replaced by 0.  Program stages that pass on subdivided regions have a limit of 14 levels, while output
stages that do not pass on subdivided regions have a limit of 15 levels.  An exception to this rule is in the simplex4 programs, where there is an explicit division by 2 in the programs, and so the corresponding limits are 13 and 14.  Note that in integer(2) arithmetic, $16384/2+16382/2=16384\neq (16384+16384)/2=(-32768)/2=-16384$, which is why in the simplex4 integration program we have not regrouped added terms into parentheses.

The recirculating  and MPI  programs make use of the observation that symmetric (or recursive) subdivision of standard simplexes, symmetric and recursive  subdivision of Kuhn simplexes,
and hypercube subdivision, all give after $\ell$ subdivisions a subregion that fits within a hypercube of side $1/2^{\ell}$ (or
$1/2^{\ell-1}$).  This observation, which is an unproved conjecture supported by our numerical results in the case of standard simplexes,
permits a doubling of the number of levels attainable within integer(2) arithmetic in the ``r'' and ``m'' programs, as follows.
At the start of the second stage of subdivision, each subregion is translated by a shift vector  and is rescaled by a factor which expands it to just fit within
 the initial lattice containing  base region.  This permits another 15 (or for recursive subdivision, 14) levels in the second stage \big(with
corresponding limits in the simplex4 programs of 14 (or 13)\big), and so the ``r'' and ``m'' programs
can subdivide to subregions that have a dimension $2^{-28}=3.725 \times 10^{-9}$ of the base region dimension.  Whether this can be attained in practice for a given dimension $p$ of course depends on available machine memory. Subdivision limits appropriate to the various cases have been incorporated into the main programs.

Because simplex points are represented in integer(2) arithmetic,
in order to apply the simplex subroutines to a starting simplex that does not have
only 0s or 1s in the vertex coordinates (for example, an equilateral triangle), one would have to change the integer(2) data type declarations to
real(4) for the programs to work correctly.  This change increases the memory requirements, and should not be made unless needed.   We note also that with the aim of conserving memory, we have used allocatable memory to store subregion information, allocating memory where needed at each level of subdivision, and deallocating memory when no longer used.

Finally, we note that the MPI programs are written using only simple MPI\_Send  and MPI\_Recv commands.  All processes simultaneously carry
out the first stage of subdivision, and then each process of rank greater than 0 takes its share of the remaining subregions after the first stage
and processes them further.  This wastes some processor time, but avoids large data transfers.  Only at the end, when all processes of rank
greater than 0 have finished, is their output combined in process 0.   Because
MPI can only pass real(8) numbers as messages,  the real(16) MPI programs  give only real(8) output. (This is one of the
reasons why the explicit
real(8) declarations are not modified in the conversion substitutions leading to real(16) programs.)  Nevertheless,
the MPI programs compute the sensitive parts of the high order integrations
in real(16), converting to real(8) only at the end when process outputs are combined.

To enhance readability of the programs, we have used indents to show the different levels of ``if'' chains, except
in one place in the MPI programs, where we have given the ``if'', ``else if'', and ``end if'' lines statement numbers 97,98,99.
We have not indented the contents of ``do'' loops, since these always begin and end with a statement number, and never with
an unnumbered ``enddo''. \big(The one exception to this is in the subroutine BestLex used for the symmetrization step in the Kuhn tiling programs
for integration over hypercubes, which has been taken verbatim
from H. D. Knoble's (1995) website.\big)  This is of course a matter of taste; our feeling was that  indenting both the ``if'' chains and ``do'' loops would result in so many levels of indents that readability of the programs would be decreased.  We also remind the reader
that the direct hypercube programs were written by minimal modification of the simplex programs, changing array arguments where needed (e.g., ``$ip+1$'' for simplexes becomes ``$2*ip$'' for hypercubes), but not changing array names.  So the array names in the direct hypercube subroutines are not the ones
that would naturally be chosen if these programs were written without reference to the simplex case.

\subsection{Inputs}

The main programs require the following inputs to be set by the user:

\begin{enumerate}

\item  {\it ithinlev} tells the program when to begin thinning subregions, by harvesting those that obey the thinning condition    of Eq. \eqref{eq:thinning}.  Thinning begins when the total level number exceeds {\it ithin}.  Thus, with $ithin=0$, thinning begins
at level 1, while if $ithin$ is greater than or equal to the maximum total level number, there is no thinning.

\item {\it ithinfun} tells the program which thinning function option to use.  As explained in Sec. X,
$ithinfun=1$ corresponds to a thinning function $f(A,B)=|A-B|$, $ithinfun=2$ to
 $f(A,B)=|A-B|/|A+B|$, and $ithinfun=3$ to  $f(A,B)=(A-B)^2$.

\item {\it isubdivision} tells the simplex programs whether to use symmetric subdivision ($isubdivision=1$) or recursive subdivision
($isubdivision=2$).   This parameter does not appear in the main programs in cube13, cube579, and cube579\_16, where there is no choice of subdivision methods.

\item{\it iaccuracy} tells the programs  to use the integration program of order $iaccuracy$. For example, in the simplex123
programs, to
select third order accuracy one sets $iaccuracy=3$, and in the simplex 579 programs, to select  seventh order integration
one sets $iaccuracy=7$.  This parameter does not appear in the main programs in simplex4, which uses only fourth order
integration.

\item{\it ip} gives the spatial dimension $p$ of the simplex or hypercube being integrated over, and can take any integer value $\geq 1$. Thus, to integrate over a three dimensional cube one would set $ip=3$.

\item{\it eps} sets the parameter $\epsilon$ appearing in the thinning condition   of Eq. \eqref{eq:thinning}. For {\it ithinfun}=1,  this gives an absolute error criterion; to
achieve a given level of relative error, one needs a rough estimate of the value of the integral as given by $outa$ or
$outb$, which can be used to readjust $eps$ by use of
Eq. \eqref{eq:pickeps}. For nonsingular integrands, the {\it eps} value when using {\it ithinfun}=3 should, as a first guess,
 be taken as the square of the {\it eps} value that one used for {\it ithinfun}=1. Note that if {\it ithinlev} is greater than the total level number, so that thinning is suppressed, the results are independent
of the value given to {\it eps}.

\item   In all programs other than the hypercube tiling program, the external function is supplied by the user in the subroutine fcn.for.  In the tiling programs, fcn.for is instead the symmetrization program for the external function supplied by the user in the subroutine fcn1.for.

\item {\it llim} sets the limit to the number of subdivisions in the programs with names not ending in ``r'' or ``m''.  It can be any integer between
1 and 15, except in the simplex4 programs, where the range is 1 to 14. In practice, the effective upper limit is set by machine memory.  Start with a low value of $llim$, and then to improve the accuracy, increase it until you get a diagnostic saying memory has been exceeded; the value of {\it llim} one less than this
 is the maximum value $llim=LMAX$ that does not exceed memory. Since the final level $l=llim$ does not further subdivide, this limit is associated with the
  number of subregions carried forward from level $l-1$ to the final level.  As $llim$ is increased the execution time will increase, and this will also impose an effective upper limit.

\item {\it llim1} and {\it llim2} in the programs with names ending in ``r'' and ``m'' set the limit to the number of subdivisions in the first and second stages of subdivision, respectively.  The maximum value of {\it llim1} is 14 and of {\it llim2} is 15,
except in the simplex4 programs, where the respective limits are 13 and 14, and also except for recursive subdivision, where the maximum value of {\it llim2} is one less than the corresponding value for symmetric subdivision.  In all cases, the built-in subdivision limits prevent the program from dividing 1 by 2, giving an integer arithmetic answer of 0.  As before, the effective upper limit will be set by machine memory and machine execution speed.  In using the ``r'' and ``m''
 programs, and setting $llim2=1$, the maximum value of $llim1$ that will not exceed memory is $llim1=LMAX-1$, with $LMAX$ the corresponding maximum determined
 as above for the single stage program.  Once $LMAX$ is determined, one can take any value $1 \leq llim2 \leq LMAX$ without exceeding memory.  Because of the staging, the numerical output depends only on the sum $llim1+llim2$, that is, one is free to redistribute the computational effort between the first and
 second stages.

\item The parameters for the higher order integration routines are contained in the main program files in subprograms with names beginning with  ``setparam''.
They are given in array constructors, and have been preset to values indicated. The program variable names have been chosen to roughly correspond to the
 symbol names in the formulas of Secs. VII and VIII.  For example, for the direct cube routines, where $\lambda$ is a free parameter, it is
 called $aalamb$; in the order 7 routine for simplex integration, $\lambda_1^i$ and $\lambda_2^i$ are the respective elements of the array constructors
 $alamb1$ ($blamb1$) and $alamb2$ ($blamb2$) corresponding to the first (second) choice of parameter values.  (In the fifth order cube and simplex routines, where only one
 pair of array constructors is needed, they are called $alamb$ ($blamb$), even though the corresponding quantity is labeled $\lambda_1^i$ in Secs. VIIC
 and IXB.)
 These presets can be changed by the user to give a different
sampling of the integrand in the integration subregions, subject to the following rules: (i) The inequalities in the comment statements must be
obeyed, to keep the sampling points inside the subregion, as discussed in  Secs. V and VI. (ii) The parameters in each array constructor must have non-degenerate values, so that
the corresponding Vandermonde equations will be solvable.  If two parameters in an array constructor are very close, solution of the Vandermonde
system will have large truncation errors, so care should be taken to keep the parameters in each array constructor reasonably well spaced.  (iii)
The ``a'' and ``b'' array constructors should have different parameter values, since these are used to give the two different integrand evaluations  used in the error estimate.

\end{enumerate}

\subsection{Outputs (and their use in making memory and running time estimates)}

Program outputs (except in the MPI case) are written to a file ``outdat.txt'' and also appear on the screen.  In the MPI case, outputs are
written to the output file specified by the system for a ``print'' statement.  A brief description of output labeling follows:

\begin{enumerate}

\item  All programs write out the user-set values of $ip$, $llim$ (or $llim1$ and $llim2$), $eps$, $ithinlev$, $ithinfun$, $isubdivision$, and $iaccuracy$.
 They do not print out the values of the parameters in the array constructors in the subprograms setparam.

\item  In all programs, $outa$ and $outb$ give two evaluations of the integral divided by the base region volume,  corresponding respectively to the two different samplings of the integrand set by the ``a'' and ``b'' parameters in the array constructors, and  ${\it |outdiff|}$ gives the difference $|outa-outb|$. The size of ${\it |outdiff|}$ gives an estimate of the likely error in the answer;
this estimate can be improved by evaluating the integral with a number of different choices of the array constructor  parameters, and
also by comparing the evaluations obtained using different program options as set by the user-set inputs.
As noted in Sec. XI, to get the value of the integral without normalization by the base region volume,
one must multiply $outa$ and $outb$ by the base region volume $V_0$.  For a standard simplex, $V_0=1/p\,!$, for a side 1
hypercube, $V_0=1$, while for a half-side 1 hypercube, $V_0=2^p$.

\item In all programs, $errsum$ gives the sum
of the absolute values of the local subinterval thinning tests,
\begin{equation}\label{eq:err2volff}
{\rm errsum}\equiv \sum_{\rm subregions} V({\rm subregion}) |f\big(I_a({\rm subregion},I_b({\rm subregion})\big)| ~~~,
\end{equation}
with $f(A,B)$ the thinning function.  As explained above, for the choice $ithinfun=1$ this gives an upper bound for ${\it |outdiff|}$,
and when $I_a({\rm subregion}-I_b({\rm subregion})$ has uniform sign over all subregions, $errsum={\it |outdiff|}$.  However, when
signs are not uniform over subregions, $errsum$ for {\it ithinfun}=1 can be much larger than the actual error, as in the two Gaussian example discussed
below.

\item In all programs, $l$ gives the level number, $ind$ gives the number of subregions carried forward to the next level,
$indmax$ gives the maximum value of $ind$ encountered over the course of the various levels that have been executed,  $fcncalls$ gives the number of function calls, $t\_\,current$ gives the current elapsed
time in seconds at the various levels of the first stage, and $t\_\,final$ gives the total elapsed execution
time in seconds.  In the approximation in which the geometric series summing the number of function calls over the various levels is
approximated by its largest term, corresponding to the highest level attained, and when there is no thinning, $fcncalls \simeq T 2^{p(llim-1)}$ for the single stage
program, and $fcncalls  \simeq T 2^{p(llim1+llim2-1)}$ for the ``r'' and ``m'' programs, with $T$ the appropriate function call value
from Table VI or VIII. \big(In the absence of thinning, the exact formula summing the geometric series is $fcncalls = T [2^{p(K+1)}-1]/[2^p-1]$, with $K=llim-1$ for the single stage program and $K=llim1+llim2-1$ for the ``r'' and ``m'' programs.\big)  When there is thinning, this gives an upper bound on the number of
function calls.

\item In the ``recirculating'' programs with main program name ending in ``r'', $t\_\,restart$ gives the time at which the second stage is initiated, in which the subregions carried forward from the first stage are subdivided sequentially.
During the second stage, the program will indicate approximately when it is 10, 20, ..., 90, 100 percent finished in sequentially processing the subregions carried forward from the first stage, by printing this information to the screen (but not by writing it to file).  In interactive mode, this permits one to gauge how long the calculation will take to finish; if it looks like
the calculation will take longer than one wishes to wait, one can stop execution and restart with different, more tractable,
parameter values.
Since these numbers are computed by integer division, the actual numbers
may be 9,19,... or other similar strings, depending on the residue modulo ten of the number of regions carried forward.
 One can also estimate the total running time by
multiplying $t\_\,restart$ by the number of subregions $ind$ carried forward to the second stage from the final level of the first stage,  further multiplied by $2^{p(llim2-llim1)}$ to  correct for a difference in the first and second stage level numbers. (Similarly, for the single stage programs, from
$t\_\,current$ and $ind$ at the output of any level $l$, the maximum running time, in the absence of thinning, to reach the level limit $llim1$ is
the product $t\_\,current$ times $ind$, further multiplied by $e^{p(llim1-l)}$.)We generally found that timing values, on a laptop, varied by one or two tenths of a second between identical runs, so estimates of total
running time become reliable only when one has proceeded to the point where several seconds have elapsed.
The final statistics include $indcount$, which gives a sum of the $ind$ values at each level of the second stage, and which indicates  the $ind$ value that would be needed if the second stage subdivisions were
carried out in the first stage by using a larger $llim1$ value.  Because of the staging strategy, the maximum $ind$ value that is required is the much smaller number $indmax$.

\item In the MPI programs with main program name ending in ``m'', $t\_\,restart$ is the time at the end of the first stage
when the subregions carried forward, numbering  $indstart$ in total, are distributed to multiple processes, and  $fcncalls$ gives the number of function calls up to this point. If $t\_\,restart$ does not appear in the output, the program has completed
execution before entering the second stage. Since MPI programs are typically run in batch mode, no intermediate statistics
are output during the second stage, but one can make a rough estimate of total second stage running time by multiplying $t\_\,restart$ by the number of subregions $indstart$ carried forward, further multiplied by $2^{p(llim2-llim1)}$ to  correct for a difference in the first and second stage level numbers, and dividing by $N_{\rm process}-1$ (process 0 serves only as an accumulation register for the output of the remaining $N_{\rm process}-1$ processes).  If this estimate
is too large, one can stop execution and restart with less ambitious parameters.  The final statistics include $indmaxprocess$, which is the maximum of the final $indmax$ over all of the processes.

\end{enumerate}

\section{Some sample results, and open questions}

\subsection{Sample results}

We turn now to some sample results which illustrate the capabilities of our numerical integration programs.  Our first example is one given in the paper on VEGAS of Lepage (1978), consisting of the
sum of two spherically symmetric Gaussians equally spaced along the diagonal of a cubical integration
volume,
\begin{equation}\label{eq:doubgauss1}
I_p=\frac{1}{2}\left(\frac{1}{a\pi^{1/2}}\right)^p\int_0^1 d^{\,p}x[e^{-\sum_{i=1}^p (x(i)-1/3)^2/a^2}
+e^{-\sum_{i=1}^p (x(i)-2/3)^2/a^2}] ~~~,
\end{equation}
with $a=0.1$.
In this form $I_p$ can be evaluated by the ``cubetile'' programs which tile a unit hypercube with Kuhn
simplexes.  In order to apply the direct hypercube ``cube'' programs, we make the change of variable $x=(1+y)/2$ to
rewrite $I_p$ as an integral over a half-side 1 hypercube,
\begin{equation}\label{eq:doubgauss2}
I_p=\frac{1}{2}\frac{1}{2^p}\left(\frac{1}{a\pi^{1/2}}\right)^p\int_{-1}^1 d^{\,p}y[e^{-\sum_{i=1}^p (y(i)+1/3)^2/(4a^2)}
+e^{-\sum_{i=1}^p (y(i)-1/3)^2/(4a^2)}] ~~~.
\end{equation}

In his paper Lepage compares numerical evaluations of $I_p$ for various $p$ with a target value of unity, which is accurate enough
for his purposes.  However, the programs given here are capable of much higher accuracy
with current computers, so we will need a high
accuracy evaluation of $I_p$ for comparison purposes.  We can get this by noting that the two Gaussians contribute
equally to $I_p$ \big(to see this, set $x \to -x$ in Eq. \eqref{eq:doubgauss2}\big), and each individual Gaussian is the $p$th power of a one dimensional integral $J$,
giving
\begin{align}\label{eq:doubgauss3}
I_p=&J^p~~~,\cr
J=&\frac{1}{2a\pi^{1/2}}\int_{-1}^1dy e^{-(y+1/3)^2/(4a^2)}~~~.\cr
\end{align}
The one-dimensional integral $J$ can be evaluated in terms of error functions or complementary error functions,
\begin{align}\label{eq:gausserrf}
J=&\frac{1}{2}[{\rm erf}\big(1/(3a)\big)+{\rm erf}\big(2/(3a)\big)]~~~,\cr
=&1-\frac{1}{2}[{\rm erfc}\big(1/(3a)\big)+{\rm erfc}\big(2/(3a)\big)]~~~,\cr
\end{align}
but it can also be evaluated numerically to 13 digit accuracy by running the ``cube'' program to a depth of twelve total levels.  Running the ``r'' version of the
programs with parameter values $ip=1$, $llim1=5$, $llim2=7$, $ithinlev=12$ (no thinning, which makes the results independent of $eps$ and $ithinfun$), and $iaccuracy=3,5,7$, we get from all three runs the result
\begin{equation}\label{eq:jvalue}
J=0.9999987857663~~~.
\end{equation}
The statistics for running time (on a MacBook Pro) and the number of function calls for these  runs  are given in Table X.  Running with $iaccuracy=1$  gave only
10 place accuracy with 12 levels, but gave 13 place accuracy when 20 levels were used (which took about a second, rather than hundredths of a second).  Thus, for this
calculation $iaccuracy=3$ is the most cost-effective program.

\begin{table} [t]
\caption{Evaluation of $J$ to 13 place accuracy using 3rd, 5th, 7th order cube routines}
\centering
\begin{tabular}{c c c c}
\hline\hline
$iaccuracy$&${\it|outdiff|}$&$fcncalls$ & $t\_\,final$\\
\hline

3&$10 ^{-16}$ &$0.2\times 10^{5}$ &$<.02$s \\
5&$10 ^{-15}$ &$0.5\times 10^{5}$ &$<.02$s \\
7&$10 ^{-13}$ &$0.9\times 10^{5}$ &$<.02$s \\

\hline
\end{tabular}
\label{table:jpower}
\end{table}

Running a program to raise $J$ to powers then gives the expected results for $I_p$ given
in Table XI, with an uncertainty of 1 in the final decimal place.
\begin{table} [t]
\caption{Evaluation of powers of $J$ to give expected values of $I_p$ to 13 place accuracy}
\centering
\begin{tabular}{ c c }
\hline\hline
$p$&$I_p=J^p$\\
\hline

1&0.9999987857663 \\
2&0.9999975715341 \\
3&0.9999963573033 \\
4&0.9999951430740 \\
5&0.9999939288462 \\
6&0.9999927146199 \\
7&0.9999915003951 \\
8&0.9999902861717 \\
9&0.9999890719498 \\

\hline
\end{tabular}
\label{table:jcalc}
\end{table}

We give in Table XII results for dimensions $p=$ 2, 3, 4, and 5 as obtained from the ``r'' version of the programs on a laptop, and in Table XIII for $p=7,9$ as obtained by running the ``m'' version on a 64 process cluster. (Laptop runs were done on a MacBook Pro and an older Dell Inspiron, and for the latter, the timings were rescaled by a factor 0.49 to give timings for a MacBook Pro. We made cluster runs with 128
 or 64 processes, and for the former, we rescaled the running time to that for 64 processes.
We invite the reader to compare the running times and accuracies summarized in Tables
XII and XIII with those that can be obtained from other  integration programs.)  For all of these runs, where thinning was used, we took $ithinfun=1$. ``Place accuracy'' indicates the decimal place where differences
first appear from the 13 place result in Table XI. To within an order of magnitude, this agrees with the difference $\rm{|outdiff|}$ between the two evaluations of the
integral given by the program.  The values of errsum for these integrals (not shown) were typically one to three orders of magnitude larger than both ${\rm |outdiff|}$ and the actual error, indicating that the local subinterval errors do not all have the same sign.
 Since ${\rm|outdiff|}$ can be smaller than the
actual error, for an unknown integral it cannot be taken as giving the error; in this
case the best way to estimate the error is to run the program with different choices of program options and to use the spread of results to estimate the error.

 From Tables XII and XIII, we see that a minimum of 5 levels is needed to get good accuracy for the double Gaussian example.  With 5 levels, the smallest hypercube
 side is $1/32=0.03125$, small enough to resolve the double Gaussian characteristic scale of $0.1$ in good detail.  On the other hand, with only 4 levels, the minimum
 side is $1/16=0.0625$, making it harder to resolve a scale of $0.1$ and limiting the accuracy to 4 significant figures.  Most of the cluster runs were done without
 thinning, and thus should characterize the accuracy attainable for any function on a unit hypercube with a characteristic scale length of $0.1$.  For serial
 runs done with thinning (Table XII), the reduction in running time was proportional to the reduction in number of function calls, and ranged from a saving
 in the range 30\% to a factor of 3.5, for values of $\epsilon$ which yield the same
 or one place less accuracy as when there is no thinning. For parallel cluster runs done with thinning (Table XIII), the reduction in running time is considerably less than
 the reduction in number of function calls.  This arises from the fact that even though the program initially distributes subregions to processes using a shuffling routine that assigns adjacent subregions in the stack to different processes, some processes get subregions (like ones near the Gaussian peak) that are ``hard''
 and so take longer to finish, as compared with processes that get ``easy'', readily thinned regions near the Gaussian tails.  Consequently, since the final time $t\_final$ records the time when all processes have finished, it is not reduced by thinning in proportion to the number of function calls.

 From Tables XII and XIII, we see that thinning with a value of $\epsilon$ equal to the error level in the runs with no thinning leads, in the double Gaussian examples, to a reduction in accuracy.  This reflects the fact that in the
 double Gaussian case, errsum is typically 2 to 3 orders of magnitude larger than $\rm{|outdiff|}$, indicating that the local errors are not of constant sign, and also errsum is
 2 to 3 orders of magnitude larger than $\epsilon$, indicating that the local thinning condition   is not satisfied in all subregions.  When local subregion errors
 alternate in sign and the thinning condition is not uniformly obeyed, there can be cancelations of errors in the output integrals, leading to improved accuracy
 and a smaller $\rm{|outdiff|}$,  but thinning can then reduce the degree of cancelation and reduce the accuracy.
 Finally, we note that the 6 level run with $iaccuracy=9$ did not give as many significant figures as the corresponding run with $iaccuracy=7$; we believe this is due to  the increased truncation errors associated with running the ninth order routine.  The cluster which we used was more than an order of magnitude slower in running quadruple precision \big(real(16)\big) as opposed to double precision \big(real(8)\big) code, so it was not feasible for us to investigate this further by
 repeating the ninth order 6 level run in quadruple precision. (We did, however, test in quadruple precision that the ninth order routines integrate polynomials of ninth degree or lower to within expected truncation errors.)

 We also studied the double Gaussian example using the ``cubetile'' programs. These integrate over a unit hypercube by integrating, over a single Kuhn simplex, the symmetrized function  that sums over corresponding points of a Kuhn tiling of the
 hypercube.  Results
 for this study are given in Table XIV.  Because of the $p\,!$ symmetrization factor in the number of function calls, the ``cubetile'' programs take longer to run than the
 ``cube'' programs for a corresponding number of subdivision levels.  Because tiling of a hypercube with Kuhn simplexes does not reduce the subregion side length, the
 $p\,!$ increase in number of subregions does not compensate for insufficient resolution when the attainable level number is not large enough.
 This can be seen from the results in Table XIV.  For $p=5$, where 5 levels can be run on the cluster in reasonable time, significantly better results are obtained from
 a 5 level ``cubetile'' run than are obtained from a 5 level ``cube'' run, at the price of a factor of 120 more function calls.  However, for $p=7$ it was not possible to do a 5 level calculation in
 reasonable cluster running time, so we had to settle for 4 levels, which as we saw above has insufficient resolution to give very high accuracy for the double Gaussian
 test problem.  The ``cubetile'' results
 in this case are better than the 4 level ``cube'' results, reflecting the factor of 5040 more function calls, but the 6 place accuracy achieved is not as good as
 what can be achieved, in less running time, by using the ``cube'' program with 5 levels.  We conclude that the ``cubetile'' programs become significantly less useful as the dimension $p$ increases, because of the $p\,!$ symmetrization factor necessitated by  Kuhn tiling.

\begin{table} [t]
\caption{Double Gaussian results using the ``cube'' program for $ip\equiv p=$2 ,3 ,4 ,5 ,7 (timings for MacBook Pro); $levels\equiv llim1+llim2$ and ``place accuracy'' compares to Table XI}
\centering
\begin{tabular}{ c c c c c c c c c c c}
\hline\hline
$ip$ & $iaccuracy$ & $levels$ &$ithinlev$ & $eps$ & $t\_\,final$ & $fcncalls$ & $(outa+outb)/2$ &${\it |outdiff|}$ & place accuracy \\
\hline
2 & 3 & 10 &  no thinning  & -- & 1.2s  & $0.31\times 10^7$  &$0.9999975715340$  & $0.4\times 10^{-12}$ & 13 \\
2 & 5 & 10 & no thinning  & -- & 3.4s  & $0.94\times 10^7$ &0.9999975715339  & $0.3 \times 10^{-14}$ & 13 \\
2 & 5 & 10 & 2  & $10^{-13}$ & 2.1s  & $0.59\times 10^7$ & 0.9999975715340 & $10^{-14}$ & 13 \\
2 & 7 & 10 & no thinning  & -- &  8.5s & $ 0.24 \times 10^8$ &0.9999975715342  & $0.5\times 10^{-12}$& 13 \\
2 & 7 & 10 & 2  & $10^{-13}$ & 2.4s  & $0.68\times 10^7$ &0.9999975715342  &$0.4 \times 10^{-12}$  & 13 \\
\hline
3 & 3 & 7 &  no thinning & -- & 1.7s& $0.39 \times 10^7$& 0.999996358&$0.2 \times 10^{-8}$ & 9 \\
3 & 3 & 9 & 2 & $10^{-13}$ &72s &$0.18 \times 10^9$ &0.999996357305 &$0.9 \times 10^{-11}$ & 12\\
3 & 5 & 7 & 2 & $10^{-9}$ & 2.8s&$0.71 \times 10^7$ &0.999996356 &$0.3\times 10^{-10}$ & 9\\
3 & 5 & 7 & 2 & $10^{-13}$& 4.1s &$0.11 \times 10^8$ &0.99999635730 &$0.2 \times 10^{-10}$& 11\\
3 & 5 & 9 & 2 &$10^{-13}$ &190s &$0.48 \times 10^9$ &0.9999963573032 &$0.3\times 10^{-13}$ &13 \\
3 & 7 & 7 & 2 &$10^{-12}$ & 13s& $0.32\times 10^8$&0.9999963573033 &$0.5\times 10^{-12}$ & 13\\
\hline
4 & 3 & 7 & 2 &$10^{-9}$ & 55s&$0.12\times 10^9$ &0.999995143 &$0.3\times 10^{-8}$ &9 \\
4 & 5 & 7 & 2 &$10^{-13}$ &310s &$0.71\times 10^9$ &0.99999514305 &$0.4 \times 10^{-10}$ & 11 \\
4 & 7 & 5 & 2 &$10^{-7}$& 3.7s &$0.83\times 10^7$ &0.99999510 & $0.3\times 10^{-8}$& 8 \\
4 & 7 & 6 & 2 &$10^{-13}$ &90s &$0.20\times 10^9$ &0.99999514308 &$0.2\times 10^{-10}$ & 11\\
 \hline
5 &3 &5 &2 & $10^{-13}$&6.4s &$0.14\times 10^8$ &0.9999940 &$0.7\times 10^{-6}$ & 7\\
5 &3 &6 &2 &$10^{-13}$ &170s &$0.37\times 10^9$ &0.99999394 &$0.5\times 10^{-7}$ & 8\\
5 &5 &5 &no thinning &-- &49s& $0.10\times 10^9$ &0.99999386 &$0.2\times 10^{-6}$ &7 \\
5& 5&5 &2 &$10^{-13}$ &29s & $0.63\times 10^8$&0.99999386 &$0.2\times 10^{-6}$ &7 \\
5 &7 &5 &no thinning &-- &240s&$0.50\times 10^9$ &0.99999393 &$0.1\times 10^{-7}$  & 8\\
5 &7 &6 & 2&$10^{-9}$& 1700s &$0.34\times 10^{10}$& 0.999993926 & $0.3\times 10^{-10}$ & 9 \\
\hline

7&3 &5 & no thinning&-- &6100s&$0.78\times10^{10}$ &0.999992 &$0.9\times 10^{-6}$ & 6\\
7 &3 &5 &2 &$10^{-9}$ &840s &$0.11\times 10^{10}$ &0.999991 &$0.9\times 10^{-6}$ &6 \\
7 &5 & 5& 2& $10^{-9}$&3500s &$0.62\times 10^{10}$ &0.999991 &$0.2\times 10^{-6}$ & 6\\
7 & 7& 4&no thinning & --&1200s &$0.21\times 10^{10}$ &0.9997 &$0.4\times 10^{-2}$ &3 \\
7&7 &4 &2 &$10^{-9}$ &340s&$0.60\times 10^9$ &0.9995 &$0.4\times 10^{-2}$ & 3 \\
\hline
\end{tabular}
\label{table:cube_laptop}
\end{table}

\begin{table} [t]
\caption{Double Gaussian results using the ``cube'' program for $ip\equiv p =$7, 9 from a 64 process cluster; $levels\equiv llim1+llim2$ and ``place accuracy'' compares to Table XI}
\centering
\begin{tabular}{ c c c c c c c c c c }
\hline\hline
$ip$ & $iaccuracy$ & $levels$ &$ithinlev$ & $eps$ & $t\_\,final$ & $fcncalls$ & $(outa+outb)/2$ &${\it |outdiff|}$ & place accuracy \\
\hline
7&3  & 5 & no thinning & -- & 11s &$0.78\times 10^{10}$  &0.999992  &$0.9\times 10^{-6}$  &6  \\
7&5  & 5 & no thinning & -- & 120s &$0.44\times 10^{11}$  &0.9999913  &$0.5\times 10^{-6}$ & 7 \\
7&7  & 5 & no thinning & --  & 760s &$0.27\times 10^{12}$  &0.99999150  & $0.3\times 10^{-7}$ &8  \\
7&1  & 6 & no thinning & -- & 2900s &$0.52\times 10^{12}$ &0.999993  &$0.4\times 10^{-6}$  & 6 \\
7&3  & 6 & no thinning & -- &4300s  &$0.10\times 10^{13}$&$0.9999915$ &$0.8\times 10^{-7}$    & 7 \\
7&5  & 6 & no thinning & -- & 17000s &$0.56\times 10^{13}$  & 0.99999150 &$0.9\times 10^{-8}$  & 8 \\
7&7  & 6 & no thinning & -- & 98000s &$0.35\times 10^{14}$  & 0.9999915003 &$0.1\times 10^{-9}$  & 10 \\
7&7  & 6 & 2           &$10^{-10}$ &37000s&$0.37\times 10^{13}$ &0.99999148&$0.1 \times 10^{-9}$&  8\\
7&9  & 5 & no thinning & -- & 4000s & $0.14\times 10^{13}$ &0.99999144 & $0.6\times 10^{-7} $ & 8\\
7&9  & 6 & no thinning & -- & 510000s &$0.18 \times 10^{15}$ &0.99999148 & $0.1\times 10^{-7}$&8\\
\hline
9&3  & 5 & no thinning & -- & 7500s &$0.25\times 10^{13}$  &0.999989  & $0.1\times 10^{-5}$ & 6 \\
9&5  & 5 & no thinning & -- & 49000s &$0.17\times 10 ^{14}$ & 0.9999888 &$0.8\times 10^{-6}$  & 7 \\
9&7  & 5 & no thinning & -- & 380000s &$0.13\times 10^{15}$  & 0.99998907 &$0.7\times 10^{-7}$  & 8 \\
9&7  & 5 & 2           &$10^{-8}$ &100000     s& $ 0.52\times 10^{13}  $ & 0.9999885   & $0.7\times 10^{-7}   $ &  7 \\

\hline
\end{tabular}
\label{table:cube_cluster}
\end{table}

\begin{table} [t]
\caption{Double Gaussian results using the ``cubetile'' program with symmetric subdivision, for $ip\equiv p =$5, 7 from a 64 process cluster; $levels\equiv llim1+llim2$ and``place accuracy'' compares to Table XI}
\centering
\begin{tabular}{ c c c c c c c c c c }
\hline\hline
$ip$ & $iaccuracy$ & $levels$ &$ithinlev$ & $eps$ & $t\_\,final$ & $fcncalls$ & $(outa+outb)/2$ &${\it |outdiff|}$ & place accuracy \\
\hline
5&7  & 5 & no thinning & -- &73s  &$ 120\times 0.27\times 10^9$  &0.999993928884  &$0.7\times 10^{-11}$  &12  \\
\hline
7&7  & 4 & no thinning & -- &12000s  &$5040\times 0.93\times 10^9$  &0.999988 &$0.1\times 10^{-5}$ & 6\\

\hline
\end{tabular}
\label{table:cubetile}
\end{table}

Our next two examples illustrate results obtained from the ``simplex'' programs to
evaluate integrals over a standard simplex.  For our first example, we consider an
integral based on the Feynman-Schwinger formula of Eq. \eqref{eq:feynschw}, with
$D_0=1$ and $D_1=...=D_p=a$,
\begin{equation}\label{eq:feynschwexample}
\frac{1}{a^p}=p\,!\int_{\rm standard~simplex}
\frac{1}{[1+(a-1)\big(x(1)+...+x(p)\big)]^{p+1} }~~~.
\end{equation}
Taking $a=0.1$ (and for $p=5$, also $a=0.01$) gives an integral that is sharply peaked on the diagonal hyperplane
$1=x(1)+...+x(p)$ bounding the simplex.  Results for this integral obtained from the
``m'' version of the program, with symmetric subdivision and no thinning on a 64 process cluster, are given in Table XV.
 ``Place accuracy'' indicates the decimal place where differences
first appear from the exact answer $a^{-p}$,  and this correlates well with the difference $\rm{|outdiff|}$ between the two evaluations of the
integral given by the program.  The values of errsum for these integrals (not shown) were nearly identical to $\rm{|outdiff|}$.

As our second simplex example, we consider the polynomial integral \big(c.f. Eq. \eqref{eq:multibeta1} \big)
\begin{equation}\label{eq:polyexample}
\frac{2^{p+1}}{(3p+2)(3p+1)...(p+2)(p+1)}=
p\,!\int_{\rm standard~simplex}[1-x(1)-...-x(p)]^2 \prod_{i=1}^p x(i)^2~~~,
\end{equation}
which is strongly suppressed at all the vertices of the simplex. Running a program
to evaluate the exact answer for this integral on the left hand side of Eq.
\eqref{eq:polyexample}, and then evaluating the integral on the right on a laptop using the ``simplex'' programs, gives the results in Table XVI.

As our final example, we consider the 1 dimensional singular integral
\begin{equation}\label{eq:pi_integral}
\int_0^1 dx \frac{1}{\sqrt{1-x^2}}=\pi/2\simeq 1.57079633~~~.
\end{equation}
In Table XVII we give results for this integral using the ``r'' version of the
``cube'' program, with $p=ip=1$, $llim1=14$ and $llim2=15$ (that is, using the maximum
allowed number of levels), $iaccuracy=5$, $ithinlev=0$
(that is, thinning starts at the outset), and $eps=10^{-10}$, as a function of the
choice of thinning function $ithinfun$.  We see that in this case, $ithinfun=3$ gives
the fastest evaluation, with $ithinfun=2$ next fastest and $ithinfun=1$ the slowest.
This differs from the double Gaussian examples, where $ithinfun=1$ gives better
results than either $ithinfun=2$ or $ithinfun=3$.

\begin{table} [t]
\caption{Feynman-Schwinger integral  for $ip \equiv p=$ 5, 7, 9 from a 64 process cluster, with no thinning;  $levels\equiv llim1+llim2$ and
``place accuracy'' compares to the exact answer $a^{-p}$}
\centering
\begin{tabular}{ c c c c c c c c c c}
\hline\hline
$ip$&$a$ & $iaccuracy$ & $levels$ & $t\_\,final$ & $fcncalls$ & $(outa+outb)/2$ &${\it |outdiff|}$ & place accuracy \\
\hline
\hline
5 &0.1&7 & 5  &0.39s  &$0.27\times 10^9$ &$0.99998\times 10^5$  &0.3   & 5 \\
5&0.01 &7 & 9   &52000s  &$0.22\times 10^{14}$  &$0.999998\times 10^{10}$  &$0.3\times 10^4$    & 6 \\
\hline
7&0.1 &7& 5  &160s    & $0.12\times 10^{12}$  &$0.99995\times 10^7$    & $0.2\times 10^3$  & 5 \\
7&0.1 &7& 6 & 20000s & $0.15\times 10^{14}$ &$0.9999995\times 10^7$ &$2.$ &  7   \\
\hline
9&0.1  &7  &5    & 56000s &$0.48\times 10^{14}$  &$0.9999\times 10^9$  &$0.3\times 10^5$   &4  \\

\hline
\end{tabular}
\label{table:feynschw}
\end{table}

\begin{table} [t]
\caption{Polynomial integral for $ip\equiv p=$ 4, 5 (timings for a MacBook Pro);
$levels\equiv llim1+llim2$ and
``place accuracy'' compares to the exact answer on the left hand side of
Eq. \eqref{eq:polyexample}}
\centering

\begin{tabular}{ c c c c c c c c c}
\hline\hline
$ip$ & $iaccuracy$ & $levels$ & $t\_\,final$ & $fcncalls$ & $(outa+outb)/2$ & exact answer & ${\it|outdiff|}$ & place accuracy \\

\hline

4&5 &6 &8.7s &$0.63\times 10^8$ &$0.88095326\times 10^{-8}$ &$0.8809532619056\times 10^{-8} $&$0.1\times 10^{-17}$ & 8\\
4&5 &8 &2200s &$0.16\times 10^{11}$ &$0.880953261905 \times10^{-8}$&$0.8809532619056\times 10^{-8}$ &$0.3\times 10^{-21}$ & 12\\
\hline
5&7 & 5&39s &$0.27\times 10^9$ &$0.21591990\times 10^{-10}$ &$0.2159199171337\times 10^{-10}$ &$0.2\times 10^{-18}$ &8 \\
5&7& 6 &1300s &$0.86\times 10^{10}$&$0.2159199171\times 10^{-10}$ &$0.2159199171337\times 10^{-10}$ &$0.7\times 10^{-21}$ &10 \\

\hline
\end{tabular}
\label{table:polycalc}
\end{table}

\begin{table} [t]
\caption{Evaluation of a one dimensional singular integral with different thinning
functions.  Running times were all less than $0.1s$; ``place accuracy'' compares
to the exact answer }
\centering
\begin{tabular}{c c c c c }
\hline\hline
$ithinfun $ & $fcncalls $ & $(outa+outb)/2 $ & ${\it |outdiff|} $ & place accuracy\\
\hline
 1 &$0.24 \times 10^6$ & $1.570778$  & $0.35\times10^{-5}$ &  5 \\
2 &$0.13 \times 10^5$ &$1.570778$   &$0.35\times 10^{-5}$  &  5 \\
3   &$0.45\times 10^4$ &$1.570777$   & $0.43\times 10^{-5}$ & 5  \\

\hline
\end{tabular}
\label{table:thinfuntest}
\end{table}

\subsection{Programming extensions and open questions}

There are a number of possible extensions of the programs that could be
pursued in the future.  (1) The MPI version of the programs could be
rewritten to include redistribution of the process workload after
each level $\ell$ of the second stage. This would make the reduction in
running time when using thinning track more closely with the reduction
in the number of function calls.  (2) The multistage strategy could
be extended to a third (or more) stages, by not harvesting the subregions
that fail to obey the thinning condition at the end of the second stage,
but instead writing them to a memory device which is then read sequentially
by a third stage, etc.  (3) One could build in an option of permuting
the simplex vertices at the start of the simplex programs, which gives a
different subdivision, and therefore a different evaluation of the integral
for use in estimating errors. (4) Finally, we remark that the same subdivision,
thinning, and staging strategies that we have used will apply with any integration
formulas that give two different estimates of the answer from each subregion,
not just the parameterized moment fitting formulas that we developed in Secs.
VII and IX.

There are also a number of mathematical questions that we have left open.
(1)  We found numerical evidence that symmetric subdivision of a
standard simplex obeys the bound of Eq. \eqref{eq:lstand1} for
reduction of side length, and that after $\ell$ symmetric
(recursive) subdivisions, the resulting subsimplexes each fit within
a hypercube of side $1/2^{\ell}$ ($1/2^{\ell-1}$).  We do not have a
proof of these conjectures, but have assumed them true in
constructing the programs.    (2) Given the regularities in the
construction of parameterized fifth, seventh, and ninth order
integration formulas for the simplex and hypercube cases, it would
be of interest to try to find a general all-orders rule for these.
(3) We have not addressed the question of analytic error estimates
for the parameterized integration formulas. (4) We have not
addressed in any systematic way the question of deciding which
thinning function is optimal for a given choice of integrand.
(5) It would be of interest to study the systematics, in the moment
fitting method, of the tradeoff between the number of parameters
that are fixed by appropriate conditions, and the number of
function calls.

\section{Acknowledgements}

I wish to thank the School of Natural Sciences computing staff,  Prentice Bisbal, Kathleen Cooper, Christopher McCafferty,
and James Stephens, for their helpful support and advice throughout this project. I also want to acknowledge several helpful conversations with
Prentice  Bisbal about Fortran language features, and to thank Susan Higgins for drawing the figures.  I am grateful to Herman D. Knoble for
permission to use in the hypercube tiling routines his program BestLex (which is in the public domain and is not covered by the copyright for this book).
This work was partially supported by the Department of Energy under grant DE-FG02-90ER40542.  I also wish to acknowledge the hospitality of the Aspen Center for Physics during the summers of 2009 and 2010.

\section{References}

Note: These references make no pretense to completeness; they list what I have found useful in constructing the algorithms discussed in this book.   For research groups with a continuing program, I have listed only recent publications that contain earlier references.
\bigskip

Cools, R. and Haegemans, A. (2003) Algorithm 824: CUBPACK: A Package for Automatic Cubature; Framework Description. {\it ACM Trans. Math.
Software} {\bf 29}, 287-296.

Dejnakarintra, M. and Banjerdpongchai (undated), D. An Algorithm for Computing the Analytical Inverse of the Vandermonde Matrix.  Searchable on-line.

Edelsbrunner, H.  and Grayson, D. R.  (2000). Edgewise Subdivision of a Simplex.  {\it Discrete Comput. Geom.} {\bf 24}, 707-719.

Genz, A. and Cools, R. (2003) An Adaptive Numerical Cubature Algorithm for Simplices  {\it ACM Trans. Math. Software} {\bf 29}, 297-308.

Good, I. J. and Gaskins, R. A. (1969)  Centroid Method of Integration. {\it Nature} {\bf 222}, 697-698.

Good, I. J. and Gaskins, R. A. (1971) The Centroid Method of Numerical Integration. {\it Numer. Math.} {\bf 16}, 343-359.

Grundmann, A. and M\"oller, N. M. (1978) Invariant Integration Formulas for the $n$-Simplex by Combinatorial Methods. {\it Siam J. Numer. Anal.}
{\bf 15}, 282-290.

Hahn, T. (2005) CUBA- a library for multidimensional numerical integration. arXiv:hep-ph/0404043.

Heinen, J. A. and Niederjohn, R. J. (1997) Comments on ``Inversion of the VanderMonde Matrix''. {\it IEEE Signal Process. Lett.} {\bf 4}, 115.

Kahaner, D. K. and Wells (1979), M. B. An Experimental Algorithm for $N$-Dimensional Adaptive Quadrature. {\it ACM Trans. Math. Software} {\bf 5}, 86-96.

Knoble, H. D. (1995), website download of BestLex subroutine.

Krommer, A. R. and Ueberhuber, C. W. (1994).  {\it Numerical Integration on Advanced Computer Systems},  Lecture Notes in Computer Science 848. (Berlin: Springer-Verlag),  p183.

Kuhn, H. W. (1960). Some combinatorial Lemmas in Topology.  {\it IBM J. Res. and Dev.} {\bf 4}, 518-524.

Lepage, G. P. (1978). A New Algorithm for Adaptive Multidimensional Integration. {\it J. Comp. Phys.} {\bf 27}, 192-203.

Lyness, J. N. (1965) Symmetric Integration Rules for Hypercubes II.  Rule Projection and Rule Extension.  {\it Math. Comp.} {\bf 19}, 394-407.

McKeeman, W. M. (1962).  Algorithm 145: Adaptive numerical integration by Simpson's rule. {\it Commun. ACM} {\bf 5}, 604.

McNamee, J. and Stenger, F. (1967). Construction of Fully Symmetric Numerical Integration Formulas. {\it Num. Math.} {\bf 10}, 327-344.

Moore, D. (1992) Subdividing Simplices, in {\it Graphics Gems III}, D. Kirk ed. (Boston: Academic Press), pp. 244-249 and pp. 534-535. See also Moore, D. (1992), Simplicial Mesh Generation with Applications, Cornell University dissertation.

Neagoe, V.-E. Inversion of the Van der Monde Matrix. {\it IEEE Signal Process. Lett.} {\bf 3}, 119-120.

Osborne, M. R. (2001) {\it Simplicial Algorithms for Minimizing Polyhedral Functions} (Cambridge: Cambridge U. Press), p. 5.

Plaza, A. (2007). The eight-tetrahedra longest-edge partition and Kuhn triangulations.  {\it Comp. \& Math. with Applications} {\bf 54}, 426-433,
Fig. 1.

Pontryagin, L. S. (1952) {\it Foundations of Combinatorial Topology}, English translation.  (Rochester: Graylock Press), pp. 10-12.

Press, W. H., Teukolsky, S. A., Vetterling, W. T., and Flannery, B. P. (1992) {\it Numerical Recipes in Fortran}, pp. 82-85.

Sch\"urer, R. (2008) HIntLib Manual, available on-line at:  mint.sbg.ac.at/HIntLib/manual.pdf~.

Stroud, A. H. (1971) {\it Approximate Calculation of Multiple Integrals}  (Englewood Cliffs: Prentice-Hall).

Ueberhuber, C. W. (1995) {\it Numerical Computation 2:  Methods, Software, Analysis} (Berlin: Springer-Verlag), pp. 161-162.

Wikipedia articles on: {\it Adaptive Simpson's method}, {\it Barycentric coordinates},  {\it Dirichlet distribution}, {\it Gaussian quadrature}, {\it Hypercube}, {\it Polynomial long division}, {\it Simplex}.

\vfill\break

\section{Contents of  programs in directories}

Each file listed in this summary contains multiple programs, each of which begins with comment lines describing its function.
\begin{itemize}
\item The simplex programs take as base region the standard simplex of Eq. \eqref{eq:stdsimplex}.  \big(When used as part of the cubetile
programs, the base region is the Kuhn simplex of Eq. \eqref{eq:kuhnsimplex}.\big)
\item The cubetile programs take as base region the side 1 hypercube of Eq. \eqref{eq:hyperintegral}.
\item The cube programs take as base region the half-side 1 (i.e., side 2) hypercube of Eq. \eqref{eq:hyperbase}.
\item The numbers after simplex or cube indicate the integration orders that are included.
\end{itemize}

\subsection{Directory simplex123}
This directory contains:

\begin{itemize}
\item The subprogram file simplexsubs123.for.
\item Main program  files  simplexmain123.for, cubetilemain123.for.
\item Recirculating main program files  simplexmain123r.for, cubetilemain123r.for.
 \item MPI parallel main program files simplexmain123m.for, cubetilemain123m.for.
\end{itemize}

\subsection{Directory simplex4}
This directory contains:

\begin{itemize}
\item The subprogram file simplexsubs4.for.
\item Main program  files  simplexmain4.for, cubetilemain4.for.
\item Recirculating main program files  simplexmain4r.for, cubetilemain4r.for.
 \item MPI parallel main program files simplexmain4m.for, cubetilemain4m.for.
\end{itemize}

\subsection{Directory simplex579}
This directory contains:
\begin{itemize}
\item The subprogram file simplexsubs579.for.
\item Main program files simplexmain579.for, cubetilemain579.for.
\item Recirculating main program files simplexmain579r.for, cubetilemain579r.for.
\item MPI parallel main program files simplexmain579m.for, cubetilemain579m.for.
\end{itemize}

\subsection{Directory simplex579\_16}
This directory contains:
\begin{itemize}
\item The subprogram file simplexsubs579\_16.for.
\item Main program files simplexmain579\_16.for, cubetilemain579\_16.for.
\item Recirculating main program files simplexmain579\_16r.for, cubetilemain579\_16r.for.
\item MPI parallel main program files simplexmain579\_16m.for, cubetilemain579\_16m.for.
\end{itemize}

\subsection{Directory cube13}
This directory contains:
\begin{itemize}
\item  The subprogram file cubesubs13.for.
\item  Main program file cubemain13.for.
\item Recirculating main program file  cubemain13r.for.
\item MPI parallel main program file cubemain13m.for.
\end{itemize}

\subsection{Directory cube579}
This directory contains:
\begin{itemize}
\item  The subprogram file cubesubs579.for.
\item  Main program file cubemain579.for.
\item Recirculating main program file  cubemain579r.for.
\item MPI parallel main program file cubemain579m.for.
\end{itemize}

\subsection{Directory cube579\_16}
This directory contains:
 \begin{itemize}
\item  The subprogram file cubesubs579\_16.for.
\item  Main program file cubemain579\_16.for.
\item Recirculating main program file  cubemain579\_16r.for.
\item MPI parallel main program file cubemain579\_16m.for.
\end{itemize}

\end{document}